\documentclass{aa}  

\usepackage{graphicx}
\usepackage{txfonts}
\usepackage[draft]{hyperref}
\begin{document}

   \title{The GAPS programme at TNG
   \thanks{Based on observations made with the Italian {\it Telescopio Nazionale Galileo} (TNG) operated by the {\it Fundaci\'on Galileo Galilei} (FGG) of the {\it Istituto Nazionale di Astrofisica} (INAF) at the  {\it Observatorio del Roque de los Muchachos} (La Palma, Canary Islands, Spain).}
  % \fnmsep \thanks{
  % Tables ~XXX to ~ XXX
  % are only available in electronic form
  % at the CDS via anonymous ftp to cdsarc.u-strasbg.fr (130.79.128.5)
  % or via http://cdsweb.u-strasbg.fr/cgi-bin/qcat?J/A+A/
  %    }
     }
   \subtitle{XLIII. A massive brown dwarf orbiting the active M dwarf TOI-5375} %# star}

   \author{J. Maldonado
          \inst{1}
          \and A. Petralia\inst{1}
	  \and G. Mantovan\inst{2}
	  \and M. Rainer\inst{3}
	  \and A. F. Lanza\inst{4}
	  \and C. Di Maio\inst{5,1}
	  \and S. Colombo\inst{1}
	  \and D. Nardiello\inst{6}
	  \and S. Benatti\inst{1}
	  \and L. Borsato\inst{6}
	  \and I. Carleo\inst{7}
	  \and S. Desidera\inst{6}
	  \and G. Micela\inst{1}
	  \and V. Nascimbeni\inst{2, 6}
	  \and L. Malavolta\inst{2}
	  \and M. Damasso\inst{8}
	  \and A. Sozzetti\inst{8}
	  \and L. Affer\inst{1}
	  \and K. Biazzo\inst{9}
	  \and A. Bignamini\inst{10}
	  \and A. S. Bonomo\inst{8}
	  \and F. Borsa\inst{3}
	  \and M. B. Lund\inst{11}
	  \and L. Mancini\inst{12,13,8}
	  \and E. Molinari\inst{14}
	  \and M. Molinaro\inst{10}
%	  \fnmsep\thanks{Just to show the usage
%          of the elements in the author field}
          }

   \institute{INAF - Osservatorio Astronomico di Palermo, Piazza del Parlamento 1, 90134 Palermo, Italy\\
              \email{jesus.maldonado@inaf.it}
    % 2	      
    \and Dipartimento di Fisica e Astronomia ``G.Galilei'', Universit\`a  degli Studi di Padova, Vicolo dell'Osservatorio 3, 35122 Padova, Italy 	      
    % 3
    \and INAF - Osservatorio Astronomico di Brera, Via E. Bianchi 46, 23807 Merate, Italy
    % 4
    \and INAF - Osservatorio Astrofisico di Catania, Via S. Sofia 78, 95123, Catania, Italy
    % 5
    \and Universit\`a degli Studi di Palermo, Dipartimento di Fisica e Chimica, via Archirafi 36, Palermo, Italy
    % 6
    \and INAF - Osservatorio Astronomico di Padova, vicolo dell'Osservatorio 5, 35122 Padova, Italy
    % 7
    \and Instituto de Astrof\'isica de Canarias, V\'ia L\'actea, 38205 La Laguna, Spain 
    % 8
    \and INAF - Osservatorio Astrofisico di Torino, Via Osservatorio, 20, I-10025 Pino Torinese To, Italy
    % 9
    \and INAF - Osservatorio Astronomico di Roma, Via Frascati 33, 00040 Monte Porzio Catone (RM), Italy
    % 10
    \and INAF - Osservatorio Astronomico di Trieste, via Tiepolo 11, 34143 Trieste
    % 11
    \and NASA Exoplanet Science Institute, Caltech/IPAC, Mail Code 100-22, 1200 E. California Boulevard, Pasadena, CA 91125, USA
    % 12
    \and Department of Physics, University of Rome ``Tor Vergata'', Via della Ricerca Scientifica 1, 00133 Rome, Italy
    % 13
    \and Max Planck Institute for Astronomy, K{\"o}nigstuhl 17, 69117 Heidelberg, Germany
    % 14
    \and INAF - Osservatorio di Cagliari, via della Scienza 5, 09047 Selargius, Italy
 %        \and
 %            University of Alexandria, Department of Geography, ...\\
 %            \email{c.ptolemy@hipparch.uheaven.space}
 %            \thanks{The university of heaven temporarily does not
 %                    accept e-mails}
             }

   \date{Received September 15, 1996; accepted March 16, 1997}

% \abstract{}{}{}{}{}
% 5 {} token are mandatory
 
  \abstract
   { Massive substellar companions orbiting active low-mass stars are rare. They, however, offer an excellent opportunity to study the main mechanisms
   involved in the formation and evolution of substellar objects.
   } 
   { We aim to unravel the physical nature
   of the transit signal observed by the TESS space mission on the 
   %a companion orbiting the %young (age $\sim$ 100 Myr)
   active M dwarf TOI-5375. % observed by the TESS space mission.
   }
   {We analysed the available TESS photometric data as well as high-resolution (R $\sim$ 115000) HARPS-N spectra.
    We combined these data to characterise the star TOI-5375 and to disentangle signals related to stellar activity from 
    the companion transit signal in the light-curve data. 
    We ran an MCMC analysis to derive the orbital solution
    and apply state-of-the-art Gaussian process regression to deal with the stellar activity signal.
%    techniques to model the signals present in the data.
   }
   {We reveal the presence of a companion in the brown dwarf / very-low-mass star boundary orbiting around the star TOI-5375. The best-fit model % describing the newly found
    %companion, 
    corresponds to a companion with an orbital period of  1.721564 $\pm$ 10$^{\rm -6}$ d, a mass of 77  $\pm$ 8 $M_{\rm J}$ and a radius of 0.99   $\pm$ 0.16 $R_{\rm J}$.
    %To the best of our knowledge, it is the most massive companion orbiting around a young M dwarf so far.
%    We derive a rotation period for the host
%    star of \bf 1.9692 $\pm$ 0.0004 d, and we conclude that the star is very close to synchronising its rotation with
%    the orbital period of the companion. %brown dwarf.
   }
   { We derive a rotation period for the host
       star of 1.9692 $\pm$ 0.0004 d, and we conclude that the star is very close to synchronising its rotation with
           the orbital period of the companion.
   }


   \maketitle

%
% _______________________________________________________________
\section{Introduction}\label{introduction}
% ---------------------------------------------------------------

 Almost thirty years after the discovery of the first exoplanets \citep{1992Natur.355..145W,1995Natur.378..355M}, 
 thousands of planetary systems have been discovered showing an astonishing diversity
 of orbits, compositions, and host stellar properties. 
 Understanding the origin of this diversity is one of the major goals of modern 
 Astrophysics. 
 Young stars are valuable targets that offer the unique opportunity 
 to provide observational constraints on the timescales involved in the formation and
 migration of planets in close orbits, the physical evolution of the planets, as well
 as atmospheric evaporation due to the strong X and UV irradiation of the host star. % {\bf references?}.
 Planets have been found in young stellar clusters \citep[e.g.][]{2016A&A...588A.118M,2017AJ....154..224R},
 young stellar kinematic groups  \citep[e.g.][]{2019A&A...630A..81B,2019ApJ...880L..17N,2022A&A...664A.163N}
 as well as single young stars \citep[e.g.][]{2020Natur.582..497P,2021A&A...645A..71C}. %   \citep[e.g.][]{2019A&A...630A..81B,2019ApJ...880L..17N}.
 However, the detection and characterisation of planets around young stars
 via radial velocity or transit methods
 are challenging
 due to the high-activity levels of the host stars %\citep[e.g.][]{2020A&A...642A.133D}
 and some previously claimed planets have been retracted after detailed analysis  \citep[e.g.][]{2018A&A...613A..50C,2020A&A...642A.133D}.
 Furthermore, high rotation rates or variability due to pulsations might also challenge the detection of
 planets around young stars. 
 As a consequence, the sample of known, well-characterised young exoplanets is still
 rather limited.

 On the other hand, M dwarfs have been recognised as promising targets to search for small, rocky
 planets with the potential capabilities of host life \citep[e.g.][]{2013ApJ...767...95D,2015ApJ...807...45D}.
 While small planets orbiting around M dwarfs seem to be abundant \citep[e.g][]{2012ApJS..201...15H,2013A&A...549A.109B,2014MNRAS.441.1545T},
 results from radial velocity surveys show that, unlike their FGK counterparts, the frequency of 
 gas-giant planets orbiting around low-mass stars is low
 \citep[][]{2006ApJ...649..436E,2006PASP..118.1685B,2008PASP..120..531C,2022A&A...664A..65P}.
 This result might be understood in the framework of core-accretion models for planet formation.
 Within this scenario, the low-mass protoplanetary disc of an M dwarf %and its short lifespan (due to the high-radiation level
 %of the host star)
 conspires against the formation of gas-giant planets. 
 However, using data from  the Transiting Exoplanet Survey Satellite \citep[TESS,][]{2015JATIS...1a4003R},
 a small number of transiting gas-giant planets orbiting around  low-mass stars has been discovered.
 Some examples include TOI-1728 b \citep{2020ApJ...899...29K}, TOI-442 b \citep{2020A&A...644A.127D} and TOI-3757 b
 \citep{2022arXiv220911160K}, among others.
 All these planets have close-in orbits and constitute valuable targets to help us in our understanding of planet formation and migration.
 While it was proposed early on that close-in gas-giant planets should form beyond the snowline and then experience
 subsequent inward migration \citep[e.g.][]{1996Natur.380..606L}, the in-situ formation of gas-giant planets has been recently revisited
 \citep{2016ApJ...829..114B,2016ApJ...817L..17B,2018A&A...612A..93M}.
 In addition to planetary companions, M dwarfs can also host brown dwarfs in close orbits.
 %In addition, 
 Recent works have discussed differences in the period-eccentricity distribution of massive ($M_{\rm P}$ $>$ 45 $M_{\rm J}$)
  and low-mass brown dwarf ($M_{\rm P}$ $<$ 45 $M_{\rm J}$) companions, as well as in the chemical composition of their respective host stars 
  \citep[][]{2014MNRAS.439.2781M,2014A&A...566A..83M,2017A&A...602A..38M}.
  It has been shown  that massive brown dwarfs have longer periods and larger eccentricities and their host stars do not show the so-called metal-rich signature.
  According to this scenario, massive brown dwarfs should be mainly formed in the same way as low-mass stars (by the fragmentation of a molecular cloud).
  On the other hand, low-mass brown dwarfs might be formed at high-metallicities by core-accretion % in metal-rich environments
  or by gravitational instability in low-metal environments \citep{2017A&A...602A..38M}.
 Furthermore, since these companions % in massive and close-in planets around low-mass stars
 are massive and close to their host stars, 
 tidal interactions between the companion and the host star are expected to constrain the evolution of the system
 \citep[e.g.][]{1988ApJ...326..256M,2008EAS....29....1M,2020MNRAS.498.2270B,2022arXiv220911375A}.

 In this paper, we report the validation and the confirmation of a transiting low-mass companion ($M_{\rm p}$ $\sim$ 77 $M_{\rm J}$) around the very active
 M0.5 dwarf TOI-5375 %with a period of 1.72 d
 by using space-based photometric observations from TESS combined with high-precision radial velocities from the ground.
  Given the high activity levels of TOI-5375, this star has a good chance to be young.
 TESS identified a candidate with a radius of 1.04 $\pm$ 0.03 $R_{\rm J}$. A companion with such a radius, if confirmed, 
 could be a giant planet, a brown dwarf or even a very low-mass star. 
 As discussed, a giant planet transiting an active
 %young (as suggested by the activity level and rotation) 
 M dwarf is of special interest.
 Therefore, we included TOI-5375 in the sample of the GAPS-Young program \citep{2020A&A...638A...5C} in order to perform the validation and the
 mass determination of the transiting candidate. 
 Our radial velocity (RV) monitoring shows that the companion is actually a massive brown dwarf or a very low mass star. The system represents a remarkable case
% The TOI-5375 planetary system constitutes a remarkable case 
 study as it is observed in the evolutionary phase in which
 the star is close to synchronising its rotation with the orbital period of the companion. %ebrown dwarf companion. 

 This paper is organised as follows. %Section ~\ref{host_star} reviews the stellar properties of TOI-5375.
 The photometric and spectroscopic data analysed in this paper are presented in Sect.~\ref{observations}.
 We note that the star is very faint for high spectral resolution observations with 4m-class telescope, resulting in low signal-to-noise ratio (S/N) spectra.
 Therefore, the extraction of RVs in this regime is discussed in this section by comparing different methods. % xxxxx  
 Section ~\ref{host_star} reviews the stellar properties of TOI-5375.
 Sect.~\ref{analysis} describes the modelling
 of the photometric and radial velocity variations. The results are discussed at length %set in the context of planetary systems
 in Sect.~\ref{discussion}. Our conclusions follow in Sect.~\ref{conclusions}.

% +++++++++++++++++++++++++++++++++++++++++++++++++++++++++++++++
\section{Observations}\label{observations}
\subsection{TESS photometry}
% +++++++++++++++++++++++++++++++++++++++++++++++++++++++++++++++

\begin{figure}[htb]
\centering
\includegraphics[scale=0.6]{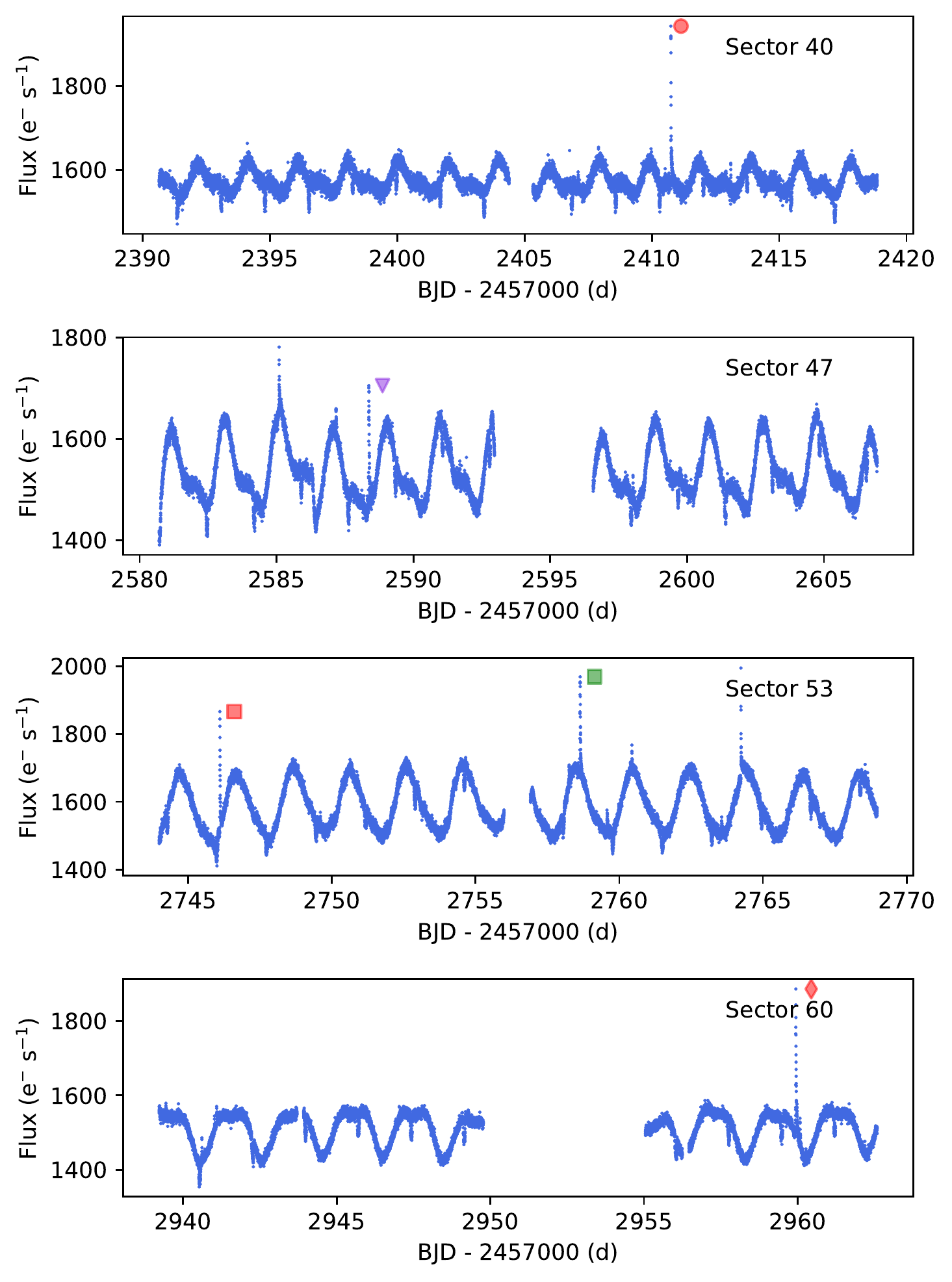}
\caption{Original TESS photometry time series.
The energetic events discussed in Sect.~\ref{flares} are shown with
different symbols according to their Sector.
Within each Sector, different colours indicate that the event occurs at a different
period (when phase-folded to the stellar rotation period).
}
\label{tess_time_series}
\end{figure}

TOI-5375 was observed photometrically by TESS %mission 
%(Transiting Exoplanet Survey Satellite, \citet{2015JATIS...1a4003R}) 
in short-cadence mode (120 seconds) in sectors 40, 47, 53, and 60.
In this work, we use the light curves extracted from all these datasets.
%{\bf It was also observed in sectors 20 and 27, but only HLSP
We use the data corrected for time-correlated instrumental signatures,
that is, the PDCSAP flux. The corresponding light curve is shown in figure~\ref{tess_time_series}.
 We note that the star was also observed in sectors 20 and 26.
However, only HLSP data is available from these sectors. Therefore, 
unless otherwise noticed, data from sectors 20 and 26 are not included
in the analysis.

\begin{figure}[htb]
\centering
\includegraphics[scale=0.5]{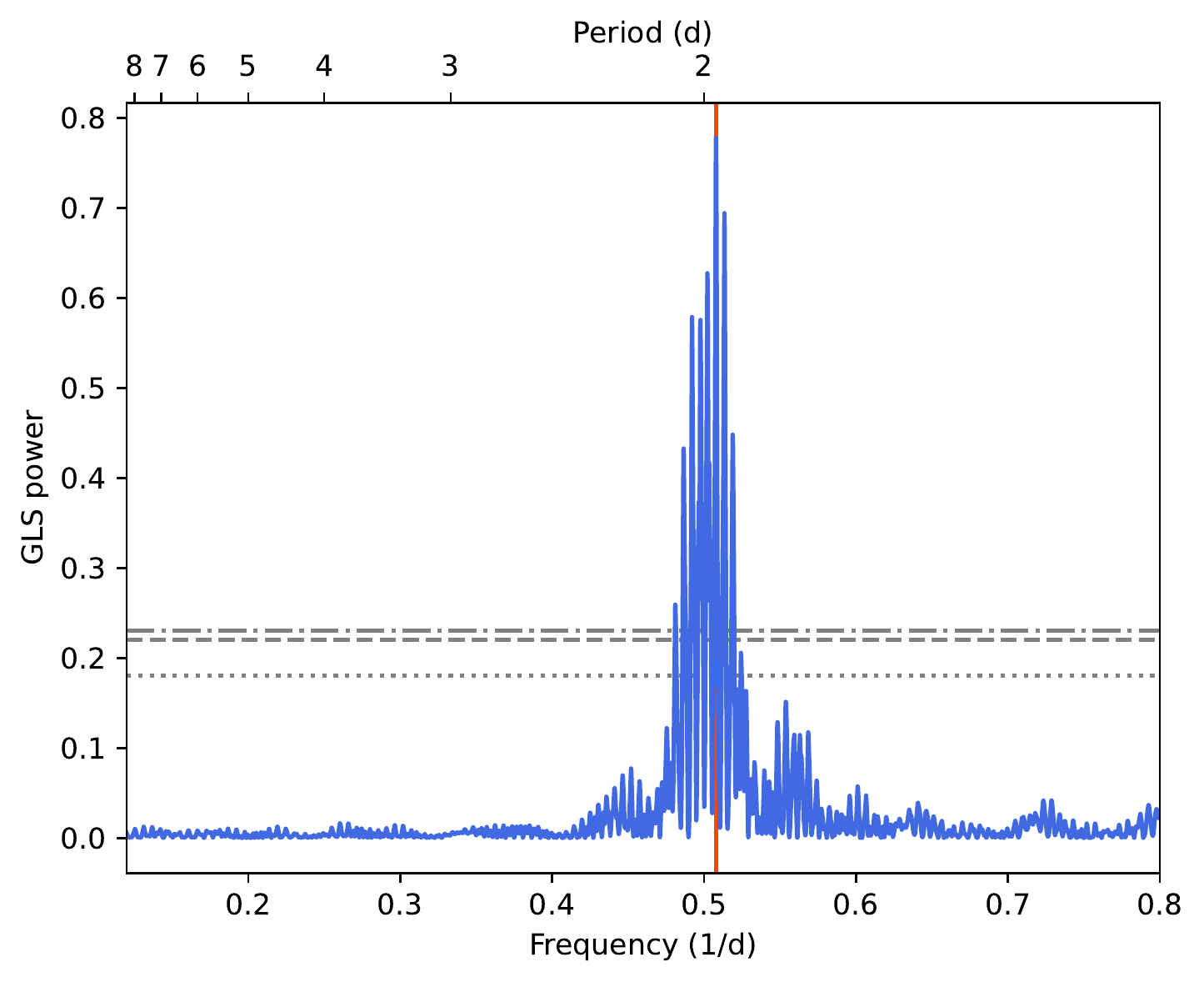}
\caption{GLS periodogram of the TESS photometry dataset.
Values corresponding to a False Alarm Probability (FAP) of 10, 1, and 0.1\% are shown with horizontal grey lines. The vertical line indicates the period
at 1.9692 d.}
\label{tess_gls}
\end{figure}

The light curve shows a clear modulation that we attribute to the rotation of the star. 
A search for periodicities 
by using the generalised Lomb-Scargle periodogram (GLS, \cite{2009A&A...496..577Z})
reveals a clear peak at  1.9692 $\pm$ 0.0004 d % 1.9691 $\pm$ 0.0003 d %%%   1.969051 $\pm$ 0.000333 d 
that corresponds
to the rotation period of the star, as shown in Fig.~\ref{tess_gls}.
Superimposed on the rotational modulation, the light curve shows periodic transit-like signals, due to a possible substellar companion.
%that might be the signal of a possible substellar companion. 

% >>>>>>>>>>>>>>>>>>>>>>>>>>>>>>>>>>>>>>>>>>>>>>>>>>>>
% Figure 2: Light curve
% <<<<<<<<<<<<<<<<<<<<<<<<<<<<<<<<<<<<<<<<<<<<<<<<<<<<

% >>>>>>>>>>>>>>>>>>>>>>>>>>>>>>>>>>>>>>>>>>>>>>>>>>>>
% Figure: TESS photometry periodogram
% <<<<<<<<<<<<<<<<<<<<<<<<<<<<<<<<<<<<<<<<<<<<<<<<<<<<

% +++++++++++++++++++++++++++++++++++++++++++++++++++++++++++++++
\subsection{Ground-Based Photometry}
% +++++++++++++++++++++++++++++++++++++++++++++++++++++++++++++++
 
 Two transits of TOI-5375 were observed on October 25th and December 19th, 2022 by the TASTE program,
 a long-term campaign to monitor transiting planets \citep{2011A&A...527A..85N}.
 The AFOSC camera, mounted on the Copernico 1.82-m telescope at the Asiago Astrophysical
 Observatory in northern Italy, was purposely defocused up to $8"$~FWHM to avoid 
 saturation and increase the photometric accuracy. Frames were acquired through
 a Sloan $r'$ filter and with a constant exposure time of 15~s. A pre-selected set of suitable
 reference stars was always imaged in the same field of view to perform differential photometry.
 Unfortunately, during the first transit, the sky transparency was highly variable through the series and thick clouds passed
 by during the second half of the transit (egress included). %, leaving us with a subset of 384 images
% with an acceptable S/N ratio.

 We reduced the Asiago data with the STARSKY code \citep{2013A&A...549A..30N}, a software
 pipeline specifically designed to perform differential photometry on defocused images. 
 After a standard correction through bias and flat-field frames, the size of the circular apertures
 and the weights assigned to each reference star are automatically chosen by the code to minimise the
 photometric scatter of our target. The time stamps were uniformly converted to the BJD-TDB standard
 following the prescription by \cite{2010PASP..122..935E}. The final (unbinned) light curves are shown
 in Fig.~\ref{tess_asiago}. 

% +++++++++++++++++++++++++++++++++++++++++++++++++++++++++++++++
\subsection{Radial velocity observations}
% +++++++++++++++++++++++++++++++++++++++++++++++++++++++++++++++

\begin{figure}[htb]
\centering
\includegraphics[scale=0.55]{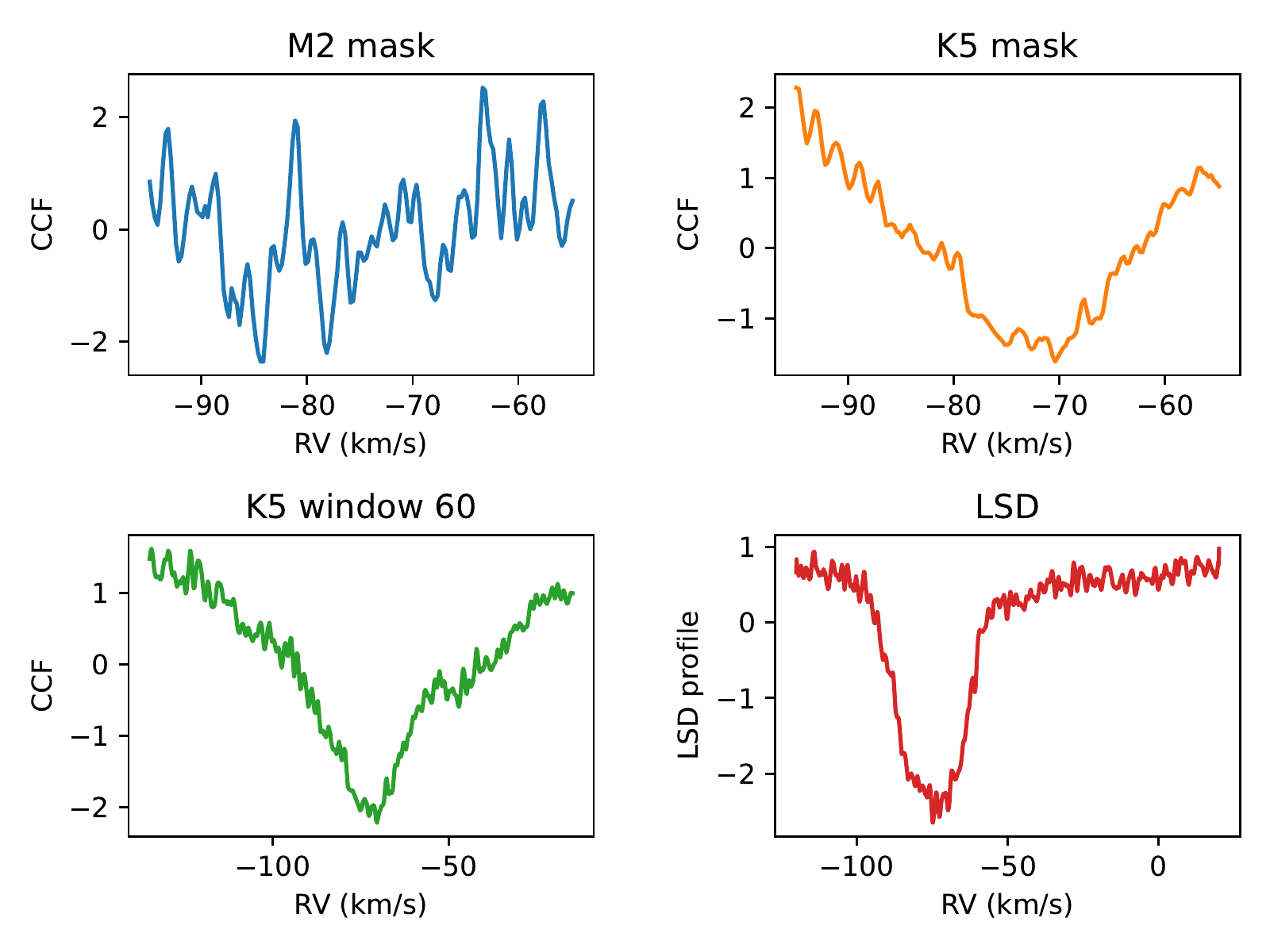}
\caption{Examples of standardised (mean subtracted and division by the standard deviation) CCF profiles obtained by the DRS using different binary masks as well
as the LSD profile.} %technique.}
\label{rv_ex_plot}
\end{figure}

 A total of 9 HARPS-N \citep{2012SPIE.8446E..1VC} observations of TOI-5375 were collected in the period between
 BJD = 2459704 d  and BJD = 2459718 d (May 4 - 18, 2022).
 HARPS-N spectra cover the wavelength range 383–693 nm with a resolving power
 of R $\sim$ 115 000.
 The observations were performed within the framework of the Global Architecture of Planetary Systems (GAPS)
 project \citep{2013A&A...554A..28C}. 
 Data were reduced using
 the HARPS-N Data Reduction Software
  \citep[DRS,][]{2007A&A...468.1115L,2021A&A...648A.103D}
% HARPS-N DRS %Data Reduction Software \citep[DRS][]{2007A&A...468.1115L,2021A&A...648A.103D}
 that takes into account the usual procedures involved in \'echelle spectra reduction
 (bias level, flat-fielding, order extraction and wavelength calibration, as well as merging of individual orders).
 Radial velocities are computed by cross-correlating the spectra of the target star
 with an optimised binary mask \citep{1996A&AS..119..373B,2002A&A...388..632P}.
 However, this procedure might not be optimal for a star like TOI-5375.
 To start with, TOI-5375 is an M dwarf-type star and the spectra of M dwarfs suffer from heavy blends.
 Side-lobes appear in the CCF that might affect the RV precision \citep[e.g.][]{2020ExA....49...73R}.
 In addition, TOI-5375 is a very faint star (V $\sim$ 14 mag). This might have an effect on the selection of an appropriate mask
 to perform the CCF. In the M2 mask, all lines have roughly a similar weight, so weak (not deep enough) lines are lost in the not-well-defined continuum
 that characterises the spectra of an M dwarf, adding noise to the CCF profile.
 In contrast, if the K5 mask is used, only deep lines are considered and the CCF profile becomes less noisy. 
 Furthermore, high levels of stellar activity might deform the core of the average line profile, while stellar rotation
 broadens the lines. Therefore, special care should be taken in selecting the half-window of the CCF in order to include
 the continuum when fitting the CCF profile \citep[see e.g.][]{2020A&A...642A.133D}. 
 In addition, since we are dealing with a very red object, the signal in the bluest part of the spectra is very low, so the analysis
 should be limited to the red part of the spectrum. 
 In order to illustrate these effects, we compare the RVs obtained with the DRS by using the K5 and M2 masks as well as the K5
 mask with a half-window of 60 km s$^{\rm -1}$ (instead of the default value of 20 km s$^{\rm -1}$). 
 The results from this comparison are given in Table~\ref{rv_extraction}, while in Fig.~\ref{rv_ex_plot} we show the corresponding CCF profiles for
 one of our observations
 (we choose the one with the
 highest S/N, $\sim$ 13 at spectral order 46, 548 - 554 nm). % taken on May 17, 2022). % corresponding to the HARPS-N order 46 (check wave).
 It is clear from the figure that the CCF profile obtained with the M2 mask is comprised only by noise, without any visible signal. %%  highly irregular with a shape far from Gaussian.
 The CCF profile obtained with the original K5 mask, although asymmetric has a Gaussian-sharp form but lacks a good continuum for a proper fitting. 
 The enlarged-window K5 mask, on the other hand, provides a better fit to a Gaussian function with a lower error.
 These calculations have been done by using the DRS version implemented through the YABI interface \citep{murdoch7974} at the Italian Centre for Astronomical Archives\footnote{https://www.ia2.inaf.it/}.

% +++++++++++++++++++++++++++++++++++++++++++++++++
% Table 2: RVs comparison
% ++++++++++++++++++++++++++++++++++++++++++++++++
\begin{table}[htb]
\centering
\caption{Comparison between the different methods used to extract the RVs.}
\label{rv_extraction}
\begin{tabular}{llccc}
\hline\noalign{\smallskip}
 Method & Mask &  $\overline{RV}$    & $\overline{\Delta{RV}}$     & rms        \\
        &      &  (km s$^{\rm -1}$) & (km s$^{\rm -1}$) & (km s$^{\rm -1}$)   \\
\hline
CCF     & M2      & -65.05           & 0.12            & 9.94                 \\
CCF     & K5      & -59.53           & 0.09            & 17.18                \\
CCF     & K5 w60  & -61.35           & 0.06            & 12.85                \\
\hline
TERRA   &         &                  & 0.04            & 0.506                \\
SERVAL  &         &                  & 0.16            & 0.99                 \\
\hline
LSD     &         & -59.63           & 0.36            & 12.42                \\
SpotCCF &         & -59.57           & 0.32            & 12.30                 \\
\hline
\end{tabular}
%\tablefoot{(a) \cite{2022yCat.1355....0G};
%(b) \cite{2022yCat.4039....0P}; (c)  \cite{2003yCat.2246....0C}.\\
%}
\end{table}

 RVs were also computed with the Java-based Template-Enhanced Radial velocity Re-analysis Application 
 \citep[TERRA][]{2012ApJS..200...15A}. TERRA measures the RVs by a least-square match of the observed spectrum to co-added
 high S/N template spectra derived from the same observations. 
 The bluest part of the spectra was excluded from the analysis and only orders with $\lambda$ $>$ 434 nm  were considered.
 It has been shown that for M dwarf stars the RVs derived by TERRA show a lower root-mean-square (rms) than the DRS and should be preferred \citep{2017A&A...598A..26P}.
 However, we note that in the case of TOI-5375 the rms of the RVs obtained by using TERRA is $\sim$ 34 times lower than the rms obtained
 when using the RVs derived by the DRS. While the CCF RVs show a rms between 9.9 and 17 km s$^{\rm -1}$ (depending on the mask considered), TERRA
 RVs show a rms of only 0.5 km s$^{\rm -1}$.  
 TERRA RVs are not consistent with the RVs variations seen in the spectra. A quick look shows that spectral lines show relative shifts between different nights
 of the order of several km s$^{\rm -1}$. 
 A possible explanation for this discrepancy might be related to the low S/N of our spectra. It is likely that TERRA identifies some of the noise as 
 ``true'' spectral features.
 In order to check this fact, we made use of the SpEctrum Radial Velocity AnaLyser tool \citep[SERVAL,][]{2018A&A...609A..12Z}.
 The concept behind SERVAL is similar to TERRA, but it constitutes an alternative implementation. 
 Again, we derive RVs that show a very low rms ($\sim$ 1  km s$^{\rm -1}$).
 Therefore, we decided to exclude the TERRA and SERVAL RVs from our analysis.

\begin{figure}[htb]
\centering
\includegraphics[scale=0.55]{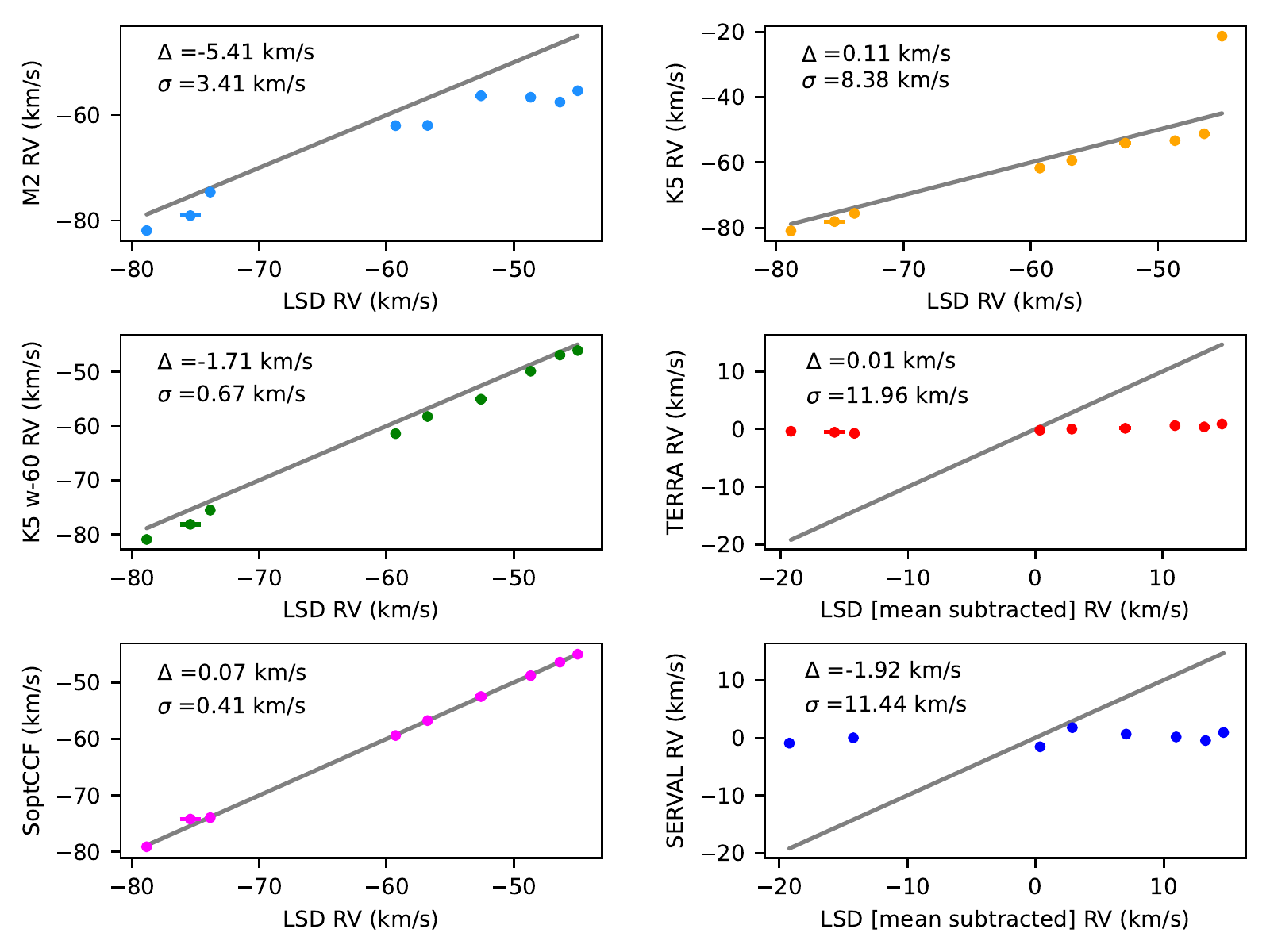}
\caption{RVs obtained using different masks and methods versus LSD RVs. For TERRA and SERVAL RVs, we subtracted the mean LSD RV for the comparison.
Each panel shows the mean value and standard deviation of the difference {\it RVs other methods - LSD RVs}.
Gray lines show the one-to-one relationship.}
\label{rv_comparison}
\end{figure}

 In order to avoid the aforementioned difficulties we decided to derive RVs
 using a self-written python package which computes the mean line profile using the least-squares deconvolution technique
% by computing the mean line profile using 
% the least-squares deconvolution technique 
 \citep[LSD,][]{1997MNRAS.291..658D}  with a stellar mask obtained from the VALD3\footnote{http://vald.astro.uu.se/} database
 \citep{1995A&AS..112..525P,2015PhyS...90e4005R} with appropriate stellar parameters for TOI-5375.
%This technique assumes that all
%spectral lines have the same profile, scaled by a certain factor,
%and that overlapping lines add up linearly. These assumptions
%%lead to a highly simplified description of the intensity and circular polarization observations in terms of the convolution of a
%known line mask with an unknown profile. With this approximation one can reconstruct an average line shape, the LSD profile,
%which is formally characterized by an extremely high
%S
%/
%N ratio. The CCF is also a kind of mean profile so ...
 The mean line profiles are then fitted with a rotational broadening function. %{\bf check with Monica}
 Further details can be found in \cite{2021A&A...649A..29R}. An example of one of the derived LSD profiles is shown in
 Figure~\ref{rv_ex_plot} where it can be compared with the CCF profiles obtained by the DRS.
 The LSD derived RVs have a rms of 12.42 km s$^{\rm -1}$ and a mean uncertainty of 0.36 km s$^{\rm -1}$.
 Figure~\ref{rv_comparison} shows a comparison between the RVs velocities obtained by using different masks and methods. 
 We note that the RVs derived with the enlarged-window K5 mask are similar to the ones derived with the LSD technique. 
 This shows that by performing an appropriate mask and half-window selection it is possible to derived precise RVs using
 the CCF method even in spectra at low S/N or for the case of fast rotators. 
 As mentioned, RVs based on the template-matching algorithm (TERRA and SERVAL) provide rather flat RV curves when compared
 with the CCF and LSD values. 
 We note that the CCF and LSD techniques combine the spectral lines to create a mean profile, thus it should be possible to achieve a reasonable S/N.
 In fact, the S/N of the CCF increases roughly as $\sqrt{N}$ where $N$ is the number of stellar lines used to compute the CCF (typically of the order of several thousands of lines).
 On the other hand, it seems that the template-matching algorithm, when all the observed spectra have a low S/N, is not able to
 generate a suitable template to compute accurate RVs. 
 Finally, RVs were also computed from the LSD profiles by using the {\tt SpotCCF} software (Di Maio et al. in prep.)  
 which takes into account the presence of multiple spots on the stellar surface. 
 It also accounts for the wings of the LSD profile by convolving the rotational profile with a Lorentzian profile. 
 The code was run without considering any spot. We find that the {\tt SpotCCF}-derived RVs are quite similar to the ones derived
 from the LSD analysis, thus confirming the reliability of our measurements. The analysis presented in the following refers to the LSD-derived RVs. 
% The analysis presented in the following refers to the LSD derived RVs. 

% >>>>>>>>>>>>>>>>>>>>>>>>>>>>>>>>>>>>>>>>>>>>>>>>>>>>
% Figure 2: Comparison CCF profiles
% <<<<<<<<<<<<<<<<<<<<<<<<<<<<<<<<<<<<<<<<<<<<<<<<<<<<

% >>>>>>>>>>>>>>>>>>>>>>>>>>>>>>>>>>>>>>>>>>>>>>>>>>>>
% Figure 2: Comparison CCF profiles
% <<<<<<<<<<<<<<<<<<<<<<<<<<<<<<<<<<<<<<<<<<<<<<<<<<<<

% +++++++++++++++++++++++++++++++++++++++++++++++++++++++++++++++
\section{The host star}\label{host_star}
% +++++++++++++++++++++++++++++++++++++++++++++++++++++++++++++++
 
 TOI-5375 (a.k.a. TIC 71268730, 2MASS J07350822+7124020, 1RXS J073508.9+712415)
 is a M0.5 dwarf located at a distance of 121.65 $\pm$ 0.17 pc from the Sun \citep{2022yCat.1355....0G}.
 Its main stellar properties are listed in Table~\ref{physical_properties}.
 Stellar effective temperature, spectral-type and metallicity are determined from a methodology based on the use of a principal component analysis
 and sparse Bayesian fitting methods\footnote{https://github.com/jesusmaldonadoprado/mdwarfs\_abundances}, % \citep{2020A&A...644A..68M},
 which make use of the same spectra used in this work for the radial velocity analysis (see Sec.~\ref{observations}).
  In brief, a relationship between the principal components and the stellar parameters is derived by training the data
 with the spectra of M dwarfs orbiting around an FGK primary or M dwarfs with effective temperature obtained from interferometric
 measurements of their radii. Further details can be found in  \citet{2020A&A...644A..68M}.
 We also compute T$_{\rm eff}$ and [Fe/H] by using photometric relationships \citep{2006MNRAS.373...13C,2012A&A...538A..25N}.
 We note that the photometric T$_{\rm eff}$ is significantly lower than the spectroscopic one. 
 This discrepancy has already been noted and discussed \citep[see e.g.][and references therein]{2015A&A...577A.132M}.
 It is related to the fact that M dwarfs emission deviates from a blackbody for wavelengths beyond $\sim$ 2 $\mu$m. 
 The photometric estimate of [Fe/H] is also lower than the spectroscopic one. In a recent work, \cite{2020A&A...644A..68M} show that the use
 of photometric estimates for M dwarfs gives a [Fe/H] distribution that is shifted towards lower values when compared with
 an otherwise similar sample of FGK stars. 
 Therefore, spectroscopic parameters are preferred, although we caution against the relatively low S/N of our spectra
 (of the order of 44 in the 604-605 nm region, measured in the coadded spectra). 
 Stellar mass is computed following the relations based on near-infrared photometry by \cite{1993AJ....106..773H}, while the stellar radius is
 obtained using the mass-radius relationship provided by \cite{2015A&A...577A.132M}. From the stellar mass and radius values, the surface gravity
 was estimated. The stellar luminosity is computed by using the bolometric corrections provided by \cite{1996ApJ...469..355F}.

%------------------------------------------------------------
% --- Table 1: Stellar parameters
%------------------------------------------------------------
\begin{table}[htb]
\centering
\caption{ Physical properties of TOI-5375.}
\label{physical_properties}
\begin{tabular}{lrl}
\hline\noalign{\smallskip}
 Parameter                 &  Value                &  Notes \\
\hline
% Other names               &  \multicolumn{2}{l}{TIC 71268730, 2MASS J07350822+7124020, 1RXS J073508.9+712415} \\   
%  # APASS 60472363, Gaia DR2 1110586978339817728, UCAC4 808-017922, WISE J073508.32+712402.3
\hline
 $\alpha$ (ICRS J2000)     &  07:35:08.22            & a       \\
 $\delta$ (ICRS J2000)     & +71:24:02.13            & a       \\
\hline
 Spectral Type                 &  M0.5              &  b  \\
 T$_{\rm eff}$ spec (K)        &  3885 $\pm$ 25     &  b  \\
 T$_{\rm eff}$ phot (K)        &  3709 $\pm$ 27     &  b  \\
 ${\rm [Fe/H]}$ spec (dex)           & +0.26 $\pm$ 0.04   &  b  \\
 ${\rm [Fe/H]}$ phot (dex)           & +0.03 $\pm$ 0.03   &  b  \\
 $M_{\star}$ ($M_{\odot}$)     &  0.64 $\pm$ 0.10   &  b  \\
 $R_{\star}$ ($R_{\odot}$)     &  0.62 $\pm$ 0.10   &  b  \\
 $\log g$ (cm s$^{\rm -2}$)    &  4.66 $\pm$ 0.15   &  b  \\
 $\log (L_{\star}/L_{\odot})$  & -1.19 $\pm$ 0.15   &  b  \\
\hline
 P$_{\rm rot}$ (d)             & 1.969226 $\pm$  0.000358 & b \\
 v$_{\rm eq}$ (km s$^{\rm -1}$)   &     15.9 $\pm$  2.6      & b  \\
%v$\sin i$ (km s$^{\rm -1}$)   &     15.9 $\pm$  2.6      & b  \\ 
% Age          (Myr)            &          $\sim$ 100        &   \\
\hline
 $\pi$ (mas)                     & 8.2202   $\pm$  0.0114   & a  \\
 $\mu_{\alpha}$ (mas/yr)         & 48.434   $\pm$  0.009    & a  \\
 $\mu_{\delta}$ (mas/yr)         & 19.069   $\pm$  0.012    & a  \\
 v$_{\rm rad}$ (km s$^{\rm -1}$) & -61.59   $\pm$  0.16   & b   \\ % -75.13  $\pm$  3.63     & a  \\
 $U$ (km s$^{\rm -1}$)           &  61.15   $\pm$  0.11     & b   \\ %70.72   $\pm$  2.56     &    \\
 $V$ (km s$^{\rm -1}$)           & -30.37   $\pm$  0.08     & b   \\ % -37.34   $\pm$  1.87     &    \\
 $W$ (km s$^{\rm -1}$)           & -5.68    $\pm$  0.08     & b   \\ %   -12.26  $\pm$  1.76     &    \\
\hline
 $B$ (mag)             & 15.418 $\pm$  0.106 & c \\
 $V$ (mag)             & 14.063 $\pm$  0.056 & c \\
\hline
 2MASS J (mag)             & 11.167 $\pm$ 0.022 & d \\
 2MASS H (mag)             & 10.480 $\pm$ 0.023 & d \\
 2MASS K$_{\rm S}$ (mag)   & 10.302 $\pm$ 0.020 & d \\
\hline
 GAIA DR3 G (mag)          & 13.410046 $\pm$ 0.003056 & a \\
 GAIA DR3 BP (mag)         & 14.336692 $\pm$ 0.006075 & a \\
 GAIA DR3 RP (mag)         & 12.448772 $\pm$ 0.005159 & a \\
\hline
 WISE1 (mag)               & 10.226 $\pm$ 0.023 &  e \\
 WISE2 (mag)               & 10.212 $\pm$ 0.021 &  e \\
 WISE3 (mag)               & 10.354 $\pm$ 0.084 &  e \\
 WISE4 (mag)               & $>$ 8.543          &  e \\
\hline
\end{tabular}
\tablefoot{(a) \cite{2022yCat.1355....0G}; (b) This work;
(c) \cite{2022yCat.4039....0P}; (d)  \cite{2003yCat.2246....0C}; (e) \cite{2014yCat.2328....0C} .\\
}
\end{table}

\begin{figure}[htb]
\includegraphics[scale=0.55]{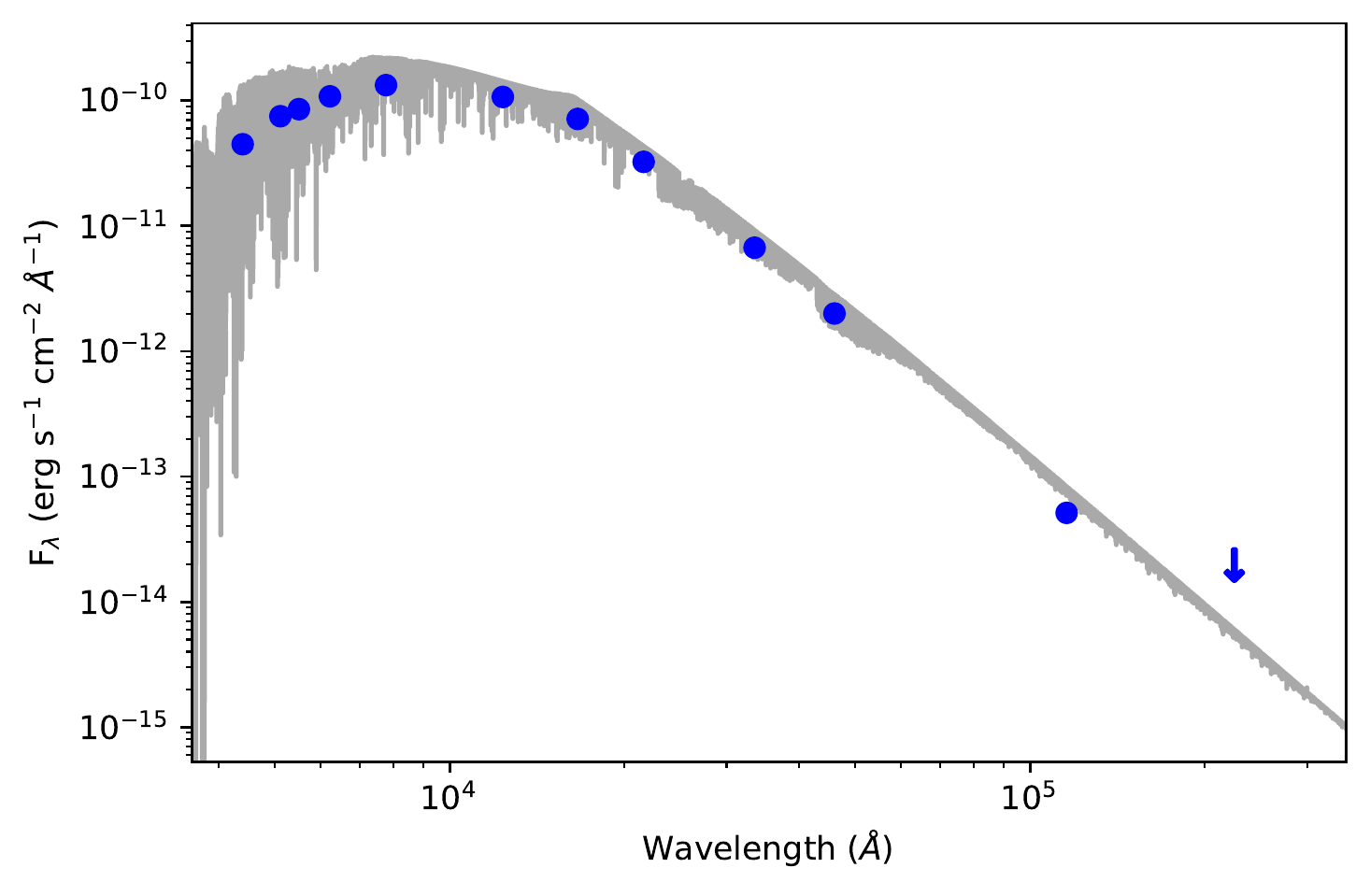}
\caption{Spectral energy distribution of TOI-5375. A PHOENIX/BT-SETTL atmospheric model
(T$_{\rm eff}$ = 3885 K, $\log g$ = 4.66 cm s$^{\rm -2}$, [Fe/H] =  +0.26 dex)
is overplotted. % with the stellar parameters 
Errors in the observed fluxes are within the symbol's size.}
\label{YO48_sed}
\end{figure}

% Galactic spatial-velocity components $(U, V, W)$ are computed from the star radial velocity, together with {\it Gaia} parallaxes and 
% proper motions \citep{2022yCat.1355....0G}. From the analysis of its velocity components we conclude that TOI-5375 is a field thin disc star. 
 The rotational period is computed from the analysis of the TESS light curve (see Sec.~\ref{observations}) while the %projected 
  equatorial velocity is estimated
 from the rotation period and the stellar radius. %, and the inclination angle {\bf derived from the modelling of the TESS light curve, see Sec.~\ref{photometry_section}.}
 The available photometry for TOI-5375 has been used to build and trace the star spectral energy distribution along with PHOENIX/BT-SETTL 
 atmospheric models \citep{2012RSPTA.370.2765A}, see Fig.~\ref{YO48_sed}.

% >>>>>>>>>>>>>>>>>>>>>>>>>>>>>>>>>>>>>>>>>>>>>>>>>>>>
% Figure 1: Spectral energy distribution
% <<<<<<<<<<<<<<<<<<<<<<<<<<<<<<<<<<<<<<<<<<<<<<<<<<<<

% --------------------------------- 
\subsection{Age estimates}\label{sect_age}
% ---------------------------------

% >>>>>>>>>>>>>>>>>>>>>>>>>>>>>>>>>>>>>>>>>>>>>>>>>>>>
% Figure 2: Age diagnostics
% <<<<<<<<<<<<<<<<<<<<<<<<<<<<<<<<<<<<<<<<<<<<<<<<<<<<
\begin{figure*}[!htb]
\centering
\begin{minipage}{0.33\linewidth}
\includegraphics[scale=0.40]{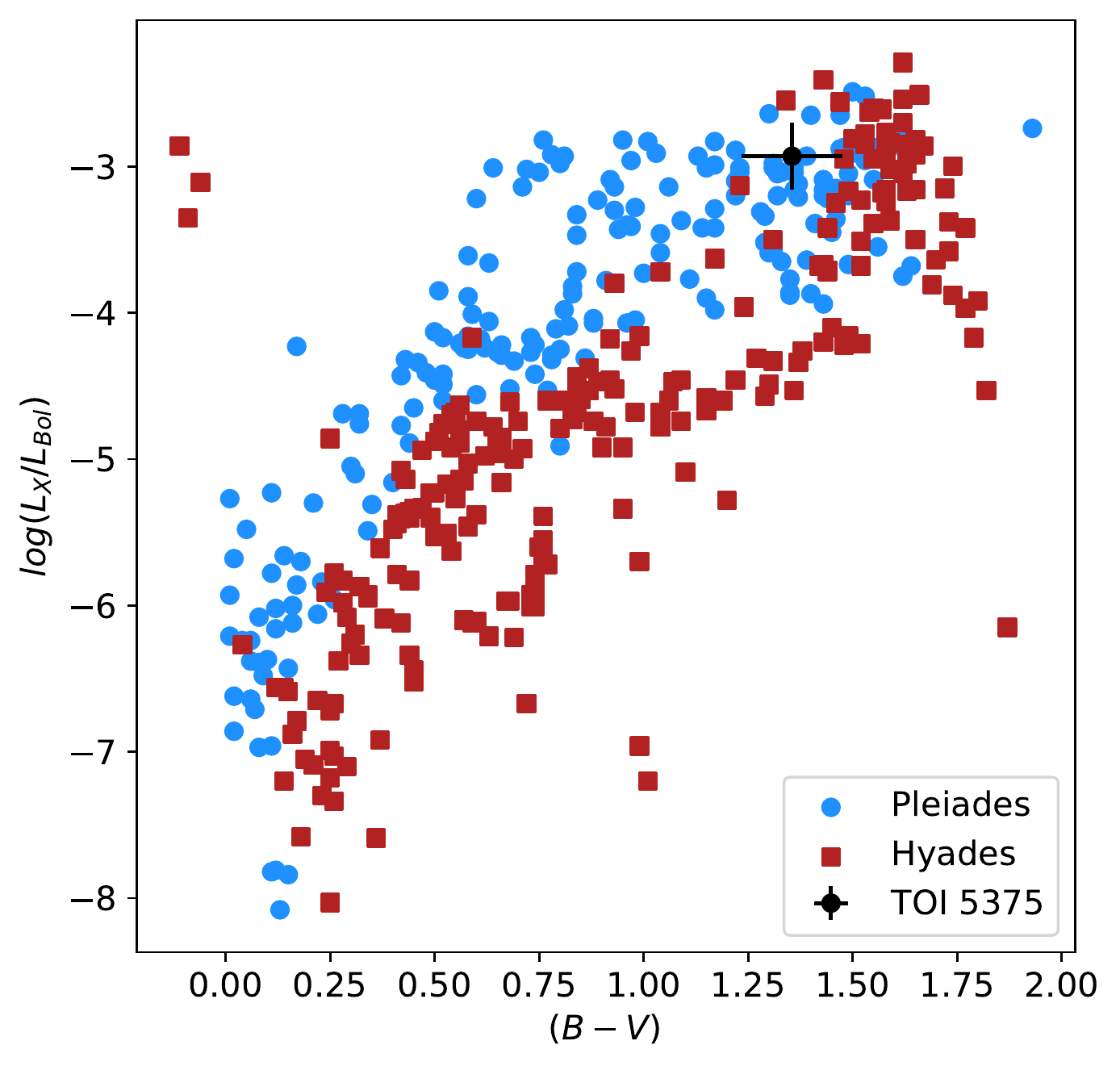}
\end{minipage}
\begin{minipage}{0.33\linewidth}
\includegraphics[scale=0.40]{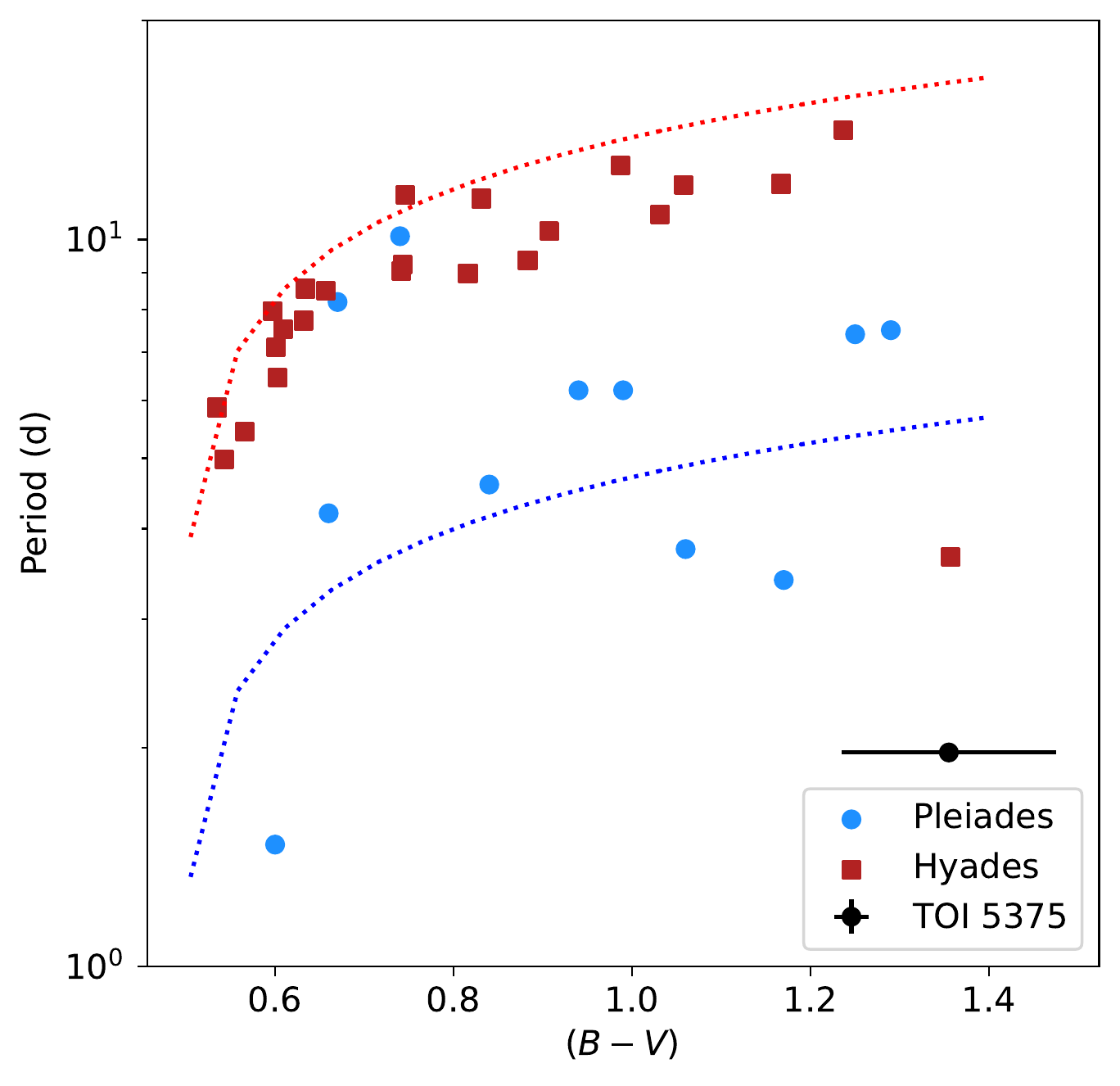}
\end{minipage}
\begin{minipage}{0.33\linewidth}
\includegraphics[scale=0.40]{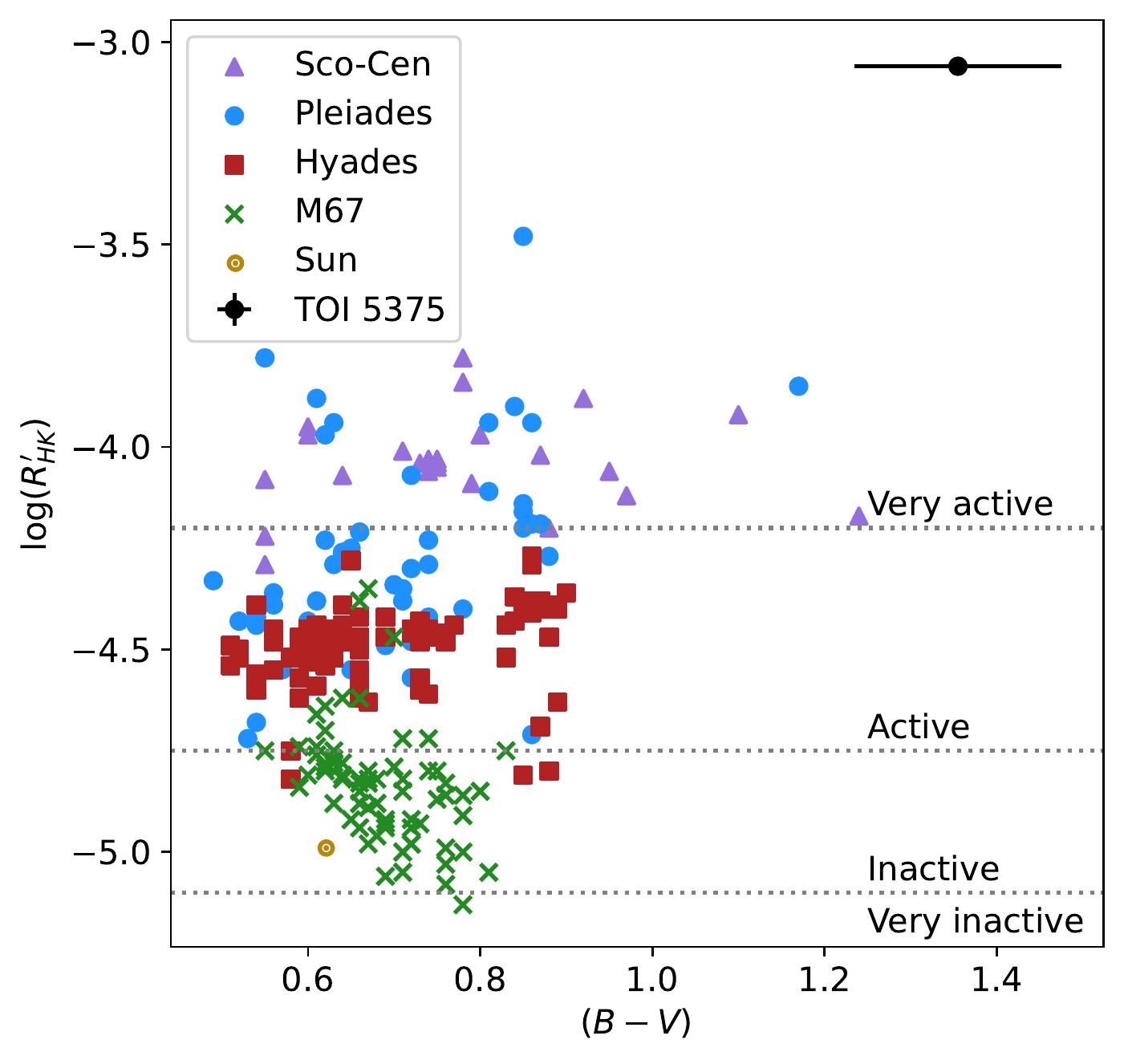}
\end{minipage}
\caption{ Age diagnostics for TOI-5375.
Left: Fractional X-ray luminosity $\log (L_{\rm X}/L_{\rm Bol})$  vs. colour index (B - V).
Centre: Rotation period as a function of the (B - V) colour.
Two gyrochrones (at the ages of the Pleiades and the Hyades) have been overplotted for comparison.
Right: $\log$(R'$_{\rm HK}$) vs. colour index (B - V).
Dashed lines are the limits for very active, active, inactive, and very inactive stars, according to \cite{1996AJ....111..439H}.
%Bottom right: Colour-magnitude diagram (M$_{\rm V}$ as a function of (B-V) colour).
}
\label{age_estimates}
\end{figure*}

 Stellar age is one of the most difficult parameters to determine in an accurate way, even for main-sequence 
 solar-type stars, as they evolve too slowly to be dated by their position in the 
 Hertzsprung-Russell Diagram.  %{\bf UVW velocities here}. 
 For solar-type stars, it is well known that stellar activity and rotation are related by the stellar dynamo
 \citep[e.g.][]{1984ApJ...279..763N} and both activity and rotation diminish as the star evolves and loses angular momentum. 
 Thus, activity and rotation tracers, such as Ca emission, X-ray, or rotation periods, are often used to estimate
 stellar ages \citep[see e.g.][]{2008ApJ...687.1264M}.
 However, in this case, the rotation of TOI-5375 is being accelerated by the tides raised by the companion
  (as we will discuss in Sect.~\ref{discussion}).
 As a consequence, the
 star can show a high level of chromospheric or X-ray emission despite being rather old. 
 Therefore, age diagnostics based on the rotation period will only provide lower limits to the age of the TOI-5375 system.  
  In the following, we consider as ``young'' those stars with an age younger than the Hyades, $\sim$ 750 Myr.

 To compute the $L_{\rm X}$ of TOI-5375 we search for an X-ray counterpart in the ROSAT All-Sky
 Bright Source Catalogue \citep{1999A&A...349..389V} and apply the count rate-to-energy flux conversion factor
 relation given by \cite{1995ApJ...450..401F}. The derived X-ray luminosity is $\log L_{\rm X}$ = 29.44 $\pm$ 0.22 erg s$^{\rm -1}$,
 while the position of TOI-5375 in the $\log (L_{\rm X}/L_{\rm Bol})$ versus colour index (B - V) plane is shown in 
 Fig.~\ref{age_estimates} (left). Bolometric corrections are derived from the (B-V) colour by interpolating in 
 \citet[][Table~3]{1996ApJ...469..355F}. Data from the Pleiades,  age $\sim$ 112 Myr \citep{1994ApJS...91..625S} and Hyades, 
  age $\sim$ 750 Myr
 \citep{1995ApJ...448..683S}
 are overplotted for comparison. The position of TOI-5375 in this diagram is consistent with the position of the Pleiades.

 Figure~\ref{age_estimates} (centre) shows the rotation period of TOI-5375 as a function of the colour index  (B-V).
 For comparison, data from the Pleiades \citep{1995PASP..107..211P} and from the Hyades \citep{1987ApJ...321..459R} are overplotted.
 Two gyrochrones (at the ages of the Pleiades and the Hyades) are also overplotted. 
 The figure seems to indicate that TOI-5375 is a young star with a fast rotation period\footnote{
 The Hyades star with a rotation period equal to 3.66 d is VB 190. \citet{1987ApJ...321..459R} discussed two possibilities to explain its peculiar behaviour.
 First, the star lies somewhat above the cluster main sequence, suggesting that it might be a binary. Then, its rapid rotation might be due to tidal coupling with the companion.
 Second, about one-third of the M-type Hyades stars are known to have projected rotational velocities larger than 10 km s$^{\rm -1}$.
 VB 190 may be a member of this population.}.

 TOI-5375 shows a clear emission in the Ca~{\sc ii} H \& K lines. Balmer lines (from H$\alpha$ to H$\epsilon$),
 as well as other known activity indexes such as the Na~{\sc i} D$_{\rm 1}$, D$_{\rm 2}$ doublet and the He~{\sc i} D$_{\rm 3}$ line
 are also shown in emission. 
 To quantify the stellar chromospheric activity we make use of the $\log$(R'$_{\rm HK}$), defined as the ratio of the chromospheric
 emission in the cores of the broad Ca~{\sc ii} H \& K absorption lines to the total bolometric emission \citep[e.g.][]{1984ApJ...279..763N}.
 We start by computing the so-called $S index$ which is measured in the Mount Wilson scale by the HARPS-N DRS %ata Reduction Software 
% \citep[DRS,][]{2007A&A...468.1115L,2021A&A...648A.103D} % \textbf{sure? shoudn't it be the new one? Dumusque et al. al 2021}
 implemented through the YABI interface. % \citep{murdoch7974} at the Italian Center for Astronomical Archives\footnote{https://www.ia2.inaf.it/}. 
 The $S index$ contains both the contribution of the photosphere and the chromosphere. 
 In order to account for the photospheric %correction
 %In order to correct for the phostospheric
 correction we used the calibration based on the colour index (B-V) developed by \cite{2015MNRAS.452.2745S}
 who updated and extended the original calibration from \cite{1984ApJ...279..763N} to the low-mass regime up to a (B-V) value of 1.90.
 The derived  $\log$(R'$_{\rm HK}$) for TOI-5375 is -3.06 $\pm$ 0.10. This value can be compared with the activity level of stars in
 clusters with well-known ages. The comparison is shown in Fig.~\ref{age_estimates} (right). Data from the Sco-Cen complex  (age $\sim$ 5 Myr),
 Pleiades, Hyades, and M67  (age $\sim$ 4000 Myr) are from \cite{2008ApJ...687.1264M}.
 TOI-5375 is located in the range of very active stars with a mean $\log$(R'$_{\rm HK}$) higher than other stars in young
 stellar associations and clusters. 

\begin{figure}[!htb]
\centering
\begin{minipage}{0.49\linewidth}
\includegraphics[scale=0.32]{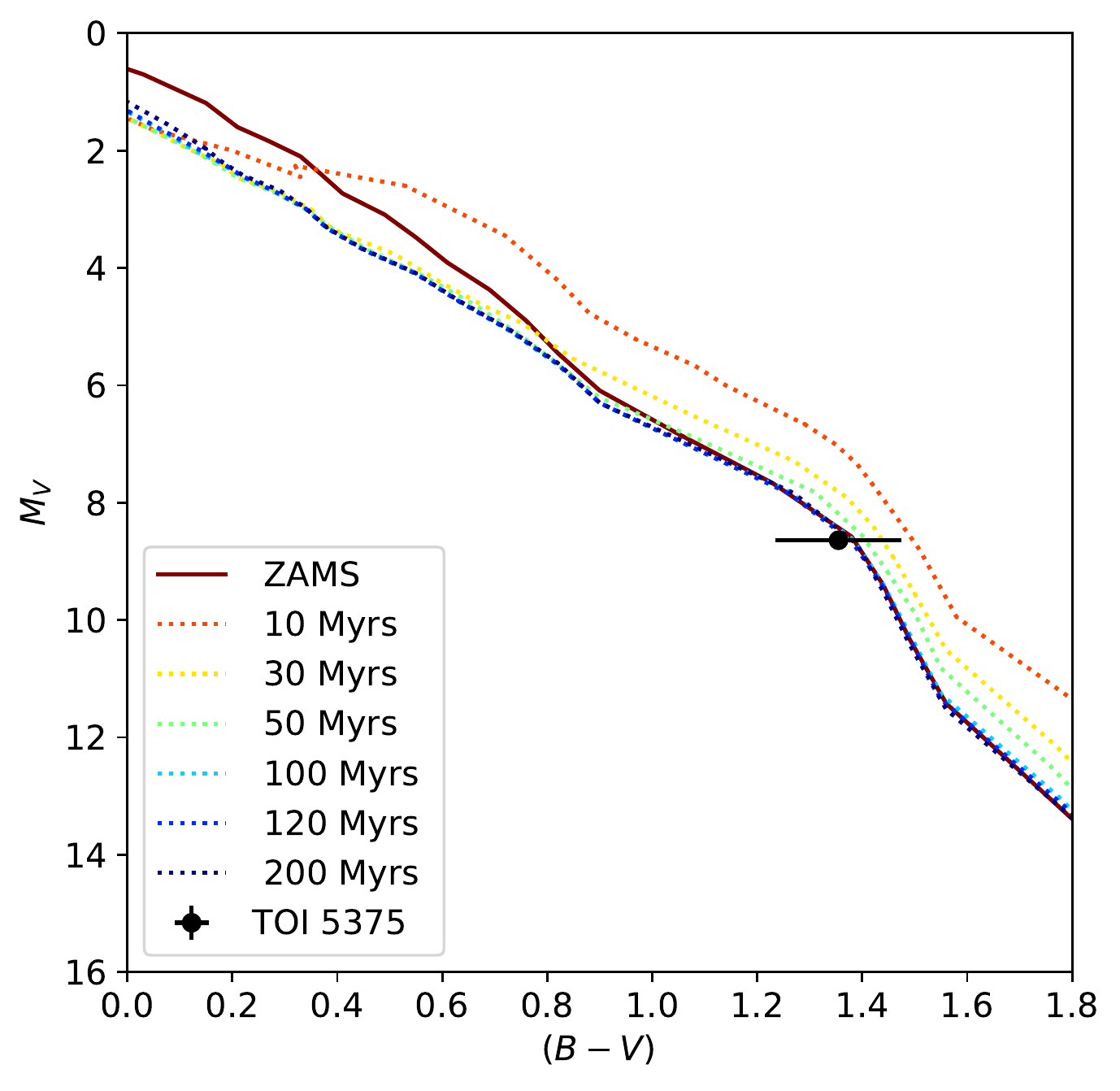}
\end{minipage}
\begin{minipage}{0.49\linewidth}
\includegraphics[scale=0.32]{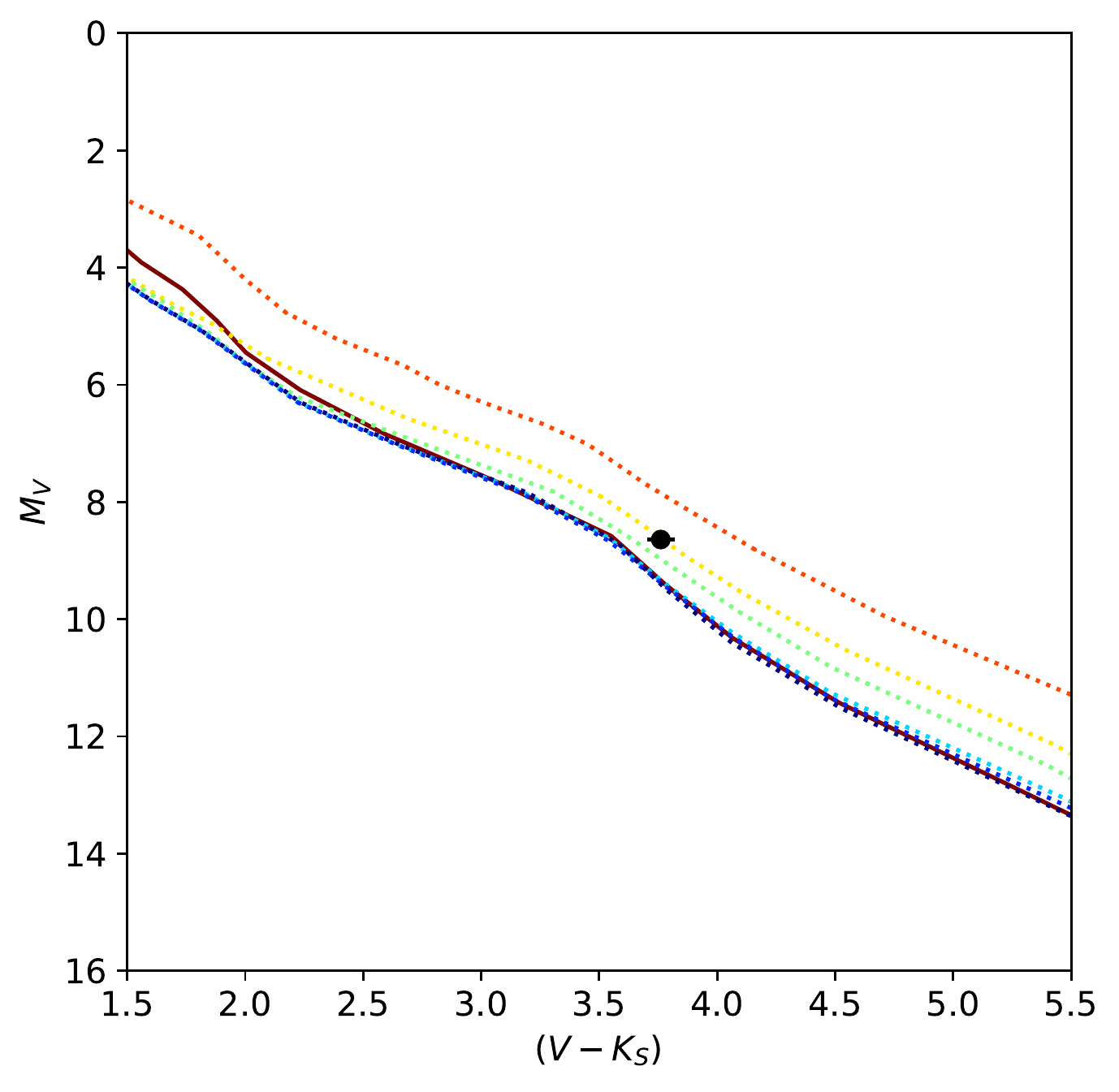}
\end{minipage}
\caption{Colour-magnitude diagram. Left: M$_{\rm V}$ as a function of (B-V) colour.
Right: M$_{\rm V}$ as a function of (V - K$_{\rm S}$) colour.}
\label{age_cmd}
\end{figure}

 As an additional age diagnostic we plot the position of TOI-5375 in a colour-magnitude diagram,
 see Fig.~\ref{age_cmd}. The ZAMS position as well as several pre-main sequence isochrones
 (computed at metallicity +0.20 dex) 
 %isocrhones
 from \cite{2000A&A...358..593S} are plotted. We use the M$_{\rm V}$ versus (B-V) (left panel) and  M$_{\rm V}$ versus (V - K$_{\rm S}$)
 \footnote{Original B, V, R, I magnitudes from the isochrones
 have been transformed into (V - K$_{\rm S}$) following the relationship by \cite{2008MNRAS.384.1178B}.}
 plots
 (right panel),
 as it has been shown that young rotating stars in the Pleiades tend to show blue excesses, that is bluer (B-V) colours, than other relatively old stars
 \citep{2003AJ....126..833S}.
 Indeed, we note that
 the position of the star in the diagram, when considering the (B-V) colour, suggests an age between  30 and 200 Myr
  (we note the large uncertainty in B magnitude),
 whilst when considering the
 (V - K$_{\rm S}$) colour, the position of TOI-5375 is more consistent with a younger, $\sim$ 30 Myr age.
 %{\bf We note that, the large uncertainty in B magnitude.}

\begin{figure}[!htb]
\centering
\includegraphics[scale=0.35]{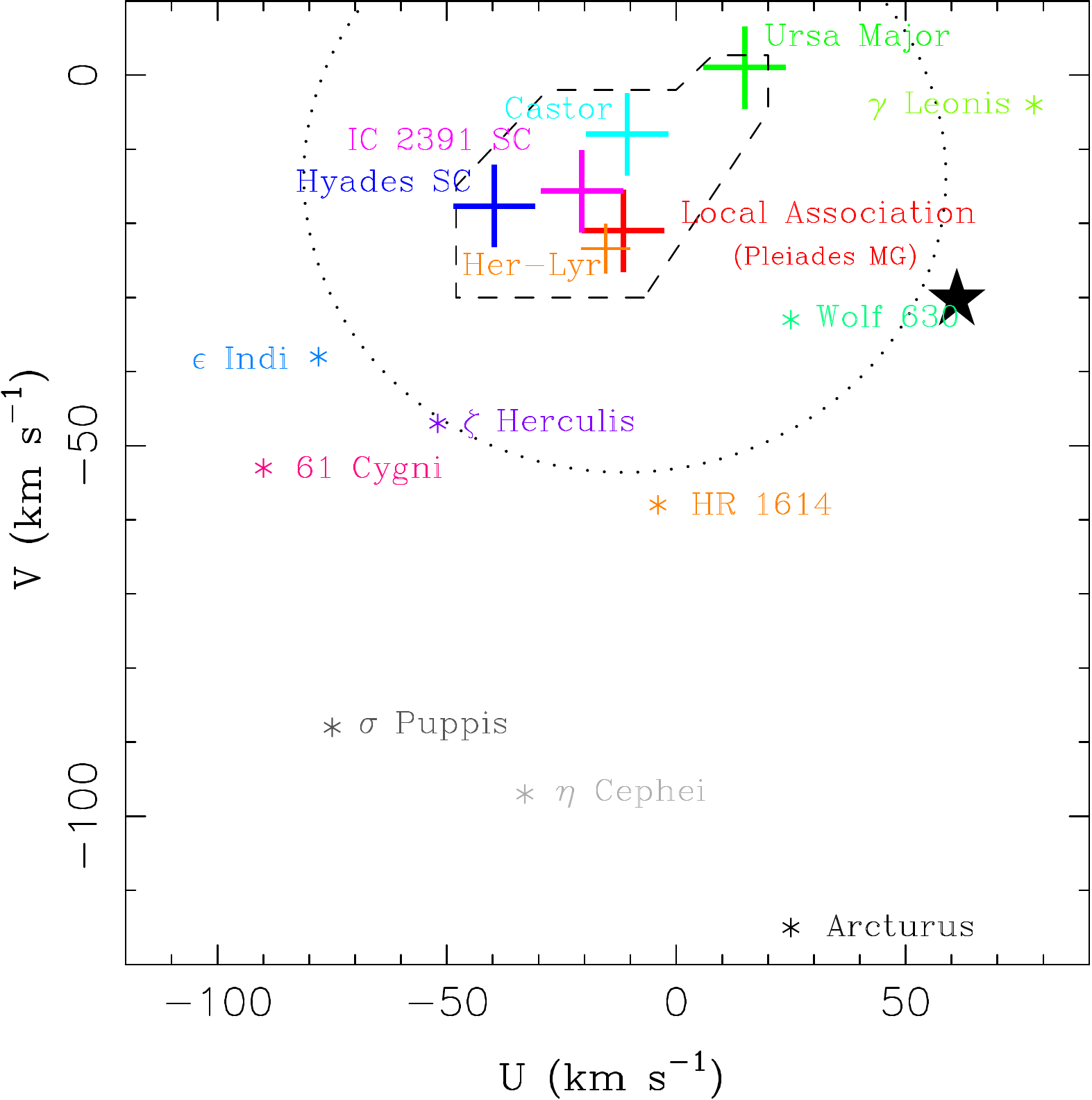}
\caption{$(U,V)$ plane showing the position of the young disc population (dashed line) as defined by \citet{1984AJ.....89.1358E,1989PASP..101..366E}  as well as the
known kinematic groups (large crosses and asterisks) in the solar neighbourhood \citep[see][and references therein]{2010A&A...521A..12M,j_maldonado_2012_3738204}. 
The dotted line represents the velocity ellipsoid determined by \citet{2009NewA...14..615F}.
The position of TOI-5375 is shown with a black star (uncertainties are within the symbol size). 
All velocities are in the Solar System reference frame. 
}
\label{uvplane}
\end{figure}

  Membership in stellar associations and young moving groups has been used as a methodology to identify young
  stars and to assign ages, especially after the release of the {\sc Hipparcos} data 
  \citep[see e.g.][and references therein]{2010A&A...521A..12M}.
  The Galactic spatial-velocity components $(U, V, W)$ of TOI-5375 are computed from the stellar radial velocity from our own orbital solution, together with {\it Gaia} parallaxes and
  proper motions \citep{2022yCat.1355....0G}.
  We follow the procedure described in \cite{2001MNRAS.328...45M}  and \cite{2010A&A...521A..12M},
  where the original algorithm \citep{1987AJ.....93..864J} is adapted to epoch J2000 in the International Celestial Reference System (ICRS) following
  Sect.~1.5 of the {\sc Hipparcos} and {\it Tycho} catalogues \citep{ESA}. 
  We made use of the full covariance matrix to compute the uncertainties. In this way, we account for the possible correlation between the astrometric parameters.
% Uncertainties were computed using the full covariance matrix, in order to account for the possible correlation between the astrometric parameters.
  The corresponding $(U, V, W)$ values are provided in Table~\ref{physical_properties}, while the position of TOI-5375 in the $(U,V)$ plane is shown in Fig.~\ref{uvplane}. 
  From the analysis of its velocity components we see that TOI-5375 is located outside the region of the young stars so
  we conclude that it is a field thin disc star.
  This conclusion is supported by the use of Bayesian methods \citep[BANYAN,][]{2018ApJ...856...23G}\footnote{http://www.exoplanetes.umontreal.ca/banyan/banyansigma.php}
  that confirm that TOI-5375
  is, with high-probability (99.9\%), a field star. 
  The high-metallicity content of the star might be an additional indication that the star is not young.
  This is because typically, the metallicity of young stars in the solar neighbourhood is close-to-solar or slightly sub-solar \citep{2017A&A...601A..70S}.

% >>>>>>>>>>>>>>>>>>>>>>>>>>>>>>>>>>>>>>>>>>>>>>>>>>>>
% Figure: Kinematics
% <<<<<<<<<<<<<<<<<<<<<<<<<<<<<<<<<<<<<<<<<<<<<<<<<<<<

 %We conclude that 
  To summarise, age estimates based on the rotation of the star seem to indicate that TOI-5375 is a young star.
 However, other age estimates based on the stellar metal content or the kinematics do not support this statement.
 The disagreement is due to the fact the stellar rotation period
% However, as its rotation period
 is being accelerated by the companion, as we shall discuss in Sect.~\ref{discussion}.
 Therefore, the only conclusion that we can draw from this analysis is that TOI-5375 is a field star that should have an age older than
 the one of stars in  young stellar associations like the Pleiades or the Hyades.

%----------------------------------------------------------------
\section{Analysis}\label{analysis}
%----------------------------------------------------------------
% ---------------------------------------------------------------
\subsection{Validation and vetting tests}
% ---------------------------------------------------------------

As mentioned, the TESS photometric light curve of TOI-5375 shows clear periodic dips
compatible with the transits of a substellar companion orbiting around this active, M dwarf.
Because of the low spatial resolution of the TESS cameras
($\thickapprox$ 21 arcsec/pixel) some of the objects initially
identified as a substellar companion might indeed be a false positive (FP).
We, therefore, perform an analysis to identify possible nearby contaminating
stars that might be the origin of blended eclipsing binaries (BEB). The analysis is based
on the use of the photometric data provided by {\it Gaia} and is described at length in
\cite{2022MNRAS.516.4432M}.

\begin{figure}[!htb]
\centering
\includegraphics[scale=0.325]{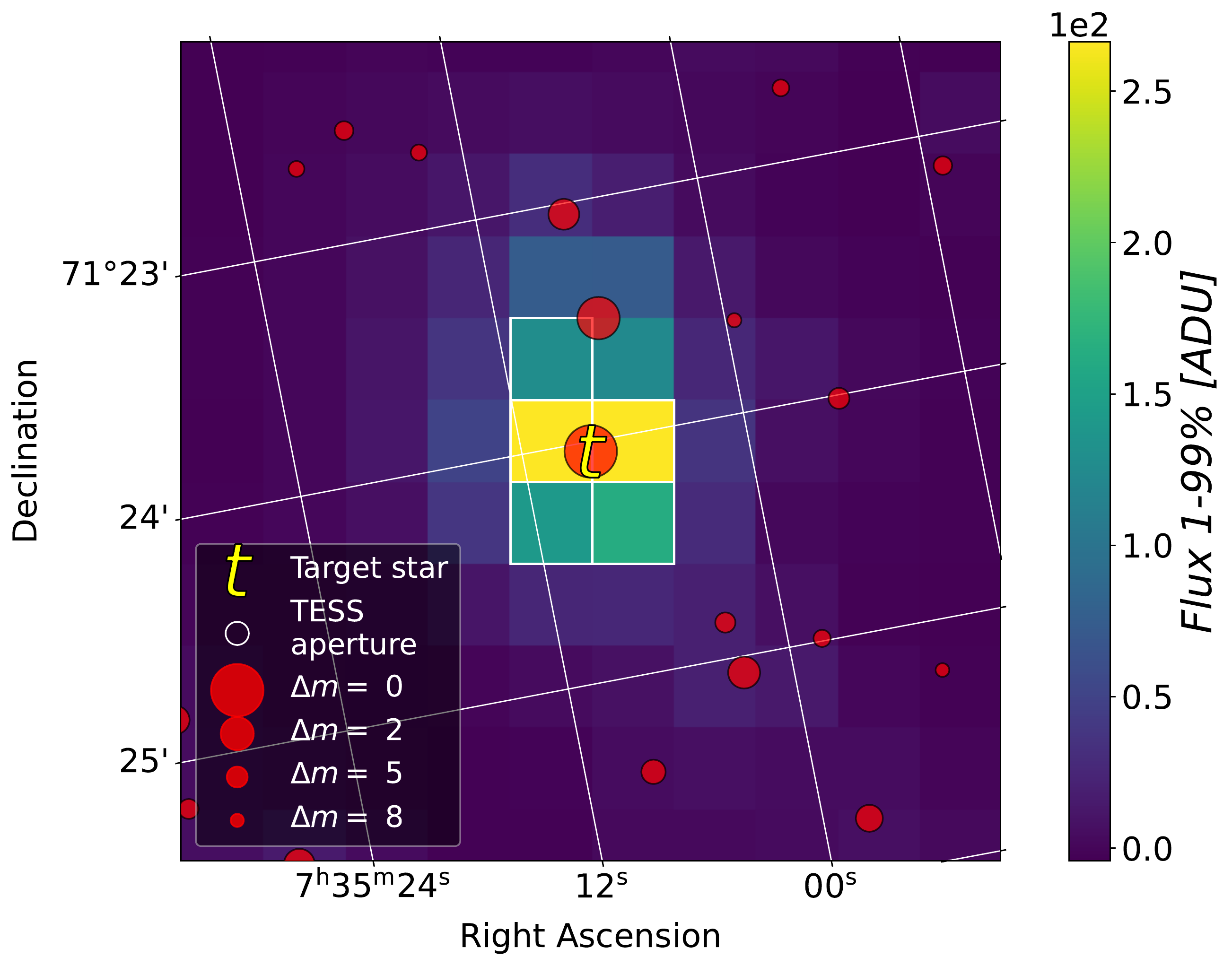}
\caption{{\it Gaia} stars identified in the TESS field.  %% of view.
The position of TOI-5375 is shown with the letter ``t''.} 
\label{validation_tests}
\end{figure}

\begin{figure}[!htb]
\centering
\includegraphics[scale=0.20]{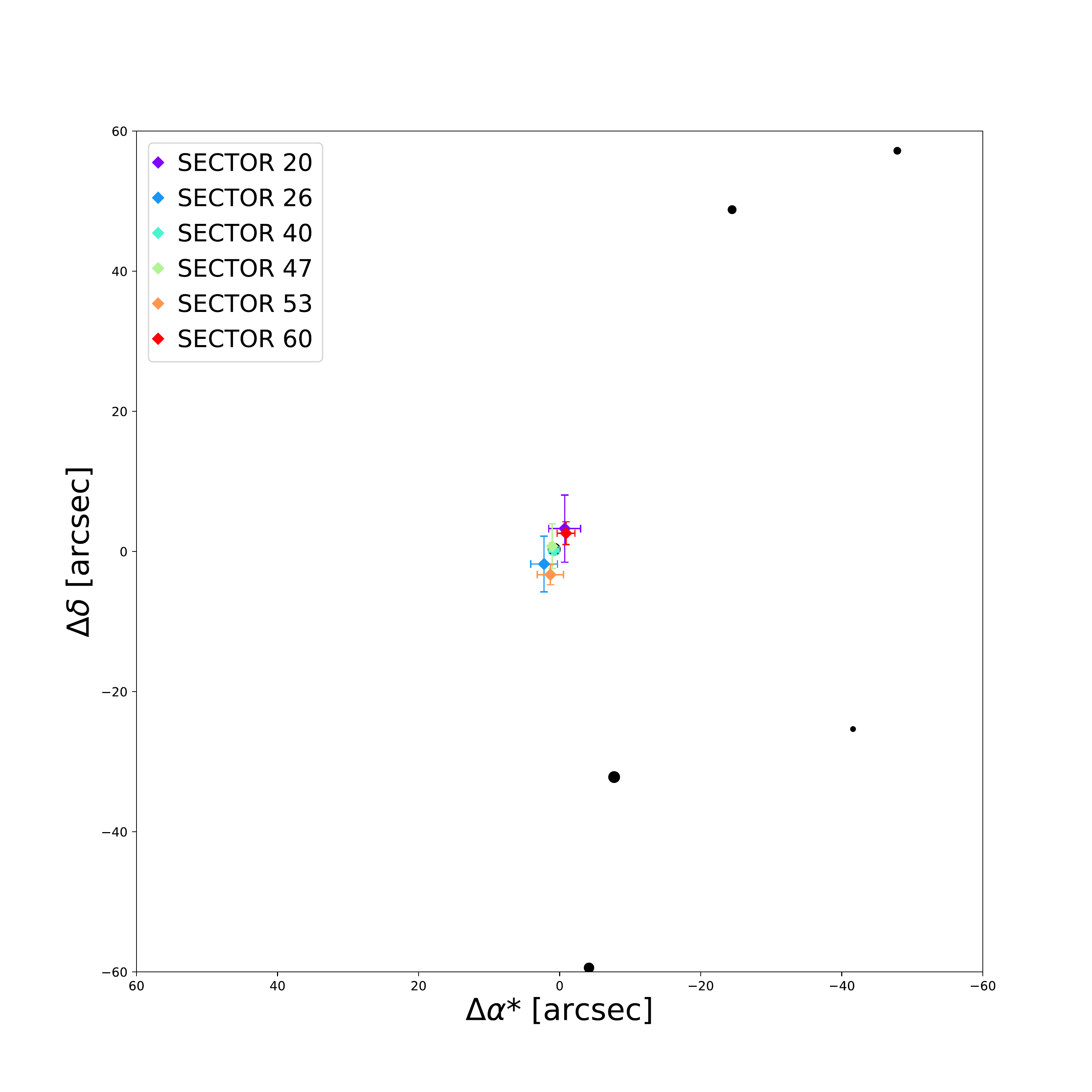}
\caption{In- or out-of-transit centroid tests calculated for each sector.
Black dots are background stars.}
\label{validation_tests2}
\end{figure}

The analysis shows that, besides TOI-5375, there is a star at 33.5 arcsec 
({\it Gaia} DR2 1110586939683894528)
separation from TOI-5375 that might reproduce the transit signal, see Fig.~\ref{validation_tests}.
To be sure that the transits are not due to  contamination by this neighbour, we perform the analysis of in- or out-of-transit centroid check
following the procedure described in \cite{2020MNRAS.495.4924N} and \cite{2020MNRAS.498.5972N}.
 Data from sectors 20 and 26 are included in this analysis. It should be noticed that the
in- or out-of-transit
centroid analysis is performed
on the full frame images and therefore it is independent from the analysis of the light curve.
The results are shown in figure~\ref{validation_tests2}. It can be seen that, within the errors, the in- or -out-of-transit mean centroids
are in agreement with the position of TOI-5375.

% ------------------------------------------------
% Validation tests
% ------------------------------------------------

% ------------------------------------------------
% Validation tests II
% ------------------------------------------------

Finally, in order to rule out the possibility that the transiting candidate is an FP
we  use  the VESPA\footnote{https://github.com/timothydmorton/VESPA} software
\citep{2012ApJ...761....6M,2015ascl.soft03011M}.
We obtain a 100\% probability of having a Keplerian transiting companion around TOI-5375,
while the probability of a FP is of the order of $\sim$ 7 $\times$ 10$^{\rm -10}$.
All these analyses confirm that TOI-5375 b is fully and statistically vetted and only requires
reconnaissance spectroscopy to be promoted to a ``statistically validated'' companion.

% --------------------------------------------------------------
\subsection{Photometric time series analysis}
\label{photometry_section} 
% ---------------------------------------------------------------

% >>>>>>>>>>>>>>>>>>>>>>>>>>>>>>>>>>>>>>>>>>>>>>>>>>>>
% Figure: Photometry corner plot
% <<<<<<<<<<<<<<<<<<<<<<<<<<<<<<<<<<<<<<<<<<<<<<<<<<<<
\begin{figure*}[htb]
\centering
\includegraphics[scale=0.25]{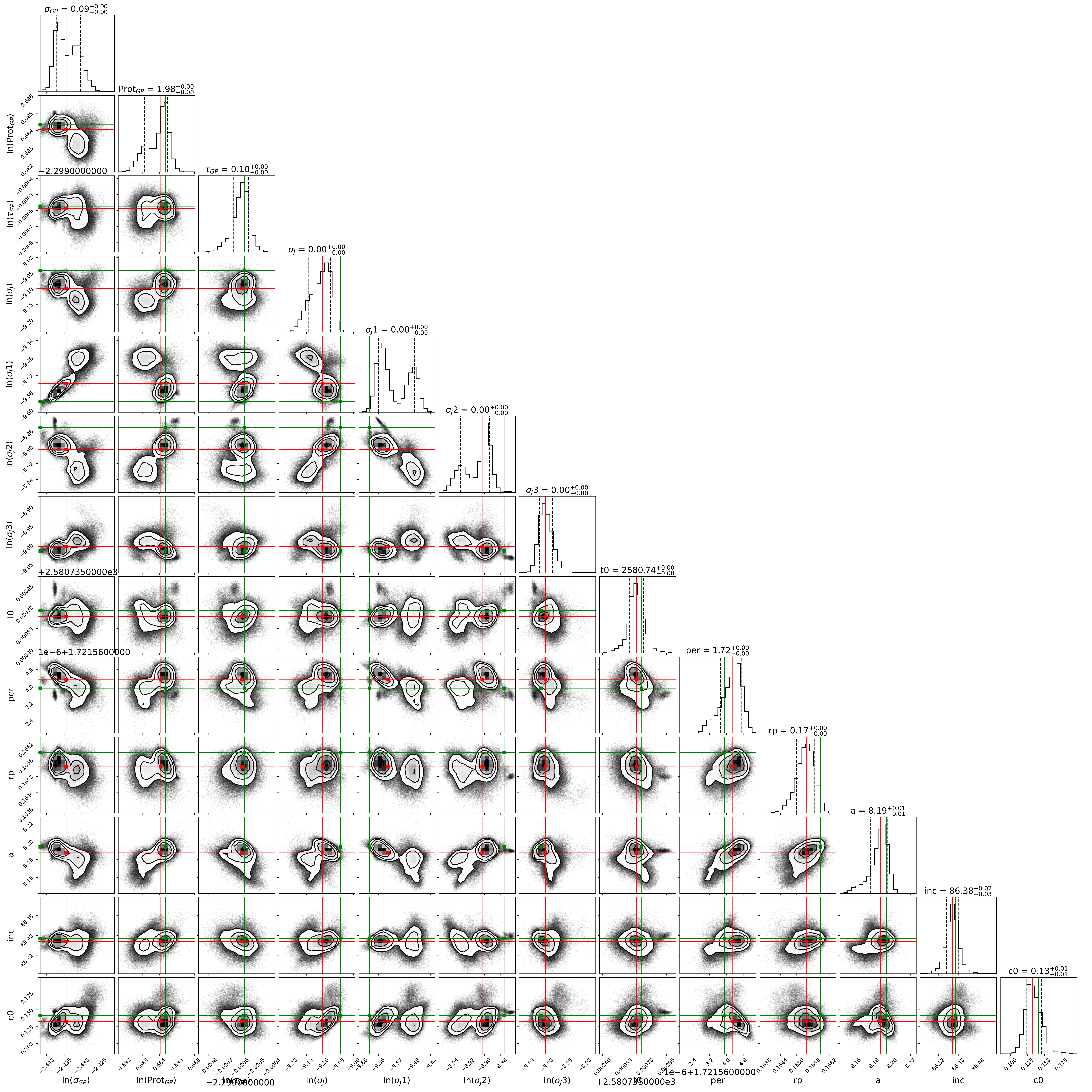}
\caption{Posterior distributions of the Bayesian fit of the photometric time series.}
\label{tess_corner}
\end{figure*}

% As mentioned, the TESS photometric light curve of TOI-5375 shows clear periodic deeps
% compatible with the the transits of a substellar companion orbiting around this young, M dwarf.
 In order to characterise the properties of TOI-5375 b, %the companion,
 the photometric data were modelled by using a Bayesian framework based on a Monte Carlo
 sampling of the parameter space. The photometric light curve contains the contribution
 of the proposed planetary signal as well as the effect of the stellar activity of the
 host star.
 The stellar signal is modelled with a Gaussian process (GP) analysis. The covariance 
 (or kernel) function adopted is a simple harmonic oscillator (SHO) kernel which is
 defined as follows

 \[
 k(i,j)=\sigma_{GP}\exp\left(\frac{-\Delta t}{\tau_{GP}}\right)\left[\cos\left(\frac{q}{P_{rot}}\Delta t\right)+\frac{P_{rot}}{q\tau_{GP}}\sin\left(\frac{q}{P_{rot}}\Delta t\right)\right]
 \]

 \[
 q=2\pi\sqrt{\left(2\pi\frac{\tau_{GP}}{P_{rot}}\right)^{2}-1}
 \]

 \noindent where $k(i,j)$ is the $ij$ element of the covariance matrix, $\Delta t = t_i - t_j$, where $t_i$ and $t_j$ are two times from the photometric data set,
 $\sigma_{GP}$ is the amplitude of the covariance, $P_{rot}$ is the rotation period, and $\tau_{GP}$ is the timescale of the exponential decay.
% The kernel is implemented in the {\tt celerite} algorithm by \cite{2017AJ....154..220F}. %\textbf{check with Antonino.}
 In order to obtain the parameters corresponding to the best transit model we use the {\tt batman} 
 package \citep{2015PASP..127.1161K} with zero eccentricity and a linear limb darkening law.
This hypothesis is justifiable thanks to the strong tidal interaction that tends to quickly damp any initial eccentricity. Of course, 
 this is assuming that there are no other bodies in the system that could re-excite the eccentricity.
% A circular orbit was fixed because, as will we discuss later, the derived planetary period for TOI-5375 b  (1.72 d) is close to the
% the rotational period of the star and therefore the star should be close to syncrhonize its rotation with the orbital period of the companion. 
 Our full model  also contains an extra white noise (jitter) term ($\sigma_{\rm j}$) per each observed
 sector. Before fitting, each sector data were normalised by their median value. 
 The (hyper)-parameter space was sampled by using the publicly available nested sampler and Bayesian inference tool
 {\tt MULTINEST V3.10}, \citep[e.g.][]{Feroz2019}  through the {\tt PYMULTINEST} \citep{Buchner2014} wrapper,
 set to run with 5000 live points and a sampling efficiency 0.3.
 Priors of the model are provided in Table~\ref{tab:priors}.

\begin{figure}[htb]
\centering
\includegraphics[scale=0.325]{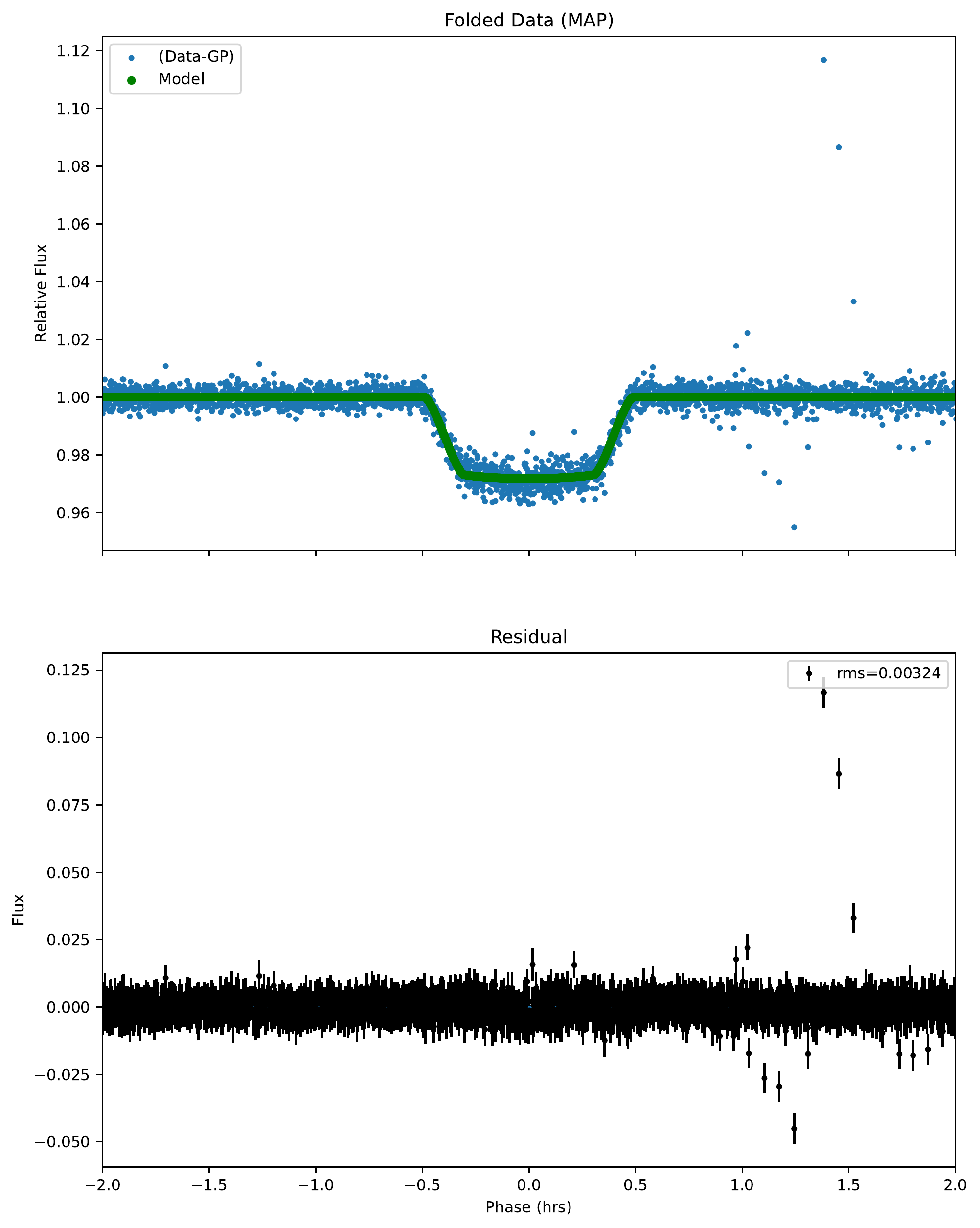}
\caption{
Photometric modelling of TOI-5375. Top: Folded transits in the light curve after the subtraction of the stellar activity signal.
The green line shows the best model fit of the transits. Bottom: Residuals of the light curve after the subtraction of the
companion transit model.}
\label{ph_fit}
\end{figure}

 The resulting posterior distributions are shown in Fig.~\ref{tess_corner} while Fig.~\ref{ph_fit}
 shows the best transit model fits all transits folded in the light curve after the subtraction of the stellar activity signal (top)
 as well as the residuals of the light curve once the planetary transit model is subtracted (bottom).
 The best-fit parameters (median values) are listed in Table~\ref{tab:priors}.
 As seen, the results are consistent with the presence of a companion orbiting around TOI-5375 with a period of 1.72 d. This value is close to,
  but significantly different from, the rotational period of the star.
 Therefore, it is likely that the star is close to synchronising its rotation with the orbital period of the companion
 (further discussion will be provided in Sect.~\ref{discussion}).
 This justifies our choice of a circular orbit.

% >>>>>>>>>>>>>>>>>>>>>>>>>>>>>>>>>>>>>>>>>>>>>>>>>>>>
% Figure: Photometry best fit
% <<<<<<<<<<<<<<<<<<<<<<<<<<<<<<<<<<<<<<<<<<<<<<<<<<<<

   We performed a search for additional companions by performing a GLS periodogram of the best-fit residuals
   \footnote{We note that two additional candidates have been identified by TESS, one using data from Sectors 40 and 47, and a second candidate
    in Sector 60.
    A careful look at their corresponding TESS reports shows that they are likely FP due to the procedure used to flatten
    the light curve used by TESS. This might happen when the star shows a high level of variability and when, as it is our case,
    flares are present in the light curve.
   }.
   The rms of the residuals is very low and no significant additional signals were found.
   %A GLS periodogram of the residuals reveals no significant signals.
   The ``best'' period is found at $\sim$ 1.97 d (that is, at the stellar rotation period)
   but with a Scargle power of $\sim$ 3$\times$10$^{\rm -6}$, which implies a FAP of 100\% 
   (we note that the analytical FAP level of 10\% corresponds to a power of  0.0014).
   % and therefore it is an spurious signal. 
   This result confirms that our treatment of the stellar activity by means of a SHO kernel does effectively remove it from the light curve. 
   It also shows that if there were more companions in the system, they would be difficult to reveal (at least with the data at hand).

%% >>>>>>>>>>>>>>>>>>>>>>>>>>>>>>>>>>>>>>>>>>>>>>>>>>>> 
%% Table of results 
%% <<<<<<<<<<<<<<<<<<<<<<<<<<<<<<<<<<<<<<<<<<<<<<<<<<<<
\begin{table*}[t]
\centering
\caption{ Best-fit values obtained for the star-planet model. Photometric fit parameters are from the analysis of the TESS light curve.}
\label{tab:priors}
\begin{tabular}{lllr}
\hline
%\hline
\noalign{\smallskip}
Parameter & Prior & Description  & Best-fit value \\
\noalign{\smallskip}
\hline
\noalign{\smallskip}
%% ---- Photometry
\multicolumn{4}{c}{\textit{Photometry fit}} \\
\hline
\noalign{\smallskip}
\multicolumn{4}{l}{\textit{GP parameters}} \\
\noalign{\smallskip}
%%+++++++++++++++++++++
$\sigma_{\rm GP}$  (e$^{\rm -}$ s$^{\rm -1}$)  &  $\mathcal{L} \mathcal{U} $(10$^{-2}$, 10$^6$)  & Amplitude of the covariance             & 0.0876$_{-0.0002}^{+0.0004}$  \\
$P_{\rm rot}$ (d)                              &  $\mathcal{N} $(1.97,10$^{-2}$)                 & Rotation period                         & 1.982$_{-0.002}^{+0.001}$     \\
$\tau$ (d)                                     &  $\mathcal{L} \mathcal{U} $(10$^{-1}$, 1$0^4$)  & Time scale of the exponential component & 100300$_{-6}^{+4}$ $\times$ 10$^{-6}$  \\
\noalign{\smallskip}
\multicolumn{4}{l}{\textit{White noise terms}} \\
\noalign{\smallskip}
$\sigma_{\rm j}$ (e$^{\rm -}$ s$^{\rm -1}$)    &  $\mathcal{L} \mathcal{U} $(10$^{-1}$, 10$^2$)  & Additional jitter term sector 40  & 11.16$_{-0.46}^{+0.31}$ $\times$ 10$^{-5}$  \\ 
$\sigma_{\rm j,1}$ (e$^{\rm -}$ s$^{\rm -1}$)  &  $\mathcal{L} \mathcal{U} $(10$^{-1}$, 10$^2$)  & Additional jitter term sector 47  & 7.21$_{-0.16}^{+0.45}$ $\times$ 10$^{-5}$  \\
$\sigma_{\rm j,2}$ (e$^{\rm -}$ s$^{\rm -1}$)  &  $\mathcal{L} \mathcal{U} $(10$^{-1}$, 10$^2$)  & Additional jitter term sector 53  & 13.60$_{-0.36}^{+0.13}$ $\times$ 10$^{-5}$  \\
$\sigma_{\rm j,3}$ (e$^{\rm -}$ s$^{\rm -1}$)  &  $\mathcal{L} \mathcal{U} $(10$^{-1}$, 10$^2$)  & Additional jitter term sector 60  & 12.29$_{-0.19}^{+0.24}$ $\times$ 10$^{-5}$  \\
\noalign{\smallskip}
\multicolumn{4}{l}{\textit{Companion parameters}} \\
\noalign{\smallskip}
$T_{\rm 0}$ (BJD - 2457000 d)        &  $\mathcal{U}$ (2579, 2582) & Time of inferior conjunction            &  2580.73564$_{-0.00005}^{+0.00005}$             \\
$P_{\rm b}$ (d)                      &  $\mathcal{U}$ (1.6, 1.9)   & Period                                  &  1.721564$_{0.0000006}^{+0.000004}$              \\
$R_{\rm b}/R_{\star}$                &  $\mathcal{U}$ (0.0, 0.3)   & Companion-to-stellar-radius ratio       &  0.165$_{-0.0004}^{+0.0003}$                    \\
$a_{\rm b}/R_{\star}$                &  $\mathcal{U}$ (7, 9)       & Semi-major-axis-to-stellar-radius-ratio &  8.187$_{-0.011}^{+0.007}$                      \\
$inc$ (degrees)                      &  $\mathcal{U}$ (70, 90)     & Orbital inclination                     &  86.38$_{-0.02}^{+0.02}$                        \\
$e$                                  &      fixed                  & Eccentricity                            &        0.00                                     \\
%%$\omega_{\rm b}$ (deg)              &      fixed                & Periastron angle                        &                           \\
$c_{\rm 0}$                          &  $\mathcal{U}$ (-1, 1)      & Limb-darkening coefficient              &  0.13$_{-0.01}^{+0.01}$                          \\
%% ---- Radial velocity
\hline
\noalign{\smallskip}
\multicolumn{4}{c}{\textit{Radial velocity fit}} \\
\hline
\multicolumn{4}{l}{\textit{White noise term}} \\
\noalign{\smallskip}
$\sigma_{\rm j}$ (km s$^{\rm -1}$)   &  $\mathcal{L} \mathcal{U}$ (10$^{-3}$, 10$^3$)         & Additional jitter    & 0.29$_{-0.20}^{+0.29}$ \\
\noalign{\smallskip}
\multicolumn{4}{l}{\textit{Companion parameters}} \\
\noalign{\smallskip}
$\gamma$ (km s$^{\rm -1}$)           &   $\mathcal{U}$ (10$^{-3}$, 10$^3$)                    & RV offset                   & -61.59$_{-0.14}^{+0.17}$        \\ 
$P_{\rm b}$ (d)                      &   $\mathcal{N}$ (1.72,10$^{-7}$)                       & Orbital period              & 1.720$_{-0.00}^{+0.00}$         \\
$T_{\rm 0, b}$ (MJD, d)              &   $\mathcal{U}$ (59703.87, 59705.87)                   & Time of periastron passage  & 59705.476$_{-0.004}^{+0.003}$   \\
$K_{\rm b}$ (km s$^{\rm -1}$)        &   $\mathcal{U}$ (10$^{-2}$, 10$^2$)                    & RV semi-amplitude           & 17.49$_{-0.22}^{+0.21}$         \\
$e_{\rm b}$                          &   fixed                                                & Orbital eccentricity        &        0.00                     \\
%\noalign{\smallskip}
\hline
\multicolumn{4}{c}{\textit{Derived quantities}} \\
\hline
%\noalign{\smallskip}
$M_{\rm b}$  ($M_{\rm J}$)           &                       & Mass            &  77     $\pm$ 8    \\
$R_{\rm b}$  ($R_{\rm J}$)           &                       & Radius          &  0.99   $\pm$ 0.16  \\
$\rho_{\rm b}$ (g cm$^{\rm -3}$)     &                       & Density         &  98     $\pm$ 49    \\ 
%19.46  $\pm$ 9.87     \\
$a_{\rm b}$ (au)                     &                       & Semi-major axis           &  0.0251 $\pm$ 0.0012     \\
T$_{\rm eq, b}$ (K)                  &                       & Equilibrium temperature   &  931 - 1107  \\
\hline
\noalign{\smallskip}
\end{tabular}
\end{table*}

% ---------------------------------------------------------------
\subsection{Ground-based photometry analysis} % and search for TTVs}
% ---------------------------------------------------------------

\begin{figure*}[htb]
\centering
\begin{minipage}{0.49\linewidth}
\includegraphics[scale=0.55]{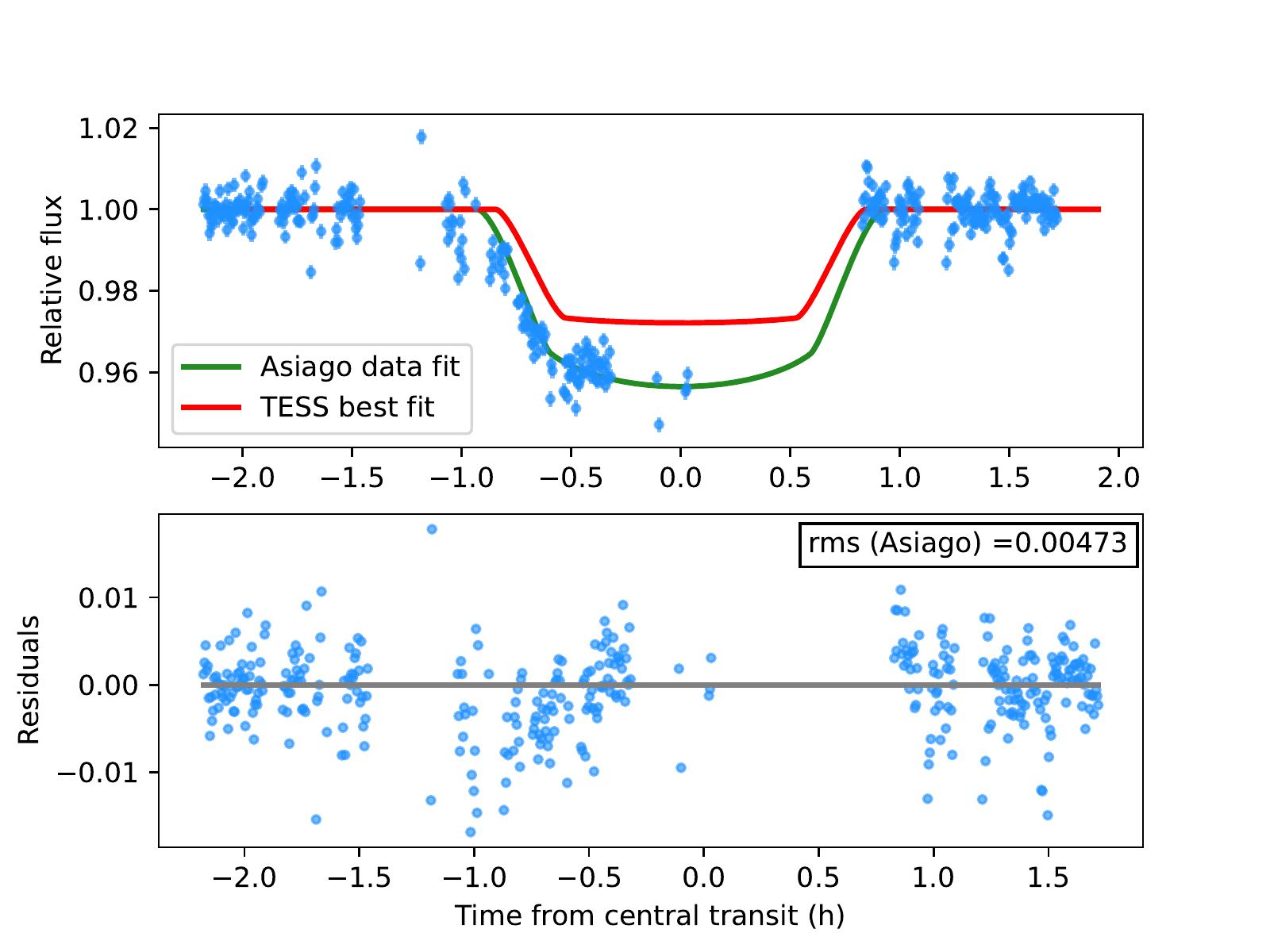}
\end{minipage}
\begin{minipage}{0.49\linewidth}
\includegraphics[scale=0.55]{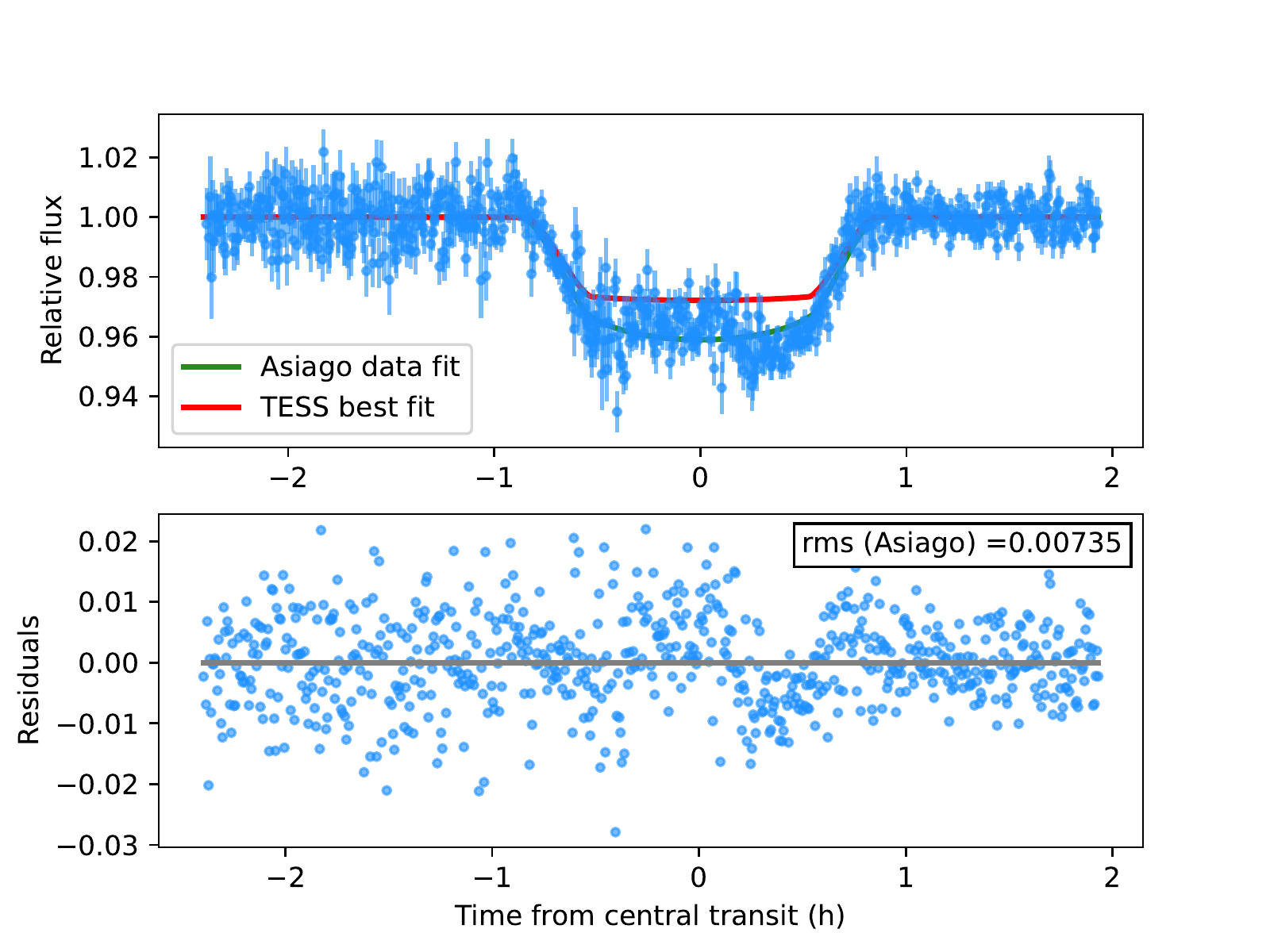}
\end{minipage}
\caption{Photometric observations from Asiago, folded on the transit model fit period of TOI-5375 b.
Left: October 25th, 2022 observations. Right: December 19th, 2022 observations.
The green line shows the best fit to the Asiago dataset, while for comparison purposes we also show in red
the best fit derived for the TESS light curves.
The residuals of the fit to the Asiago ground-based photometry are shown in the bottom panel.}
\label{tess_asiago}
\end{figure*}

 We also fit the ground-based photometry dataset obtained from the Asiago observatory.
 These observations confirm the transit of a companion around TOI-5375 and are additional proof that 
 TOI-5375 b is real and not a false positive due to an BEB. We followed a similar analysis as described  in Sect.~\ref{tessasiago}, see below,
 but we did not consider the dilution factor
  (which defines the total flux that falls into the photometric aperture from contaminants divided by the flux contribution of the 
 target star)
 and a second-order polynomial trend. We treated the stellar limb-darkening contribution by setting the Sloan $r’$ filter in {\tt PyLDTk}.
 From the analysis of the ground-based photometry, 
 we derive a  companion-to-stellar-radius ratio, $R_{\rm b}/R_{\star}$, of  0.1850 $\pm$ 0.0012 %    of 0.1944 $\pm$ 0.0021
 which is larger than the value derived from the analysis of the TESS dataset, see Fig.~\ref{tess_asiago}.
 This difference can be explained by several reasons. 
% To start with, the Asiago observations performed on October 25th
% do not properly sample the full transit (there is a lack of points in the second half of the transit).
% This effect might lead to an overestimation of the transit depth and makes it difficult to determine
% accurate transit ingress/egress times as well as to evaluate the limb darkening.
 To start with, this effect can be due to the fact that the TESS and ground-based observations are taken at different wavelengths 
 (atmospheres absorb larger at some wavelengths and less at others).

Other explanations can be due to stellar activity.
 A visual inspection of the individual TESS transits reveals that some of them show ``bumps'' during the transit that
 can be related to the presence of stellar spots.
 In some cases, there are also negative ``bumps'', likely due to the occultation of faculae.
 We also note that the TESS and the Asiago datasets are not simultaneous.
 To clarify this point we analyse each individual TESS transit as well as the ones observed from the Asiago Observatory.
 In order to analyse the space and the ground-based photometry in a homogeneous way, we use
 the TESS SAP light curve using a dilution factor to take into account the possible contamination
 by nearby stars. In addition, the transit depth is fitted simultaneously to both TESS and Asiago datasets.

% >>>>>>>>>>>>>>>>>>>>>>>>>>>>>>>>>>>>>>>>>>>>>>>>>>>>
% Figure: TESS/Asiago comparison plot
% <<<<<<<<<<<<<<<<<<<<<<<<<<<<<<<<<<<<<<<<<<<<<<<<<<<<

\begin{figure}[!htb]
\centering
\includegraphics[scale=0.45]{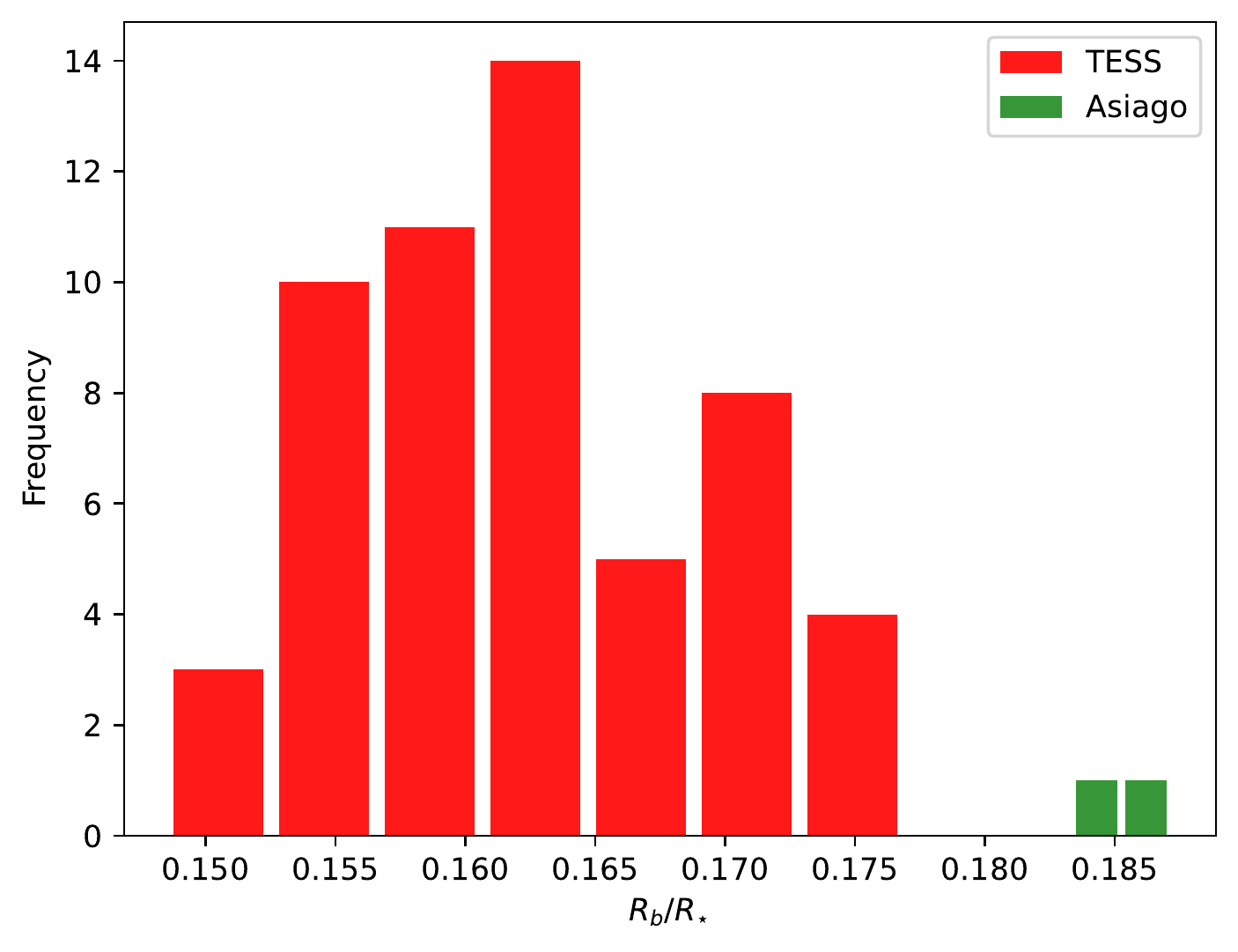}
\caption{Histogram of the derived $R_{\rm b}/R_{\star}$ values from each individual TESS and Asiago transit.}
\label{tess_asiago_histogram}
\end{figure}

 Figure~\ref{tess_asiago_histogram} shows the histogram of the derived $R_{\rm b}/R_{\star}$ for each individual transit.
 Excluding one clear outlier, the values of $R_{\rm b}/R_{\star}$ obtained from the analysis of the individual TESS transits show a rms of 0.053 covering the
 range between 0.151 and 0.172. The apparent variation in the companion radius due to the effects of starspots can be written as:

 \begin{equation}
 \frac{\Delta R}{R}\backsimeq-\frac{1}{2}A_{\lambda}\left(f_{i}-f_{0}\right)
 \end{equation}

 \noindent where $A_{\lambda}$ is the contrast which is a function of the bandpass, $f_{0}$ is the filling factor due to un-occulted spots,
 and $f_{i}$ is the filling factor of spots occulted during the transit. Following \cite{2012A&A...539A.140B}, for an early-M dwarf like TOI-5375, and taking into account that the TESS passband
 covers the range from 600 to 1000 nm and is centred on the Cousins $I$-band, %\footnote{https://heasarc.gsfc.nasa.gov/docs/tess/docs/TESS\_observatory\_guide\_v1.1.pdf},
 we consider $A_{\lambda}$ = 0.924. 
 Then, in the extreme case of $f_{i}$ = 0, we derive
 a value of $f_{0}$ of the order of 0.285 which is a reasonable value for a very active star like this case.
 Therefore, we conclude that the spread in $R_{\rm b}/R_{\star}$ from one transit to another can be explained by the stellar activity of the host star.
 On the other hand, the difference in transit depths between the TESS and the Asiago observations is likely due to a combination of activity and the
 effect of observing at two different bandpass.

  Another possibility could be due to an additional companion (of redder colour) spatially 
 unresolved in {\it Gaia} observations. It should be noted
 that very close binaries ($P$ $<$ 3 days) have very often tertiary companions \citep{2006A&A...450..681T}. 
 Also, an unresolved companion might explain the slight offset in (V-K) colour with respect to the main sequence.
 However, this hypothesis is quite speculative, and we do not have indications of an additional component in
 our spectra.

% >>>>>>>>>>>>>>>>>>>>>>>>>>>>>>>>>>>>>>>>>>>>>>>>>>>>
% Figure: Histogram
% <<<<<<<<<<<<<<<<<<<<<<<<<<<<<<<<<<<<<<<<<<<<<<<<<<<<

% ------------------------------------------------------
\subsection{Independent joint analysis: TESS and ground-based photometry}\label{tessasiago}  
% ------------------------------------------------------

 In order to have an independent check of our results we perform an additional test by analysing the TESS light curve using the 
 {\tt PyORBIT} software\footnote{https://github.com/LucaMalavolta/PyORBIT} \citep{2016ascl.soft12008M,luca} finding similar results.
 In this case, we analyse the ground-based photometry
  simultaneously with the TESS data.
  Furthermore, to test whether or not the difference in transit depth between the Asiago and the TESS data could be due
  to a possible contamination of the TESS aperture,
 we used the Simple Aperture Photometry (SAP) fluxes over Presearch Data Conditioning (PDC) ones. In this way, we conserve
 the contribution of stellar contamination from neighbour stars and later perform our correction for stellar dilution.
% We did the latter by measuring a dilution factor, which defines the total flux that falls into the photometric aperture from
% contaminants divided by the flux contribution of the target star.
We followed \cite{2022MNRAS.516.4432M} to calculate the
 ``fraction of light of the stellar PSF which falls into a given photometric aperture'' (k parameter) for each star,
 and we then got the dilution factor (and its associated error).
 
 We selected from the SAP light curves each transit event and an out-of-transit portion twice as long as the transit duration % in- and an out-of-transit twice as along as the transit duration
 (almost two hours, both before the ingress and after the egress). To perform the selections, we 
 created a mask that flags each transit and cuts portions of the light curves accordingly.
 For each TESS transit, we fitted the following: the central time of transit ($T_{\rm 0}$), the planetary-to-stellar radius ratio,
 the impact parameter, b, the stellar density ($\rho_{\star}$, in solar units), the quadratic limb-darkening (LD) law with the parameters q$_{\rm 1}$ and q$_{\rm 2}$ introduced in
 \cite{2013MNRAS.435.2152K} - where u$_{\rm 1}$ and u$_{\rm 2}$ have been estimated using {\tt PyLDTk}\footnote{https://github.com/hpparvi/ldtk}
 \citep{2013A&A...553A...6H,2015MNRAS.453.3875P} and considering the TESS filter, the dilution factor, a second-order polynomial trend
 (with c$_{\rm 0}$ as the intercept, c$_{\rm 1}$ as the linear coefficient, and c$_{\rm 2}$ as the quadratic coefficient), and a jitter term
 (different for each sector, to take into account possible TESS systematics and short-term stellar activity noise) to be added in quadrature to the errors
 of TESS light curves. We imposed a Gaussian prior on the stellar density, limb-darkening coefficients, and dilution factor, while we kept fixed the period,
 eccentricity, and argument of pericentre. %This analysis has been done simultaneously with the observed ground-based photometry (see below).  

 We performed a global optimisation of the parameters by running a Differential Evolution algorithm, {\tt pyDE} \citep{storn}
 and then a Bayesian analysis of each selected light curve around each transit. 
 To perform the latter, we used the affine-invariant ensemble sampler \citep{2010CAMCS...5...65G} for Markov Chain Monte Carlo (MCMC) implemented within the {\tt emcee} package \citep{2013PASP..125..306F}.
 We modelled each transit with {\tt batman} \citep{2015PASP..127.1161K}. We fit each transit with $4n_{dim}$ walkers (with $n_{dim}$ being the model’s dimensionality) for 20000 generations with {\tt pyDE} and then with 300000 steps
 with {\tt emcee}  – where we applied a thinning factor of 200 to reduce the effect of the chain auto-correlation. We discarded the first 25000 steps (burn-in) after checking the convergence of
 the chains with the Gelman–Rubin (GR) statistics \citep[][threshold value \^{R} = 1.01]{1992StaSc...7..457G}.
 From this analysis, we derive a companion-to-stellar-radius ratio, $R_{\rm b}/R_{\star}$, of  0.1699 $\pm$ 0.0011, similar to the value
 obtained in the previous analysis ($R_{\rm b}/R_{\star}$, of 0.165 $\pm$ 0.002) . The small difference is likely due to the fact that this time, we analyse the ground-based photometry
 simultaneously with the TESS data. 
 The best-fitting parameters and the corresponding plots are shown in Appendix ~\ref{app_tess_asiago}.

% ---------------------------------------------------------------
\subsection{Search for TTVs and other variability effects}
% ---------------------------------------------------------------

\begin{figure}[!htb]
\centering
\includegraphics[scale=0.65]{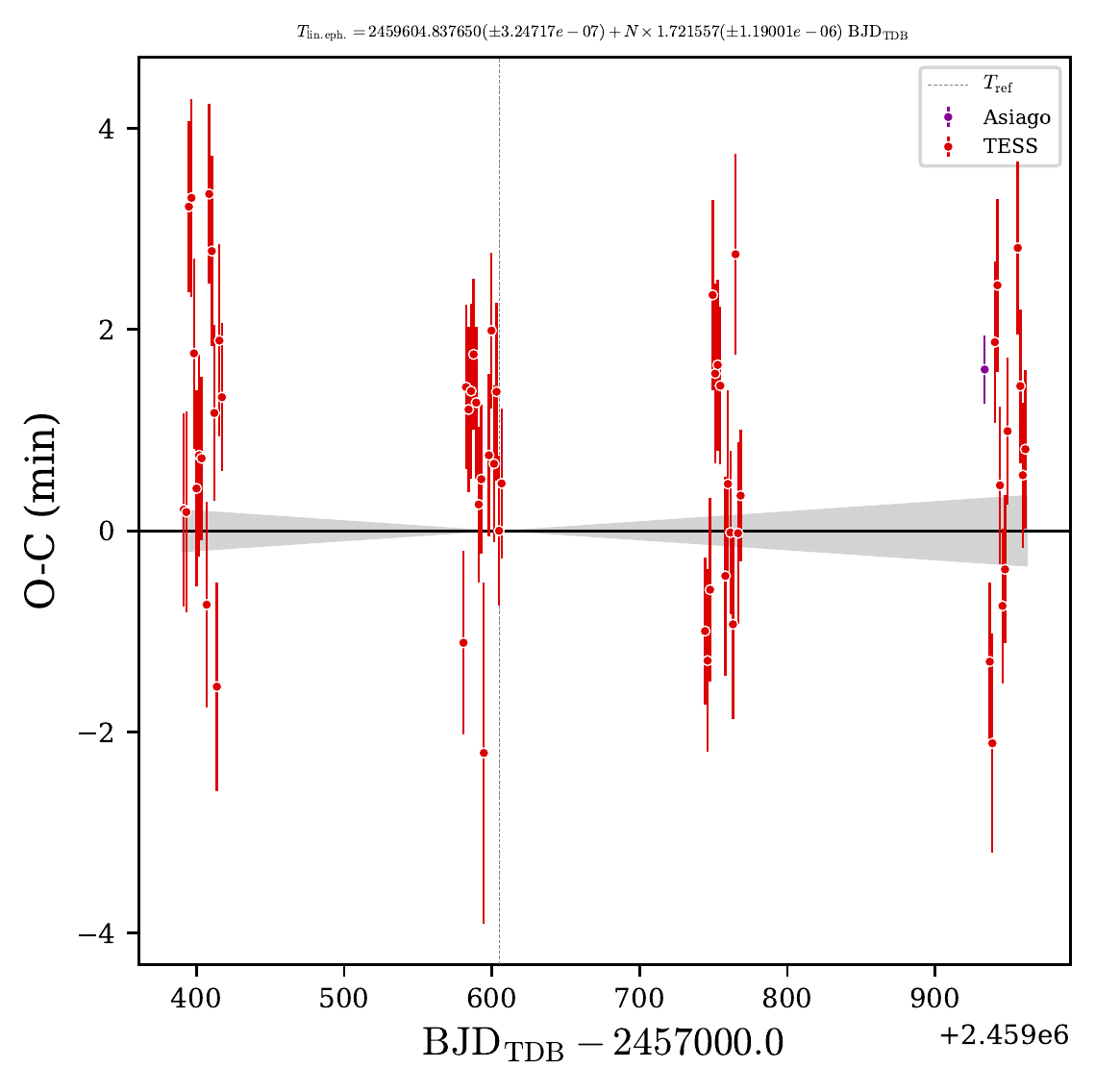}
\caption{ Residuals of the timings of mid-transits versus a linear ephemeris.}
\label{ttv_plot}
\end{figure}

Finally, even though the residuals of the fit of the TESS photometry do not reveal additional signals, % orbiting around TOI-5375 we perform
a search for Transit Timing Variations (TTVs) due to a possible additional companion orbiting around TOI-5375 was performed
\citep[see e.g.][]{2019MNRAS.484.3233B}. 
To this purpose we analyse each individual TESS transit as well as the second transit observed from the Asiago Observatory 
(the first transit is not included, since, as we have seen, its sampling is not complete).
%In order to analyse the space and the ground-based photometry in an homogeneous way, we use
%the TESS SAP light curve using a dilution factor to take into account the possible contamination
%by nearby stars. In addition, the transit depth is fitted simultaneously to both TESS and Asiago datasets. 
Once the transit times were derived, we fit them with a straight line to obtain a linear ephemeris:

\begin{eqnarray}
%\begin{equation}
T_{\rm 0} = 2459604.837650 \, (\pm 3\times10^{-7}) \\ 
          +  N \times 1.721557 \, (\pm 2\times 10^{-6}) \, BJD_{TDB}
%\end{equation}
\end{eqnarray}

 \noindent where $N$ is the epoch of the transit. The corresponding fit is shown in Fig.~\ref{ttv_plot}.
% We note that in this fit the transits from Asiago are not included. % since, as we have seen, the sampling of
% the transit is not complete. 
 A search for periodicities in the residuals was performed by means of a GLS periodogram. The periodogram
 identifies a significant peak at  12.72  $\pm$ 0.03 d with a FAP lower than 1\%. The semi-amplitude of the signal is $\sim$ 1 minute. 
 As said, some of the TESS transits are affected by stellar activity. 
% A visual inspection of the individual TESS transits reveals that some of them show ``bumps'' during the transit that
% can be related to the presence of stellar spots. 
 Indeed, according to the simulations performed by \citet[][Fig.~8]{2013A&A...556A..19O}, 
 for a transit with $R_{\rm b}/R_{\star}$ $\sim$ 0.15, TTVs of the order of one minute can be explained by the presence
 of stellar spots with a filling factor of $\sim$ 0.6\%. 
 Given the very active nature of TOI-5375, our conclusion is that the $\sim$ 1 minute amplitude TTV signal seen in this star
 is most likely related to stellar activity. 
 We note that the synodic period between the stellar rotation and the orbital period is  13.59 days, which is close to the apparent TTV period of 12.7 days. This reinforces the interpretation in terms of surface activity features of the apparent TTVs.

% >>>>>>>>>>>>>>>>>>>>>>>>>>>>>>>>>>>>>>>>>>>>>>>>>>>>
% Figure: TTV variations
% <<<<<<<<<<<<<<<<<<<<<<<<<<<<<<<<<<<<<<<<<<<<<<<<<<<<

 Finally, we estimate the photometric flux variations due to other variability effects.
 In particular, we consider the Doppler beaming (the reflex motion of the star induced by its planetary companion 
 produces Doppler variations in its photometric flux), the ellipsoidal light variations due to tidal effects on the star from the planet
 as well as the reflected light in the companion surface.
 Following \cite{2003ApJ...588L.117L} we derive flux variations, $\Delta F/F_{\rm 0}$, of the order of
 2.8 $\times$ 10$^{\rm -4}$, 9.4 $\times$ 10$^{\rm -5}$, and 2.4 $\times$ 10$^{\rm -4}$ for Doppler beaming,
 ellipsodial light variations, and reflected light, respectively.
 The calculation of Doppler beaming was done in the TESS band ($\lambda$ $\sim$ 7850 \AA), while for the 
 reflected light effect a geometric albedo of 2/3 was considered. 
 All these effects are extremely small compared with the intrinsic variability of the star (the TESS light curve
 shows a rms of 26.7, 54.35, 68.68, and 43.96 e$^{\rm -}$ s$^{\rm - 1}$ in Sectors 40, 47, 53 and 60, respectively).

% ---------------------------------------------------------------
\subsection{High energy events in the TESS light curve}\label{flares}
% ---------------------------------------------------------------

\begin{figure*}[htb]
\centering
\begin{minipage}{0.33\linewidth}
\includegraphics[scale=0.425]{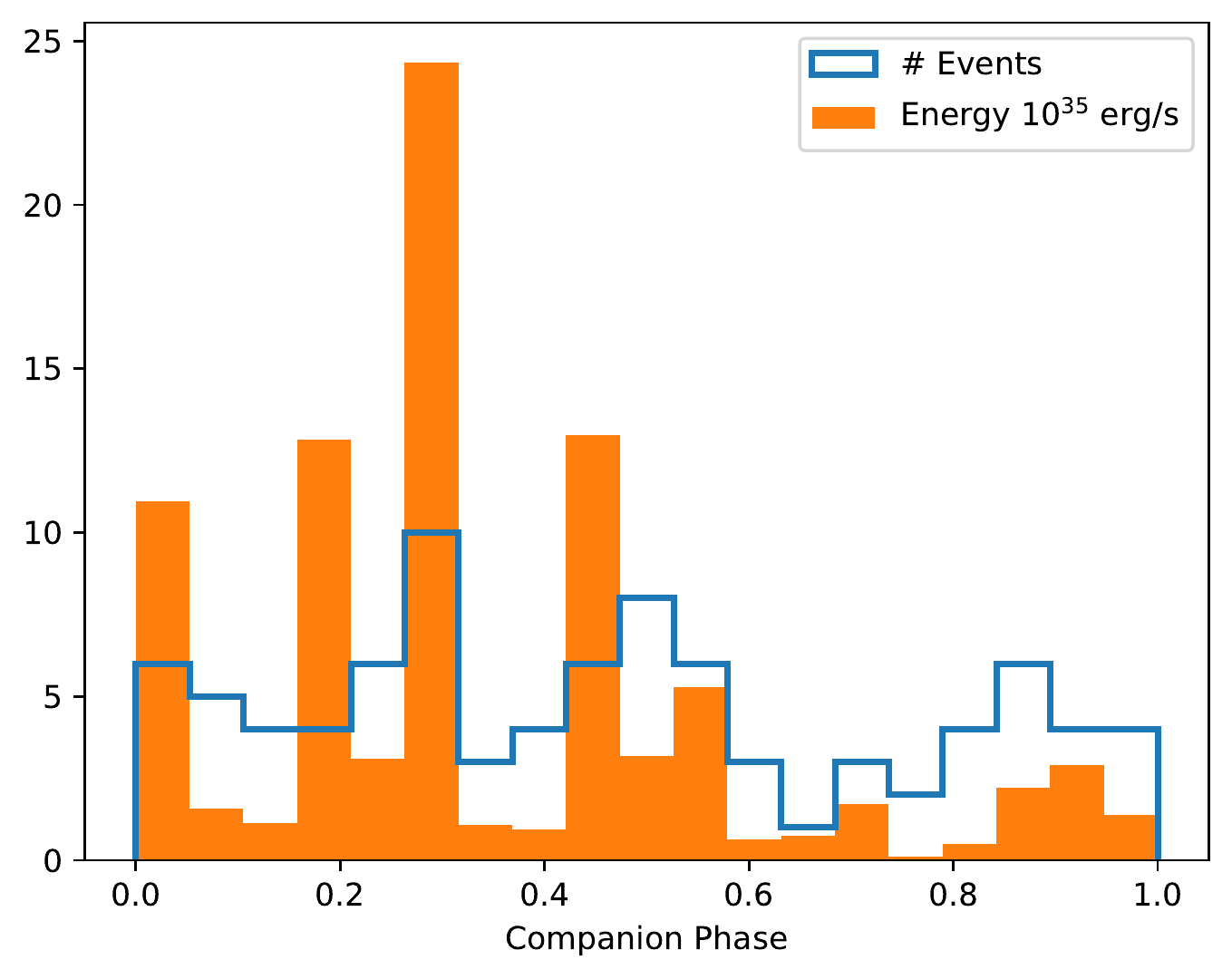}
\end{minipage}
\begin{minipage}{0.33\linewidth}
\includegraphics[scale=0.425]{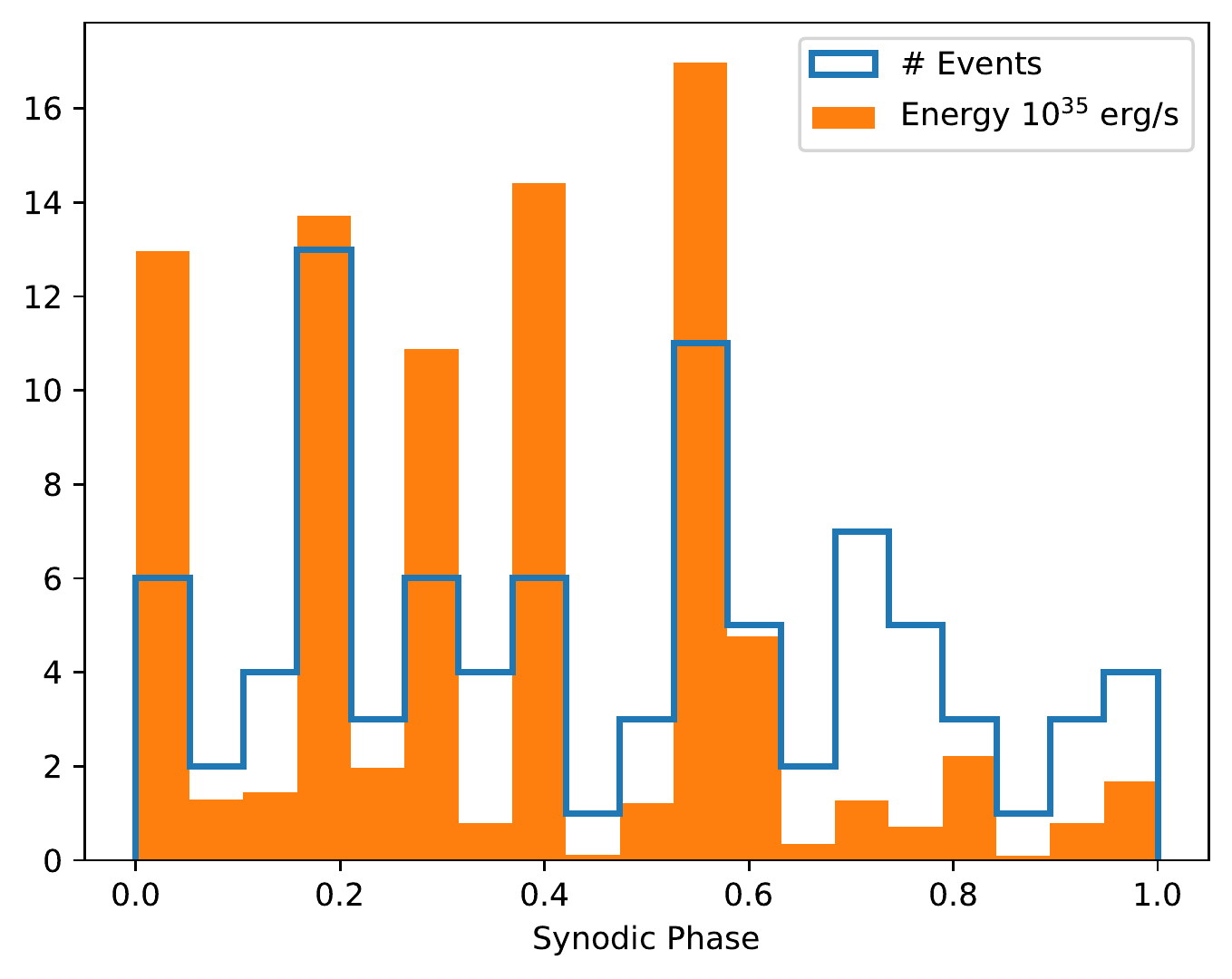}
\end{minipage}
\begin{minipage}{0.33\linewidth}
\includegraphics[scale=0.425]{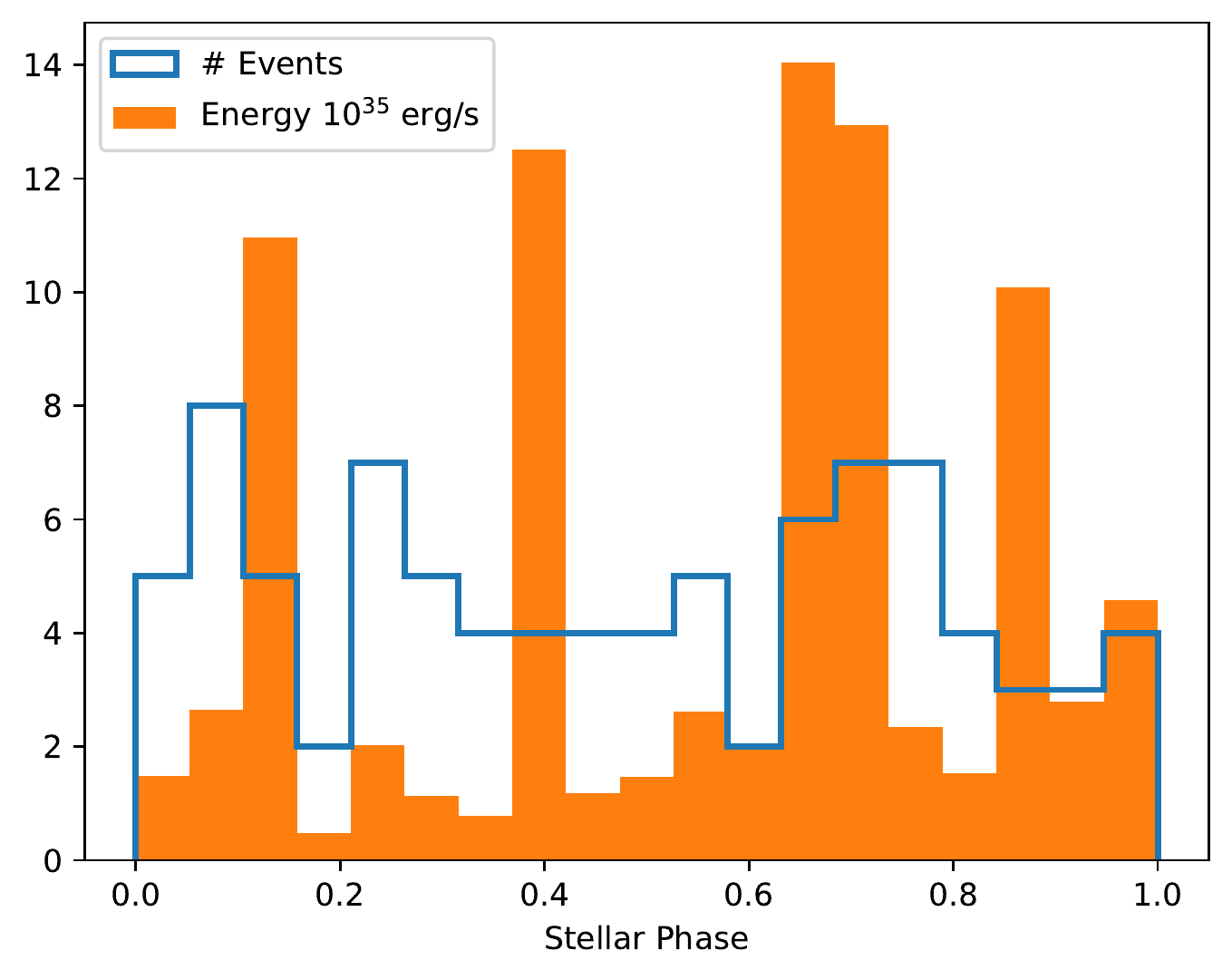}
\end{minipage}
\caption{Histogram of impulsive events (blue empty line) and amount of energy emitted (orange filled histogram) phase-folded to the companion orbital period (left), the synodic period (centre)
and the stellar period (right).
}
\label{flares_histograms}
\end{figure*}

 Many energetic events or flares seem to be present in the TESS light curve (see Fig.~\ref{tess_time_series}). %data.
 This is in agreement with the high rotation level of this star but it might also
 be  evidence of a possible interaction between TOI-5375 and its companion. 
 In order to test this hypothesis we use the methodology described by \cite{2022A&A...661A.148C}
 to identify the impulsive events and derive their amplitudes and the amount of energy emitted.
 We then phase-fold the times at which the events take place with the companion period (1.72 d), the rotation
 period (1.97 d) as well
 as the synodic period of the TOI-5375 system. 
 The synodic period is defined as $1/P_{\rm synodic} = 1/P_{\rm companion} - 1/P_{\rm rotation}$, that is,
 P$_{\rm synodic}$ = 13.59 d. 
 Figure~\ref{flares_histograms} shows the number of events (blue empty histogram) and
 the released energy (orange filled histogram) phase-folded with the companion period (left) and with the synodic
 period (centre), and the stellar period (right).
 In order to test whether the distribution of the released energy is or not uniform we made use of a Kolmogorov-Smirnov (KS) test
 \footnote{Computed with the Python {\tt scipy} ks\_1samp routine.}.
 The results are shown in Table~\ref{ks_test}. In all cases, we test the null hypothesis that the phase-folded released energy distribution
 is drawn from a uniform distribution. 
 The results from the K-S test show that the distribution of events is far from uniform which supports the star-companion
 interaction scenario.
 When phased to the stellar rotation period, four main peaks of events can be seen at phases
 $\sim$ 0.1, 0.4, 0.7 and $\sim$ 0.9.
 The first peak is driven by an energetic event in Sector 53 (see Fig.~\ref{tess_time_series}, green square).
 The peak at phase $\sim$ 0.4 corresponds to an energetic event in Sector 47 (shown as a purple triangle in 
 Fig.~\ref{tess_time_series}), which is followed by many more, less energetic, events.
 The peak at phase $\sim$ 0.7 is driven by two high-energetic events in Sectors 60 and 53
 (marked with a red diamond and a red circle in Fig.~\ref{tess_time_series}), again followed by several less energetic
 events. Finally, it can be seen a fourth peak close to a phase 0.9 
 corresponding to a high-energy event in Sector 40 (see Fig.~\ref{tess_time_series}, red circle).
 On the other hand, if the distribution of events is folded with respect to the period of the orbital companion 
 (Fig.~\ref{flares_histograms}, left) or the synodic period (Fig.~\ref{flares_histograms}, centre),
 the impulsive events
 are mainly at phase values between 0.0 and 0.6 and are driven by the five energetic events already discussed.

% +++++++++++++++++++++++++++++++++++++++++++++++++
% Table K-S tests
% ++++++++++++++++++++++++++++++++++++++++++++++++
\begin{table*}[htb]
\centering
\caption{Results of the K-S tests performed in this work.}
\label{ks_test}
\begin{tabular}{lccc}
\hline\noalign{\smallskip}
 Distribution phase-folded to         & K-S statistic & $p$-value                    & H$^{\ddag}_{\rm 0}$ \\
\hline 
Companion's orbital period (1.72 d)   & 0.74          & 4.30 $\times$ 10$^{\rm -11} $ &                   1 \\
System's synodic period    (13.59 d)  & 0.68          & 2.30 $\times$ 10$^{\rm -9}  $ &                   1 \\
Stellar rotation period    (1.97 d)   & 0.89          & 5.30 $\times$ 10$^{\rm -19} $ &                   1 \\
\hline
\end{tabular}
\tablefoot{$^{\ddag}$ 0: Accept null hypothesis; 1: Reject null hypothesis.}
\end{table*}

% +++++++++++++++++++++++++++++++++++++++++++++++++++++
% Histogram of energetic events/flares
% +++++++++++++++++++++++++++++++++++++++++++++++++++++

% ---------------------------------------------------------------
\subsection{RV time series analysis}
% ---------------------------------------------------------------

\begin{figure}[!htb]
\centering
\includegraphics[scale=0.5]{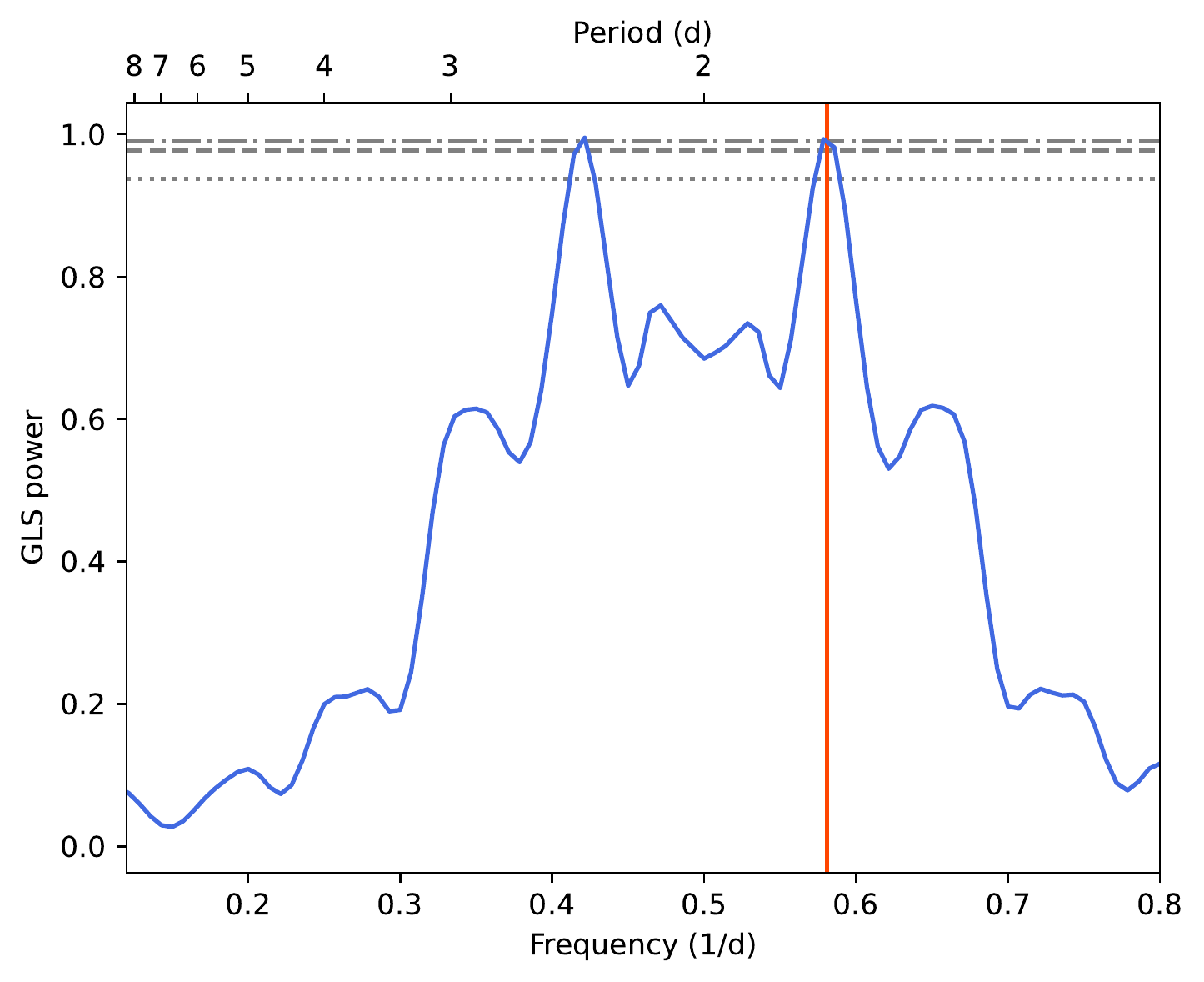}
\caption{GLS periodogram of the RV dataset.
Values corresponding to a FAP of 10, 1, and 0.1\% are shown with horizontal grey lines. The vertical line indicates the period of TOI-5375 b at 1.72 d.}
\label{harpn_gls}
\end{figure}

 Table~\ref{rv_timeseries} provides the RVs time series for TOI-5375 analysed in this work
 which, as discussed earlier, are those obtained by the LSD method. 
% while the corresponding plot is shown in figure~\ref{rv_fit} (top).
 The RV data show a rms of 12.42 km s$^{\rm -1}$, which 
 is approximately 35 times the mean error of the measured RVs (0.36 km s$^{\rm -1}$).
 The GLS periodogram (see Fig.~\ref{harpn_gls}) reveals a double peak structure with significant frequencies at
 f$_{\rm 1}$ = 0.4202 $\pm$ 0.0008 d$^{\rm -1}$ (P = 2.380 $\pm$ 0.004 d), and 
 f$_{\rm 2}$ = 0.5810 $\pm$ 0.0006 d$^{\rm -1}$ (P = 1.721 $\pm$ 0.002 d).
 For frequencies values larger than 0.5 d$^{\rm -1}$ the periodogram is clearly a mirror copy of the one
 between 0 and 0.5 d$^{\rm -1}$ so both peaks should be related.
 The significance of these frequencies was tested by using a bootstrapping analysis to compute the FAPs
 %alarm probability (FAPs) 
 values \citep[see e.g.][]{2001A&A...374..675E}.
% For frequencies values larger than 0.5 d$^{\rm -1}$ the periodogram is a mirror copy of the one
% between 0 and 0.5 d$^{\rm -1}$ so both peaks should be related. 
 The signal at 1.72 d agrees well with the periodicity found 
 for the transit of TOI-5375 b
 in the light curve analysis. On the other hand, the period at 2.38 d is likely due to the observing sampling 
 as our observations have
 a gap of $\sim$ 6 d (|f$_{\rm 1}$ - f$_{\rm 2}$| $\sim$ 1/6 d$^{\rm -1}$).

% >>>>>>>>>>>>>>>>>>>>>>>>>>>>>>>>>>>>>>>>>>>>>>>>>>>>
% Figure: RV photometry periodogram
% <<<<<<<<<<<<<<<<<<<<<<<<<<<<<<<<<<<<<<<<<<<<<<<<<<<<

% --------------------------------------------------------------
% Table with RV values 
% --------------------------------------------------------------
\begin{table}[!htb]
\centering
\caption{RVs (with their corresponding uncertainties) of TOI-5375 used in this work.}
\label{rv_timeseries}
\begin{tabular}{lcc}
\hline\noalign{\smallskip}
 MJD             & RV                & $\Delta$RV        \\
    (d)          & (km s$^{\rm -1}$) & (km s$^{\rm -1}$) \\
\hline
59703.87173611112  &    -46.37 & 0.42 \\
59705.87925925944  &    -59.28 & 0.33 \\
59706.89432870364  &    -52.57 & 0.45 \\
59707.880509259645 &    -75.41 & 0.82 \\
59713.876840277575 &   -48.67  & 0.36 \\
59714.874594907276 &   -78.84  & 0.33 \\
59715.87714120373  &   -44.97  & 0.18 \\
59716.87711805524  &   -73.85  & 0.17 \\
59717.86982638901  &   -56.77  & 0.15 \\
\hline
\end{tabular} 
\end{table}

\begin{figure}[!htb]
\centering
\includegraphics[scale=0.275]{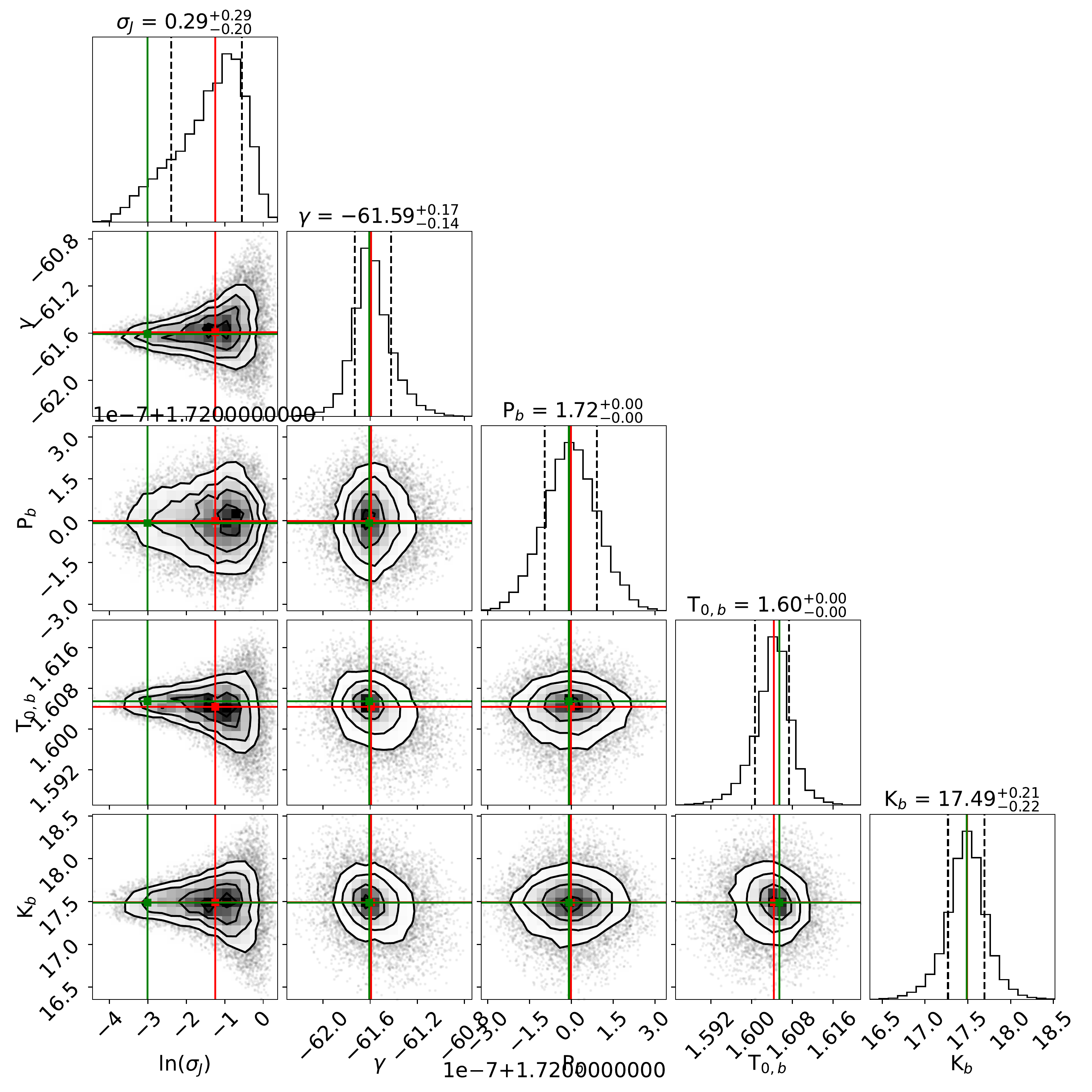} %YO48_rv_model_february01.pdf}
\caption{Posterior distributions of the Bayesian fit of the RVs.}
\label{rv_corner}
\end{figure}

\begin{figure}[htb]
\centering  %0.25
\includegraphics[scale=0.55]{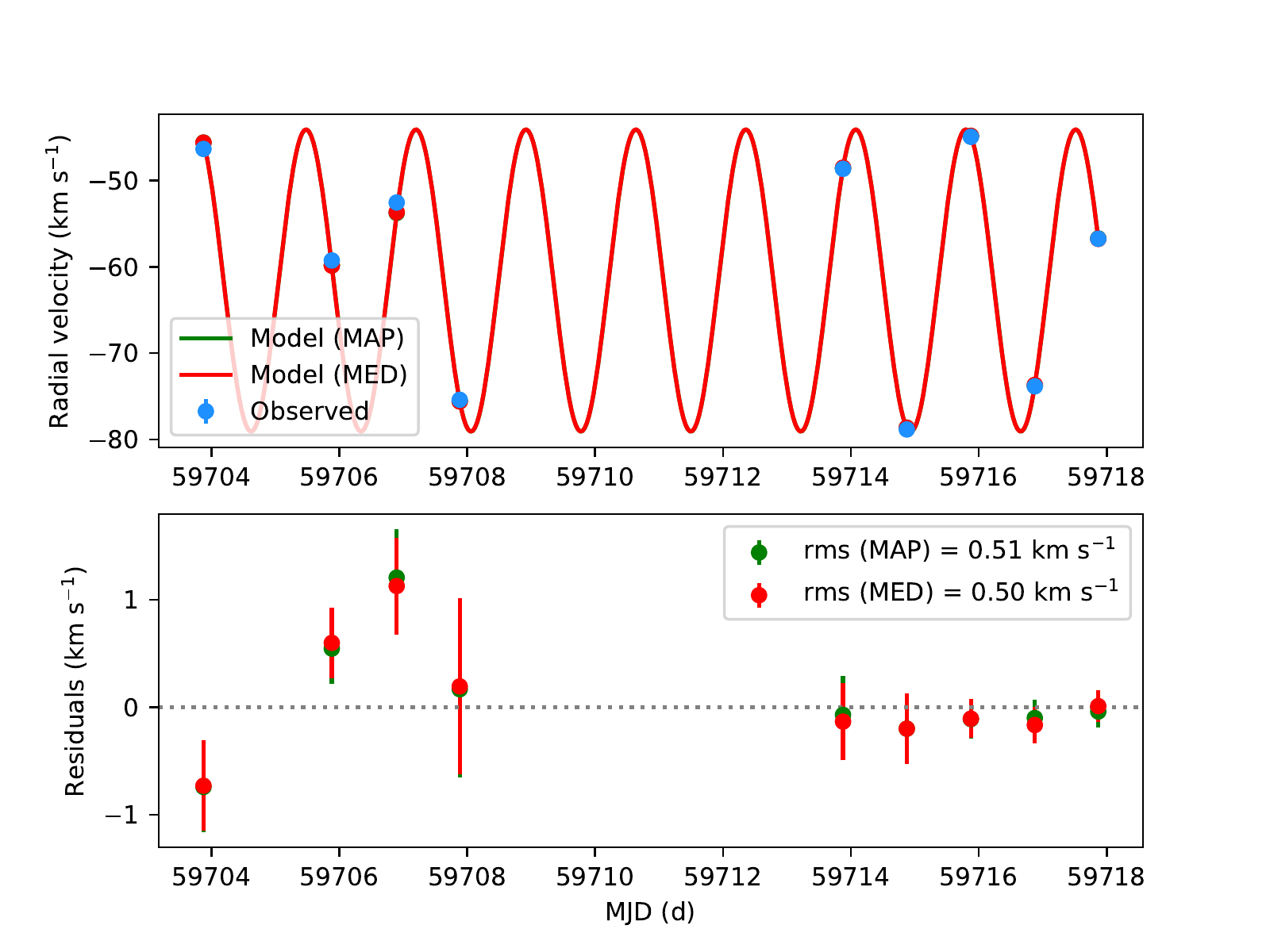} %YO48_rv_rainer_revised_february02_fit_ver.pdf} 
%monica_best_fit.pdf}
\caption{%TOI-5375 RVs time series (top), 
Best-fit planetary model for TOI-5375 b (top), and RV residuals (bottom).
%The X axis has been shifted so time equal to zero corresponds to the first observation. 
}
\label{rv_fit}
\end{figure}

 We have applied the same Bayesian framework as for the photometric curve to the RV data but, in this case, the model contains only a Keplerian signal and the number of live points has been set to 2000.
 Given the small number of RVs points, we do not perform a simultaneous analysis of the RV with the photometry. Instead,
 we take advantage of the photometry fit and set tight priors  in the RV fitting on all parameters but in K$_{\rm b}$, where
 a wide prior was considered, see Table~\ref{tab:priors}.
 The resulting posterior distributions are shown in Fig.~\ref{rv_corner} while Fig.~\ref{rv_fit}
 shows the best-fit planet model (top) and the corresponding RV residuals (bottom). %, and the RV curve folded
 %at the best-fit orbital period of the companion (bottom). 
 The best-fit parameters are also listed in Table~\ref{tab:priors}.
 No significant signals are found in the analysis of the RV residuals. 
 We note that the resiudals show a larger dispersion during the first observing season. 
  A possible explanation for this could be related to the high variability and activity level of TOI-5375.

% >>>>>>>>>>>>>>>>>>>>>>>>>>>>>>>>>>>>>>>>>>>>>>>>>>>>
% Figure: RV corner plot
% <<<<<<<<<<<<<<<<<<<<<<<<<<<<<<<<<<<<<<<<<<<<<<<<<<<<

% >>>>>>>>>>>>>>>>>>>>>>>>>>>>>>>>>>>>>>>>>>>>>>>>>>>>
% Figure: RV best fit
% <<<<<<<<<<<<<<<<<<<<<<<<<<<<<<<<<<<<<<<<<<<<<<<<<<<<

 For the sake of completeness, we also measured the main activity indexes of chromospheric activity,
 namely the Ca~{\sc ii} H \& K index, H$\alpha$, and Na~{\sc i} D$_{\rm 1}$, D$_{\rm 2}$ \citep[see e.g.][]{2019A&A...627A.118M}.
 All indexes show a symmetric periodogram around 0.5 d$^{\rm -1}$ with several peaks, all with a significance below the 10\%
 analytical FAP value. 
 Only the H$\alpha$ index shows a dominant peak, close to the rotation period at $\sim$ 1.99 d.
  The time series of the different activity indexes as well as their corresponding periodograms
 are shown in Appendix~\ref{app_activity}.
 %with an analytical 
 %FAP $\sim$ 10\%. 

%----------------------------------------------------------------
\section{Discussion}\label{discussion}
%----------------------------------------------------------------
 
\begin{figure}[htb]
\includegraphics[scale=0.55]{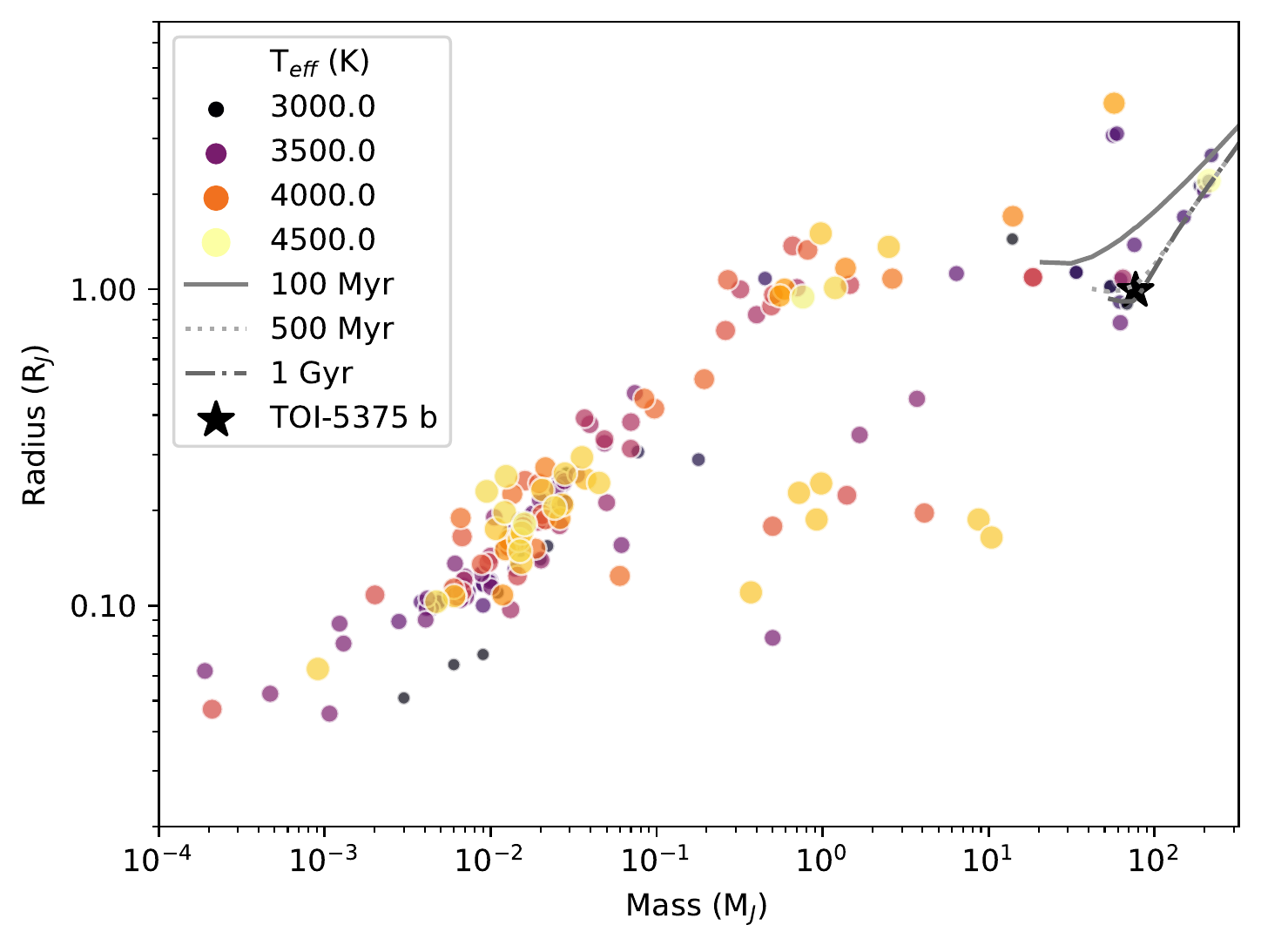} %#YO48-2022-12-13-radius-mass-teff.pdf}
\caption{
Known substellar companions and low-mass stars (radius versus mass) around late-K and M dwarfs
(As listed at http://exoplanet.eu/).
Different colours and symbol sizes indicate the
effective temperature of the host star. Models for substellar objects and low-mass stars with solar metallicity from \citet{2015A&A...577A..42B}
are overplotted with grey lines. 
}
\label{mass_radius_fit}
\end{figure}

 From the best values of $K_{\rm b}$ and $P_{\rm b}$ derived in the previous section,
 we derived a mass for TOI-5375 b of 77 $\pm$ 8 ($M_{\rm J}$) and a semi-major axis $a$ = 0.0251 $\pm$ 0.0012 au.
 On the other hand, from the analysis of the light curve, we obtain a radius of 0.99 $\pm$ 0.16 ($R_{\rm J}$). 
 TOI-5375 b is therefore an object
 in the boundary between the brown dwarfs and
 %substellar companion %in %the 
% the lower limit for
 the very-low-mass stars.
% brown dwarf regime. 
 This is a remarkable result, as massive substellar
 companions orbiting around M dwarfs %as well as orbiting around young stars
 are rare (see Sect.~\ref{introduction}). % {\bf references}  
%  Indeed, it seems the most massive substellar companion (at the type of writing) orbiting both types of stars. 
 Figure~\ref{mass_radius_fit} %(left) 
 shows the position of TOI-5375 b in the planetary radius versus planetary mass diagram.
 For comparison purposes, the location of the known substellar companions around late-K and M dwarf stars
 (with known values of planetary radius and mass) are shown.
 In order to discuss our results in a broader context, we also show the location of several known
 brown dwarfs and low-mass stars orbiting around a late-K or M dwarf primary \citep[][]{2020AJ....160..133M,2022arXiv221202502C}.

% It can be seen that TOI-5375 b is the most massive known companion orbiting around a low-mass star.
 The figure shows two regimes of the mass-radius relationship.
 Up to planetary masses of the order of $\sim$ 0.2 $M_{\rm J}$, the planetary radius steadily increases with the planetary mass.
 For higher masses, the radius remains constant and even shows a slight decrease with increasing planetary mass. 
 TOI-5375 b is clearly
 located
 in the region of the diagram where $R_{\rm p}$ is roughly constant ($\sim$ 1 $R_{\rm J}$) with $M_{\rm p}$.
 The two regimes of the mass-radius relationship are related to the boundary between objects 
 with a dominating hydrogen and helium composition and objects that consist of
 different composition and different mass-radius relationships \citep[see e.g.][and references therein]{2017A&A...604A..83B}.

% >>>>>>>>>>>>>>>>>>>>>>>>>>>>>>>>>>>>>>>>>>>>>>>>>>>>
% Figure: Mass/Radius relationship for low-mass stars / young stars
% <<<<<<<<<<<<<<<<<<<<<<<<<<<<<<<<<<<<<<<<<<<<<<<<<<<<

%%%%%%%%%%%%%%%%%%%%%%%%
% Discussion about tidal forces
%%%%%%%%%%%%%%%%%%%%%%
 
 The TOI-5375 system constitutes an interesting case  study that can help us to shed light on %into %our understanding of the
 the
 tidal interactions between close companions and their host stars. With an orbital period for TOI-5375 b of 1.72 d, a rotational
 period for the host star of 1.97 d, %and given its young age of $\sim$ 100 Myr, 
 the system is likely being observed in the short time frame
 %currently observed in the short phase
 in which the star is close 
 to synchronising its rotation with the orbital period of the companion. %brown dwarf. 
 Tidally driven orbital circularisation and synchronisation is a well-known phenomenon in binary systems containing low-mass 
 stars \citep[e.g.][]{1989A&A...223..112Z}, whilst the evidence for similar phenomena in systems harbouring short period hot Jupiter planets have been recently
 discussed \citep{2014MNRAS.440.1470B,2016A&A...588L...6M,2017AJ....154....4P,2017ApJ...836L..24W,2019AJ....157..217B,2020ApJ...888L...5Y}.

 Taking into account the properties of the TOI-5375 system, tidal dissipation should be mainly driven by inertial waves
 \citep{2007ApJ...661.1180O}.
 Under these circumstances,
 and following the model described in \cite{2020MNRAS.498.2270B}, we derive a tidal quality factor due to inertial waves of $Q'_{\rm IW}$ = 3.9 $\times$ 10$^{\rm 5}$.
 %This implies that the orbit of the brown dwarf companion decays very quickly.
 We use the formulae of \cite{2010A&A...516A..64L} which takes into account
 the rotation of the star in the evolution of the system parameters to derive the characteristic time-scales 
 (the e-folding time $\tau$ of a given parameter $q$ is defined as $\frac{1}{\tau}=\frac{1}{q}\frac{dq}{dt}$) for the TOI-5375 system. 
 We derive an estimated time-scale for tidal synchronisation of the stellar spin of
 only $\sim$ 3.36 Myr and a circularisation time-scale of $\sim$ 6.76 Myr.
 Such a synchronisation timescale is much shorter than the angular momentum loss timescale of the stellar magnetised wind 
 (of the order of 100 Myr given the fast rotation of the star), thus supporting the statement of Sect.~\ref{sect_age}  
 that we cannot use gyrochronology to estimate the age of this star. As a matter of fact, the rotation period of TOI-5375
 is expected to be longer than about 10 days from the central plot in Fig.~\ref{age_estimates}, given that it is likely to belong to the thin disk stars, that is, 
 it is probably older than the Hyades. Conversely, it shows a rotation period as short as 1.97 days
 that we interpret as the consequence of the tidal acceleration given an estimated present synchronisation timescale much shorter than the expected magnetic braking timescale.
The characteristic time-scale for the semi-major axis decay is $\sim$ 115 Myr, while the obliquity of the orbit has a decay time-scale of $\sim$ 0.72 Myr.
The total angular momentum of the system is about 1.61 times larger than the minimum value required to reach a final stable tidal equilibrium with the star and the low-mass companion fully synchronised on a circular orbit. After that stage, the orbit will slowly decay as a result of the angular momentum losses in the magnetised stellar wind of the star \citep[see][]{2015A&A...574A..39D}.

Given the mass ratio between the star and its companion, $M_{\star}$ / $M_{\rm b}$ $\sim$ 8.7, it is likely
that the TOI-5375 system was formed as a low-mass binary system. 
One possibility is that TOI-5375 and its companion formed from the same molecular cloud that
fragmented into two objects. Another possibility is that TOI-5375 formed a protoplanetary disc
susceptible to gravitational instabilities. 
A relationship between the disc mass and the stellar mass of the form $M_{\rm disc}$ $\propto$  $M_{\star}^{\rm 1.2}$ has been suggested \citep{2011A&A...526A..63A}.
Thus, the relatively low-mass protoplanetary disc around an M dwarf star like TOI-5375
 avoided the formation of two stellar objects of equal mass.
%could only produce very unequal mass ratio objects, as in this case. %it is our case.
Disc fragmentation has been shown to successfully explain the formation of relatively close binaries, and due to
the interactions with their natal disc, significant migration can also occur \citep[e.g.][]{2011ASPC..447...47K}.
Simulations show that discs with a moderate size and mass (100 au, 0.25 $M_{\odot}$) might fragment
around late-K/early-M dwarfs forming mainly brown dwarfs, but also low-mass hydrogen-burning stars in generally close-in orbits 
\citep{2011MNRAS.413.1787S}.
Therefore, another piece of evidence supporting the disc fragmentation might come from the fact that TOI-5375 and its companion are very close and
the star is to align its stellar spin with the binary orbit.
However, we should note that the current spin-orbit alignment of the star may not be primordial, but it could have been established recently because the present timescale for the obliquity damping by tides is only 0.72 Myr (see above).

%----------------------------------------------------------------
\section{Conclusions}\label{conclusions}
%----------------------------------------------------------------

 In this work, we report the discovery and confirmation of a transiting companion in the 
 brown dwarf / very-low-mass star boundary, TOI-5375 b, orbiting an active %young (age $\sim$ 100 Myr)
 M0.5 dwarf with a period of 
 1.72 d. We make use of the available TESS photometric data and high-resolution HARPS-N spectra to model the signals
 present in the data. 
 We derive a rotational period of 1.97 d for the host star and conclude that the orbit of the companion
 decays quickly and that the system should be in the phase in which the star is very close to synchronising its rotation with
 the orbital period of the companion. %brown dwarf.

 TOI-5375 b, with a mass of 77  $\pm$ 8 $M_{\rm J}$ and a radius of 0.99 $\pm$ 0.16 $R_{\rm J}$,
% is one of the most massive companions found around an M dwarf and
 joins the population of massive substellar / low-mass stellar companions in close-in orbits around
 low-mass stars that is currently emerging.
 Further studies of the properties of these objects will help us to explore
 %These planets constitute valuable objects in which to test 
 our models of low-mass companion formation and migration as well
 as to study how the tidal interactions with their host stars affect their evolution.

 We can measure the change in the orbit of TOI-5375 b by computing the associated shift in transit arrival time (transits should occur earlier as the
 companion's orbit decay). We obtain a shift in the transit time of 20.53 seconds in a timespan of 10 yr.
 In our analysis, we have combined data from  four TESS sectors containing a total of  53 transits.
 The precision in the estimated transit-mid time (from the simultaneous analysis of all transits)
 is of the order of  5$\times$10$^{\rm -5}$ d or $\sim$ 4.32 s, while the relative error in the
 transit's depth is of the order of $\sim$0.2\%.
% New TESS observations of TOI-5375 are expected in Sector 60, but even with these observations we will not achieve
% enough timing precision.
 Therefore, while TOI-5375 is a good candidate to observe transit timing variations due to the effects of tidal waves, a long-term monitoring observation from ground-based facilities
 of this star would be needed (unfortunately the star is too faint to be observed by the CHEOPS satellite).

% {\bf cheops?, Borsato et al. 2021, shows that for stars with G $>$ 11 (TOI-5375 has a magnitude G of 13.4) CHEOPS might rise a precision on T0 of the order of 2 minutes, combining
% multiple observations. This is by far larger than the 21.25 seconds that we expect in 10 yr. In addition, the nominal duration of CHEOPS is 3.5 yr. 
% In this time we will espect a transit shift of only 2.6 seconds !!} 
 The TOI-5375 system can also be an interesting target for ongoing and future space missions aimed at the characterisation of the atmospheres of substellar companions like
 the {\it James Webb} Space Telescope \citep{2006SSRv..123..485G} or
 {\it Ariel} \citep{2018ExA....46..135T}.
  We computed the transmission spectroscopy metric (TSM) as well as the emission spectroscopy metric (ESM) following the procedure defined in \citep{2018PASP..130k4401K}.
 The derived TSM value, 0.94, is significantly low, but we caution that TSM values
 were defined for small, rocky planets, so it is unclear whether or not they are still valid
 for a massive brown dwarf like TOI-5375 b or not.
 On the other hand, we obtain an ESM value of 125, which suggests
 that TOI-5375 b is indeed a good target for atmospheric characterisation. 
 
\begin{figure}[htb]
\centering
\includegraphics[scale=0.625]{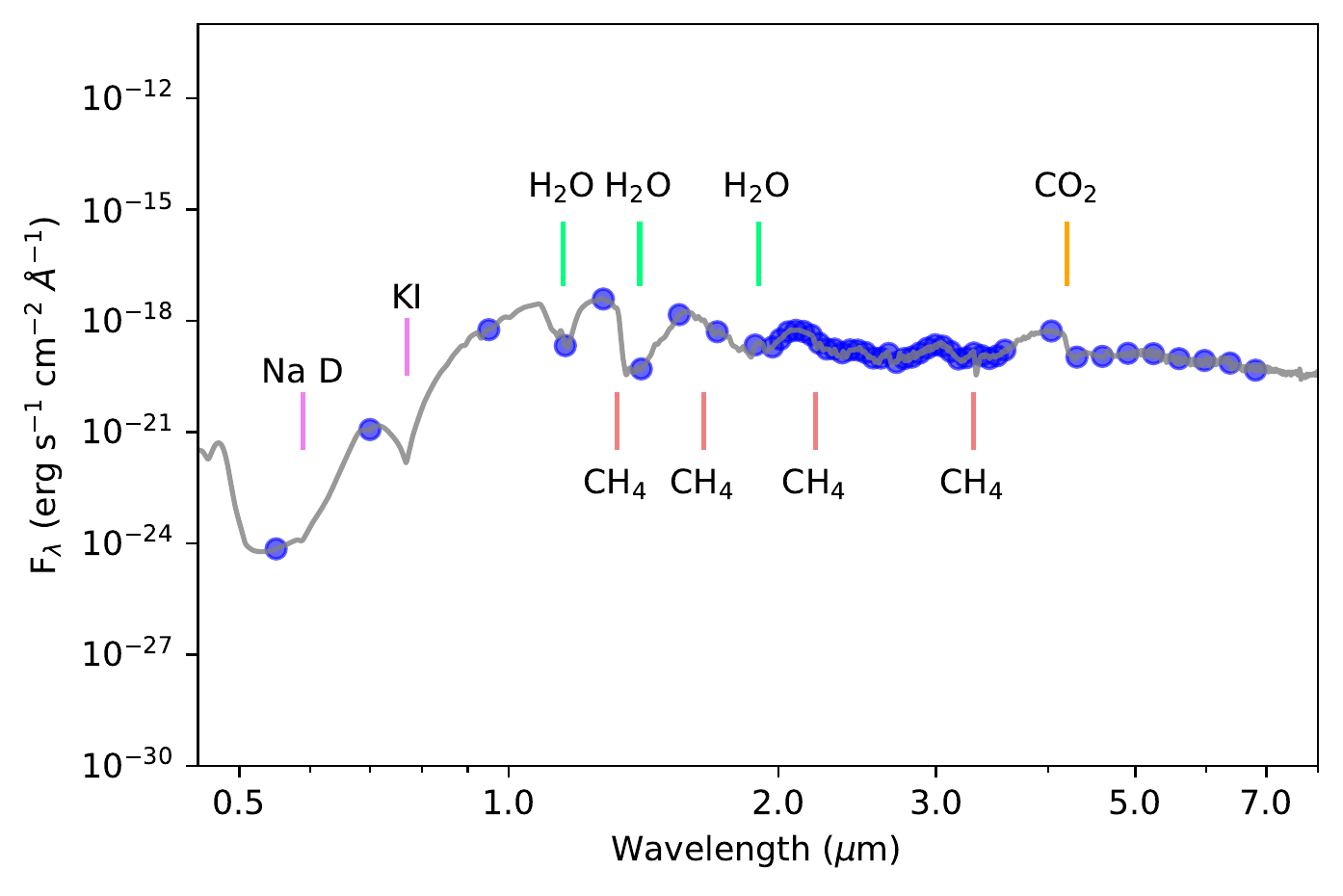} 
\caption{
Simulated observed spectrum by the {\it Ariel} mission of TOI-5375 b generated from a PHOENIX/BT-SETTL atmospheric model.
The grey line shows the spectra degraded to a resolution of R = 50, while the blue circles indicate the expected fluxes at {\it Ariel} wavelengths. 
Some characteristic spectral features of a brown dwarf are indicated \citep[see e.g.][]{2002ApJ...573..394B,2006ApJ...637.1067B,2012ApJ...760..151S}.
}
\label{ariel_spectra}
\end{figure}

 In order to test what we might expect in the future, we generate a synthetic spectrum of TOI-5375 b 
 as observed by {\it Ariel}. For this purpose, we made use of a PHOENIX/BT-SETTL atmospheric model \citep{2012RSPTA.370.2765A} 
 with parameters T$_{\rm eff}$ = 1000 K, $\log g$ = 4.5 cm s$^{\rm -2}$, and [Fe/H] = +0.00 dex. We convolved the original model with a Gaussian filter in order to match 
 the expected resolution for {\it Ariel} Tier 2 observations (R $\sim$ 50 at 1.95 - 3.95 $\mu$m) and computed the expected fluxes at {\it Ariel} wavelengths.
 The synthetic spectrum is shown in Fig.~\ref{ariel_spectra}.
 The comparison of the spectral properties of TOI-5375 b with other brown dwarfs and stars
 will help us to discover whether their atmospheric properties are equal or deviating, which may be a signspot for different formation scenarios.

   We acknowledge that while this manuscript was under the referee review process, an independent analysis of the TOI-5375
  system has been published \citep{2023arXiv230316193L}.
  These authors found similar results to ours. Nevertheless, we have intentionally maintained our analysis completely independent
  of these authors.

% >>>>>>>>>>>>>>>>>>>>>>>>>>>>>>>>>>>>>>>>>>>>>>>>>>>>
% Figure: Example of ARIEL spectra for YO48
% <<<<<<<<<<<<<<<<<<<<<<<<<<<<<<<<<<<<<<<<<<<<<<<<<<<<

%+++++++++++++++++++++++++++++++++++++++++++++++++++++++++++++++++++
\begin{acknowledgements}
%+++++++++++++++++++++++++++++++++++++++++++++++++++++++++++++++++++

  This paper includes data collected by the TESS mission. Funding for the TESS mission is provided by the NASA's Science Mission Directorate.
  Based on observations collected at Copernico 1.82m telescope (Asiago Mount Ekar, Italy)
  INAF - Osservatorio Astronomico di Padova. 
  This work has been supported by the PRIN-INAF 2019 ``Planetary systems at young ages (PLATEA)'' and ASI-INAF agreement n.2018-16-HH.0.  
  J.M. acknowledge support from the \emph{Accordo Attuativo ASI-INAF n. 2021-5-HH.0,
  Partecipazione alla fase B2/C della missione Ariel (ref. G. Micela).}
  We sincerely appreciate the careful reading of the manuscript and the constructive comments of an anonymous referee.

\end{acknowledgements}

%<<<<<<<<<<<<<<<<<<<<<<<<<<<<<<<<<<<<<<<<<<<<<<<<<<<<<<<<<<<<<<<<<<<<
%                      Bibliografia
%>>>>>>>>>>>>>>>>>>>>>>>>>>>>>>>>>>>>>>>>>>>>>>>>>>>>>>>>>>>>>>>>>>>>
%
\bibliographystyle{aa}
\bibliography{yo48_toi5375.bib}

%%<<<<<<<<<<<<<<<<<<<<<<<<<<<<<<<<<<<<<<<<<<<<<<<<<<<<<<<<<<<<<<<<<<<<
%%                      Apendix
%%>>>>>>>>>>>>>>>>>>>>>>>>>>>>>>>>>>>>>>>>>>>>>>>>>>>>>>>>>>>>>>>>>>>>
\begin{appendix}

\section{Results from the joint analysis: TESS and ground-based photometry}\label{app_tess_asiago}

\begin{figure}[htb]
\centering
\includegraphics[scale=0.325]{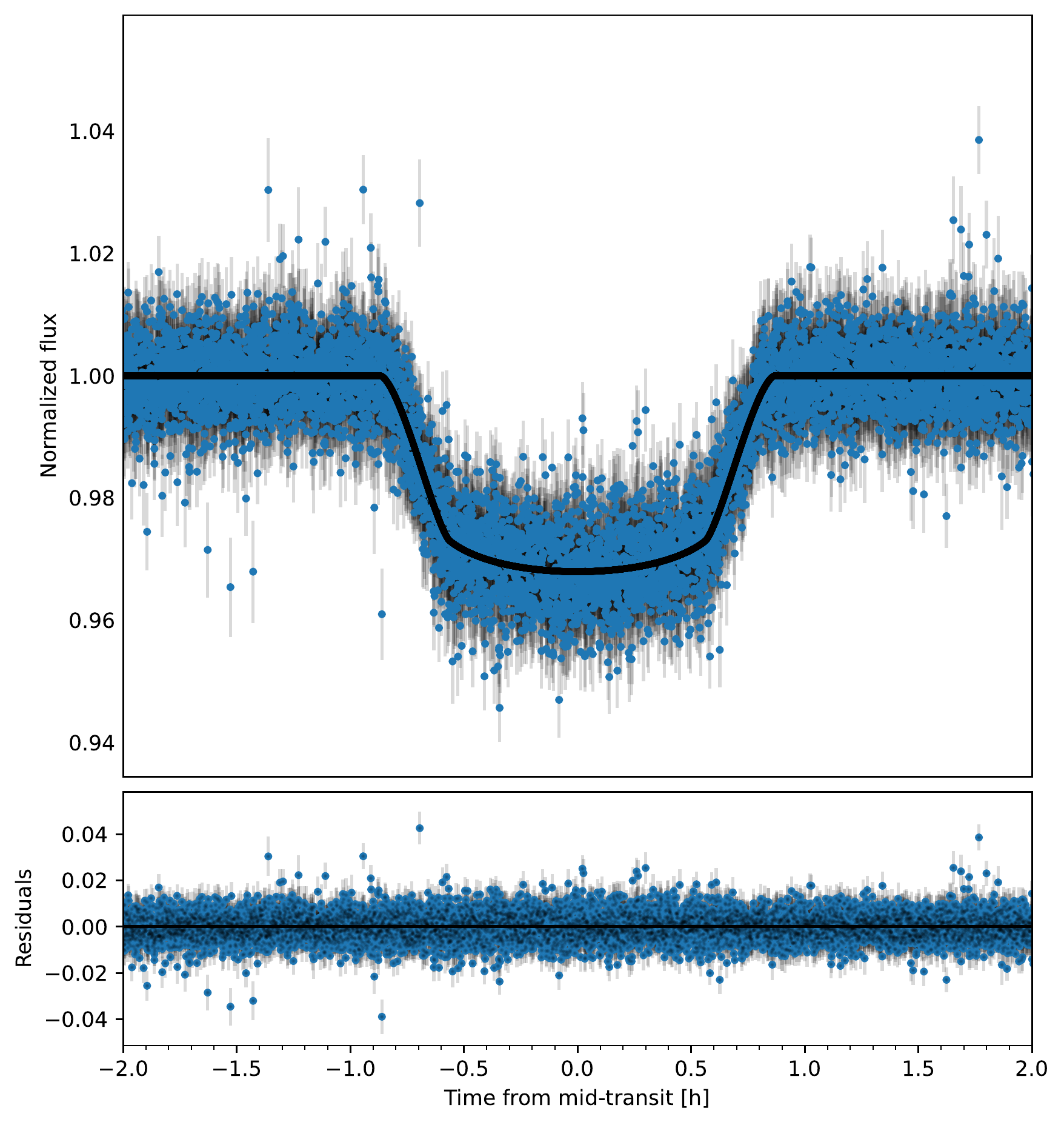}
\caption{
Photometric modelling of TOI-5375 (TESS and Asiago photometry). Top: Folded transits in the light curve after the subtraction of the stellar activity signal.
The black line shows the best model fit of the transits. Bottom: Residuals of the light curve after the subtraction of the
companion transit model.}
\label{joint_ph_fit}
\end{figure}

{ Table~\ref{padova_fit} provides the best-fit values obtained for the simultaneous fit of the TESS SAP light curve and the Asiago observations,
as described in Sect.~\ref{tessasiago}. The best transit model fit is shown in Figure~\ref{joint_ph_fit}}.

\begin{table*}[t]
\centering
\caption{ Best-fit values obtained for the simultaneous fit of the TESS SAP light curve and the Asiago observations.} % as described in Sect.~\ref{tessasiago}.}
\label{padova_fit}
\begin{tabular}{llr}
\hline
%\hline
\noalign{\smallskip}
Parameter                    & Description                              & Best-fit value \\
\noalign{\smallskip}
\hline
\noalign{\smallskip}
$\sigma_{\rm ground,1}$       & Jitter Asiago dataset 1                          &             0.0040$_{-0.0002}^{+0.0002}$   \\
$R_{\rm b}/R_{\star}$,Ground 1& Companion-to-stellar-radius ratio, Asiago data 1 &             0.191$_{-0.002}^{+0.002}$   \\
$\sigma_{\rm ground,2}$       & Jitter Asiago dataset 2                          &             0.0040$_{-0.0003}^{+0.0003}$   \\
$R_{\rm b}/R_{\star}$,Ground 2& Companion-to-stellar-radius ratio, Asiago data 2 &             0.188$_{-0.002}^{+0.002}$   \\
\hline
$\sigma_{\rm 0}$              & TESS jitter term sector 40               &              0.76$_{-0.51}^{+0.56}$  \\
$\sigma_{\rm 1}$              & TESS jitter term sector 50               &              6.11$_{-0.12}^{+0.12}$  \\
$\sigma_{\rm 2}$              & TESS jitter term sector 53               &              0.86$_{-0.54}^{+0.48}$  \\
$\sigma_{\rm 3}$              & TESS jitter term sector 60               &              2.07$_{-0.29}^{+0.26}$  \\
\hline
$P_{\rm b}$ (d)               & Period                                   &              1.721556$_{-0.000001 }^{+0.000001}$       \\
$b$                           & Impact parameter                         &              0.43$_{-0.03}^{+0.02}$        \\
$T_{\rm 0}$ (BJD, d)          & Time of inferior conjunction             &              2459580.736510$_{-0.000096}^{+0.000095}$  \\
$R_{\rm b}/R_{\star}$,TESS    & Companion-to-stellar-radius ratio, TESS  &              0.170$_{-0.001}^{+0.001}$        \\
$\rho_{\star}$ ($\rho_{\odot}$) & Stellar density                          &            2.57$_{-0.09}^{+0.09}$\\
$c_{\rm 1}$, TESS             & Limb-darkening coefficient TESS          &              0.21$_{-0.05}^{+0.06}$  \\
$c_{\rm 2}$, TESS             & Limb-darkening coefficient TESS          &              0.35$_{-0.06}^{+0.07}$  \\
$c_{\rm 1}$, Ground           & Limb-darkening coefficient Asiago        &              0.42$_{-0.07}^{+0.08}$  \\
$c_{\rm 2}$, Ground           & Limb-darkening coefficient Asiago        &              0.35$_{-0.03}^{+0.04}$  \\
$d$, TESS                     & Dilution factor TESS                     &              0.0756$_{-0.0005}^{+0.0005}$  \\
\hline
$a_{\rm b}/R_{\star}$         & Semi-major-axis-to-stellar-radius-ratio  &              8.28$_{-0.09}^{+0.09}$     \\
$inc$ (degrees)               & Orbital inclination                      &              87.00$_{-0.21}^{+0.22}$    \\
$R_{\rm b}$  ($R_{\rm J}$)    & Companion Radius                         &              1.05$_{-0.03}^{+0.03}$     \\
$a_{\rm b}$ (au)              & Companion Semi-major axis                &              0.024$_{-0.008}^{+0.001}$  \\
\hline
\noalign{\smallskip}
\end{tabular}
\end{table*}

% >>>>>>>>>>>>>>>>>>>>>>>>>>>>>>>>>>>>>>>>>>>>>>>>>>>>
% Figure: Photometry best fit
% <<<<<<<<<<<<<<<<<<<<<<<<<<<<<<<<<<<<<<<<<<<<<<<<<<<<

% --------------------------------------------
\section{Activity indexes}
\label{app_activity}
% --------------------------------------------

\begin{figure}[htb]
\includegraphics[scale=0.60]{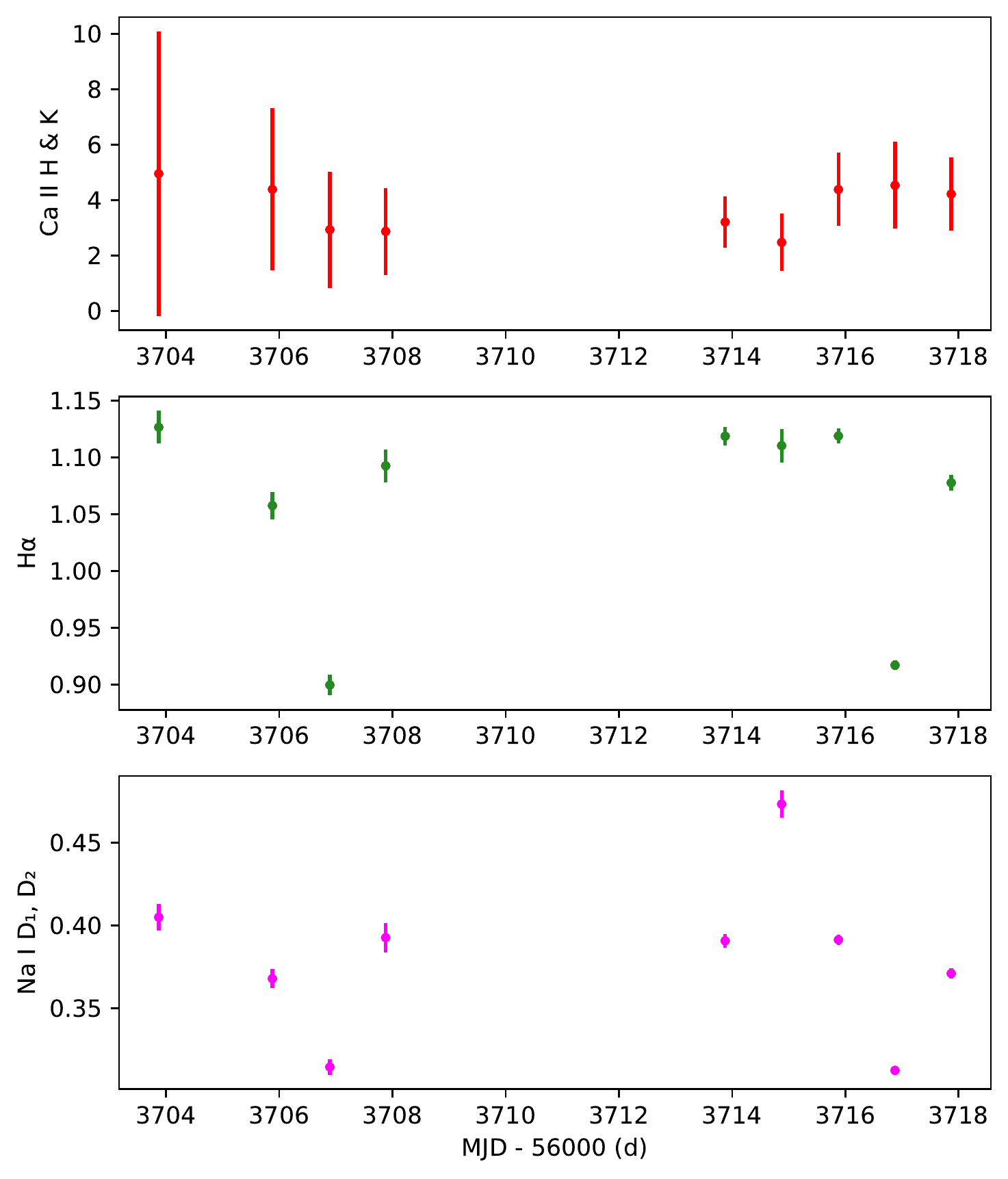}
\caption{
From top to bottom: Ca~{\sc ii} H \& K index, H$\alpha$ index, and Na~{\sc i} D$_{\rm 1}$, D$_{\rm 2}$ index time series for TOI-5375.
}
\label{act_time}
\end{figure}

\begin{figure}[!htb]
\centering
\includegraphics[scale=0.65]{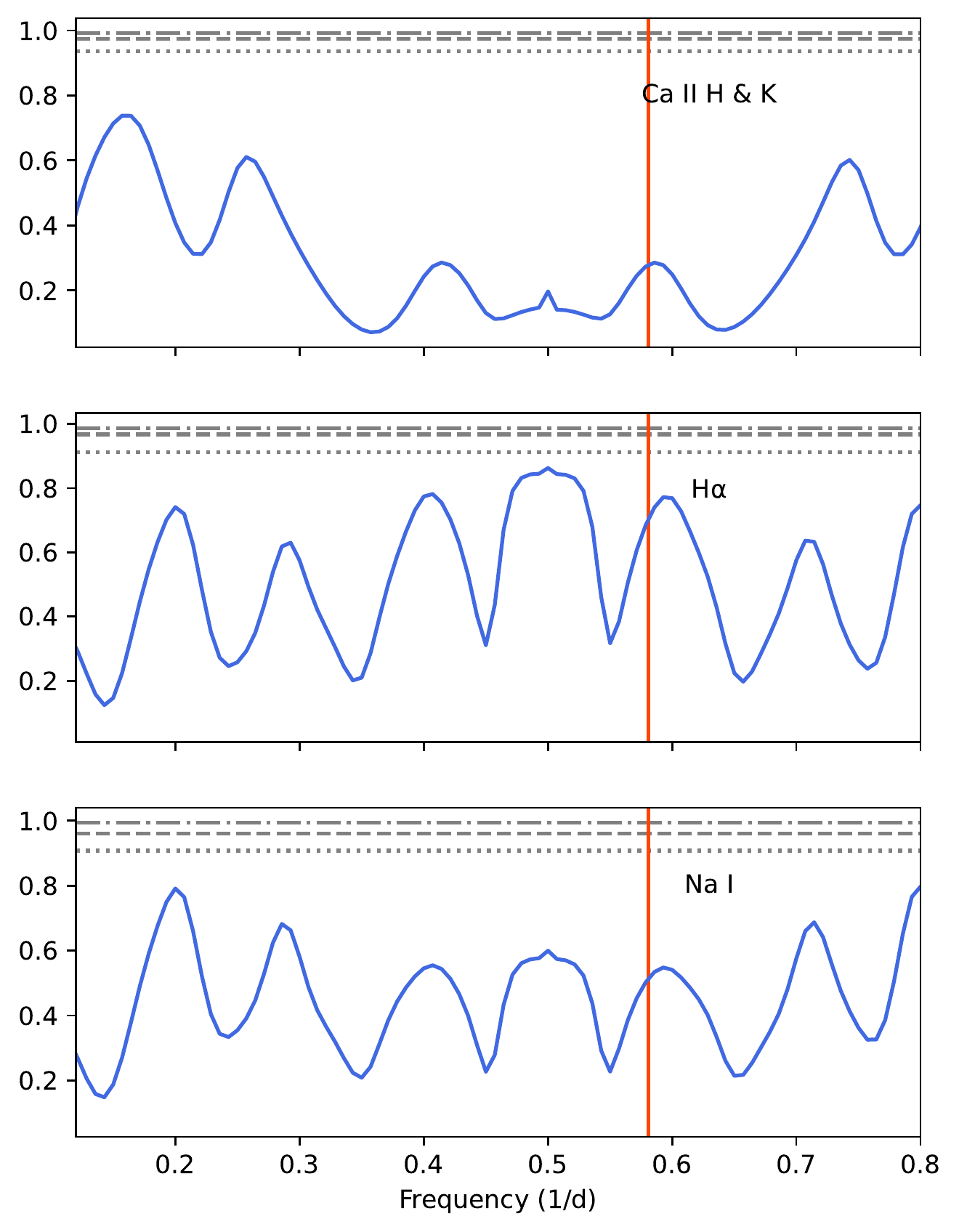}
\caption{
From top to bottom: GLS periodogram of the Ca~{\sc ii} H \& K, H$\alpha$, and Na~{\sc i} activity indexes.
The vertical line indicates the period of TOI-5375 b at 1.72 d.
Values corresponding to a FAP of 10\%, 1\%, and 0.1\% are shown with
horizontal grey lines.
}
\label{activity_indices}
\end{figure}

{ Figure~\ref{act_time} shows the temporal variation of the different activity indexes
measured in this work, while the periodogram analysis is given in Fig.~\ref{activity_indices}.
Table~\ref{activity_indexes} provides the measured values with their corresponding uncertainties.
}

\begin{table*}[!htb]
\centering
\caption{ Activity indexes measured for TOI-5375.}
\label{activity_indexes}
\begin{tabular}{lccc}
\hline\noalign{\smallskip}
 MJD (d)        &        Ca~{\sc ii} H \& K     &               H$\alpha$       & Na~{\sc i}  \\
\hline
59703.871736    &       4.9593  $\pm$   5.1386  &       1.1264  $\pm$   0.0145  &       0.4049  $\pm$   0.0082  \\
59705.879259    &       4.3944  $\pm$   2.9213  &       1.0575  $\pm$   0.0122  &       0.3679  $\pm$   0.0058  \\
59706.894329    &       2.9384  $\pm$   2.1017  &       0.8998  $\pm$   0.0093  &       0.3147  $\pm$   0.0048  \\
59707.880509    &       2.8771  $\pm$   1.5704  &       1.0925  $\pm$   0.0144  &       0.3926  $\pm$   0.0090  \\
59713.876840    &       3.2140  $\pm$   0.9153  &       1.1185  $\pm$   0.0080  &       0.3907  $\pm$   0.0041  \\
59714.874595    &       2.4812  $\pm$   1.0379  &       1.1102  $\pm$   0.0148  &       0.4731  $\pm$   0.0083  \\
59715.877141    &       4.3896  $\pm$   1.3210  &       1.1187  $\pm$   0.0067  &       0.3913  $\pm$   0.0032  \\
59716.877118    &       4.5392  $\pm$   1.5677  &       0.9172  $\pm$   0.0042  &       0.3126  $\pm$   0.0023  \\
59717.869826    &       4.2266  $\pm$   1.3254  &       1.0776  $\pm$   0.0071  &       0.3711  $\pm$   0.0032  \\
\hline
\end{tabular}
\end{table*}

% >>>>>>>>>>>>>>>>>>>>>>>>>>>>>>>>>>>>>>>>>>>>>>>>>>>>
% Figure: Activity indexes time series 
% <<<<<<<<<<<<<<<<<<<<<<<<<<<<<<<<<<<<<<<<<<<<<<<<<<<<

%% >>>>>>>>>>>>>>>>>>>>>>>>>>>>>>>>>>>>>>>>>>>>>>>>>>>>
%% Figure 2: Indices de actividad
%% <<<<<<<<<<<<<<<<<<<<<<<<<<<<<<<<<<<<<<<<<<<<<<<<<<<<

\end{appendix}
% +++ 
\end{document}